\documentclass{article}

\usepackage{PRIMEarxiv}

\usepackage[utf8]{inputenc} 
\usepackage[T1]{fontenc}    
\usepackage{hyperref}       
\usepackage{url}            
\usepackage{booktabs}       
\usepackage{amsfonts}       
\usepackage{nicefrac}       
\usepackage{microtype}      
\usepackage{lipsum}
\usepackage{fancyhdr}       
\usepackage{graphicx}       
\graphicspath{{media/}}     

\title{Behavioral changes during the pandemic worsened income diversity of urban encounters}

\author{
  Takahiro Yabe \\
  Institute for Data, Systems, and Society, \\
  MIT \\
  \texttt{tyabe@mit.edu} \\
   \And
  Bernardo Garc\'ia Bulle Bueno \\
  Institute for Data, Systems, and Society, \\
  MIT \\
  \texttt{bernard0@mit.edu} \\
   \AND
   Xiaowen Dong \\
   Department of Engineering Science, \\
   University of Oxford \\ 
   \texttt{xdong@robots.ox.ac.uk
} \\
   \And
   Alex `Sandy' Pentland \\
   Institute for Data, Systems, and Society \& Media Lab, \\
   MIT \\ 
   \texttt{sandy@media.mit.edu} \\
   \And
   Esteban Moro \\
   Institute for Data, Systems, and Society \& Media Lab,\\
   MIT \\ 
   Departamento de Matemáticas \& GISC, \\ Universidad Carlos III de Madrid \\
   \texttt{emoro@mit.edu} \\
}

\begin{document}
\maketitle

\begin{abstract}
Diversity of physical encounters and social interactions in urban environments are known to spur economic productivity and innovation in cities, while also to foster social capital and resilience of communities. 
However, mobility restrictions during the pandemic have forced people to substantially reduce urban physical encounters, raising questions on the social implications of such behavioral changes. 
In this paper, we study how the income diversity of urban encounters have changed during different periods throughout the pandemic, using a large-scale, privacy-enhanced mobility dataset of more than one million anonymized mobile phone users in four large US cities, collected across three years spanning before and during the pandemic. 
We find that the diversity of urban encounters have substantially decreased (by $15$\% to $30$\%) during the pandemic and has persisted through late 2021, even though aggregated mobility metrics have recovered to pre-pandemic levels. 
Counterfactual analyses show that while the reduction of outside activities (higher rates of staying at home) was a major factor that contributed to decreased diversity in the early stages of the pandemic, behavioral changes including lower willingness to explore new places and changes in visitation preferences further worsened the long-term diversity of encounters. 
Our findings suggest that the pandemic could have long-lasting negative effects on urban income diversity, and provide implications for managing the trade-off between the stringency of COVID-19 policies and the diversity of urban encounters as we move beyond the pandemic. \end{abstract}


\section*{Introduction}

Cities are the central drivers of economic productivity and innovation owing to its capacity to foster dense social connections through physical encounters \cite{balland2020complex,eagle2009inferring, pan2013urban}. 
Among the various characteristics of social connections and network structures, empirical studies have shown that the diversity of networks are significant predictors of economic growth and recovery \cite{eagle2010network,yuan2022factors}.
Moreover, integrated community networks and the inherent social capital have been shown to be crucial for resilience to shocks such as natural hazards \cite{aldrich2015social,yabe2019mobile}. The lack of community support could lead to inequitable access to urban amenities and services, ultimately affecting social, economic, and health outcomes of people living in urban areas \cite{abramson1995changing}.
However, in addition to rising inequality and segregation \cite{van2015inequality}, the COVID-19 pandemic and the consequential countermeasures including mobility restrictions have posed significant challenges for maintaining both the quantity and quality of such physical encounters in cities.

Large scale location data (e.g., CDRs\cite{blondel2015survey,gonzalez2008understanding}, credit card data \cite{dong2020segregated}, and social media \cite{wang2018urban}) have been used to understand the nature of physical encounters of people in cities \cite{stopczynski2018physical,sun2013understanding}. 
Recently, such mobility datasets have been used to measure and understand the diversity of encounters in cities, by measuring the homophily of co-locations at points-of-interest (POIs) during daily routines.
A study using mobile phone data in 10 American cites revealed that peoples' mobility behavior, as opposed to their residential locations, account for $55$\% of urban segregation (which is an inverse metric of diversity) \cite{moro2021mobility}. Another study using Foursquare data revealed that people mostly visit places in their own socioeconomic class, occasionally visiting venues from higher classes \cite{hilman2022socioeconomic}. 
Compared to analysis limited to quantifying static residential segregation measures using census data \cite{browning2017socioeconomic}, such studies based on mobility data have provided a more comprehensive understanding of income segregation in urban environments. 

Various non-pharmaceutical interventions imposed by governments and agencies have enforced people to substantially change their lifestyles and reduce daily activity patterns, reducing trips to urban amenities such as restaurants, bars, and entertainment establishments \cite{sevtsuk2021impact}.
While such behavior changes have had significant impacts on our physical health and activities \cite{tison2020worldwide,hunter2021effect} and mental wellbeing \cite{campion2020addressing}, studies have also suggested their impact on social encounters in urban environments, affecting the social fabric of cities we live in \cite{yoon2021covid}. The aftermath of the pandemic has brought also significant changes in behavior in our cities, including less use of public transportation \cite{de2022overview}, more hours working from home \cite{pewwork}, and higher usage of online food and goods delivery services \cite{janssen2021changes}.
Despite the rich literature on the mobility dynamics and its impact on disease spread using location data collected via mobile phones \cite{oliver2020mobile,aleta2020modelling}, little is understood on how much longitudinal effects the pandemic has had on the quantity and quality of our encounters in urban environments. 
Measuring the dynamics and potential causes of fluctuations in the diversity of urban encounters across different periods of the pandemic could be valuable in understanding the long-term impacts of the pandemic on cities, and for developing resilient policies to better prepare for future outbreaks. 


\begin{figure}[!t]
\centering
\includegraphics[width=\linewidth]{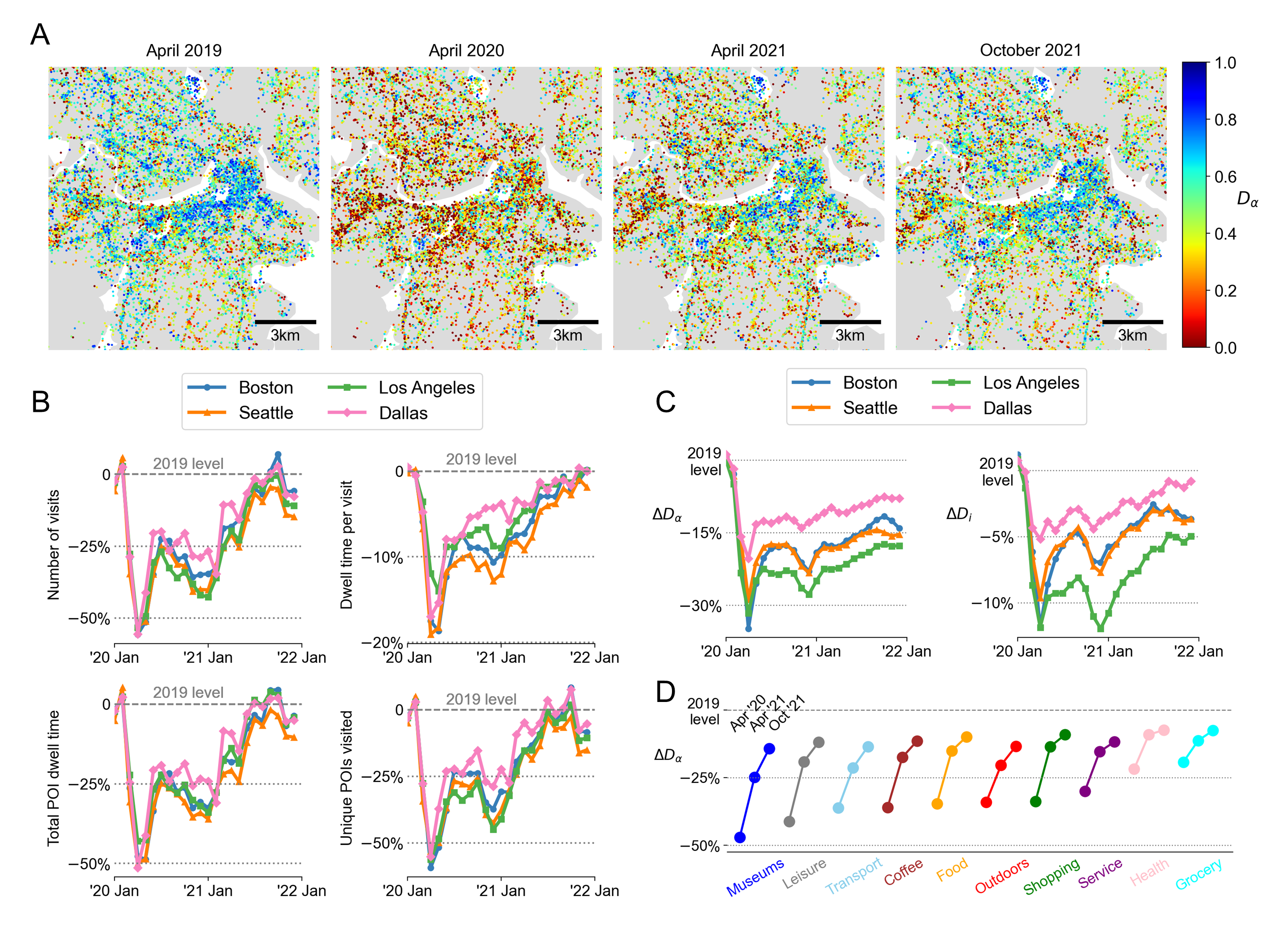}
\caption{\textbf{Diversity of urban encounters have decreased during COVID-19.} A) Income diversity of encounters in places in the Boston and Cambridge area decreased during the pandemic. Diversity gradually recovers with reopening, albeit not fully compared to pre-pandemic levels. B) Aggregate mobility metrics, such as the daily number of visits per individual, daily amount of time spent at POIs, and number of visited unique POIs have all returned back to pre-pandemic levels by late 2021. C) Despite the recovery in mobility statistics, the diversity in encounters experienced at places and by individuals have decreased and have not recovered back to pre-pandemic levels. D) Income diversity decreased in all major place categories both in the short-term (e.g., April 2020) and long-term (e.g., October 2021) in all cities. Grocery stores consistently experienced the least effects of the pandemic while museums, leisure, transport, and coffee places saw the largest decrease.}
\label{fig:fig1}
\end{figure}

\section*{Results}

Using a large and longitudinal dataset of individual GPS location records in four major metropolitan areas in the US across more than three years, we analyze how experienced income diversity of urban encounters have changed during different periods of the COVID-19 pandemic. Specifically, we analyze the dynamics of income diversity of encounters at the level of individual places (points-of-interest; POIs) and individual users in cities. We seek to identify behavioral changes that were at the cause of such long-term changes, and we further unravel the sociodemographic, economic, and behavioral characteristics that explain the spatial heterogeneity in decreased diversity. Mobility data was provided by Spectus, who supplied anonymized, privacy-enhanced, and high-resolution mobile location pings for more than 1 million devices across four U.S. census core-based statistical areas (CBSAs) (Supplementary Table S2). All devices within the study opted-in to anonymized data collection for research purposes under a GDPR and CCPA compliant framework. Post-stratification techniques were implemented to ensure the representativeness of the data across regions and income levels (Supplementary Note 2 and Supplementary Figure S6). Our second data source is a collection of 433K verified places across four CBSAs, obtained via the Foursquare API (Supplementary Table S1). Robustness of the results on income diversity against the choice of places dataset was checked using the ReferenceUSA Business Historical Data \cite{GW2P3G2014} (Supplementary Note 1.3 and Supplementary Figure S2).  

To analyze the income diversity of urban encounters, each anonymized individual user in the dataset was assigned a socio-economic status (SES) proxy, estimated from their home census block group (CBG) (Supplementary Note 1.4 and Supplementary Figure S3). The approximate home area of each individual user was estimated by Spectus at the granularity of CBGs using their most common location during the nighttime, between 10 p.m. and 6 a.m. every week. Individuals were then categorized into four equally sized SES quantiles according to the median household income of their home CBG. The results on decreased income diversity were robust against the number of quantile categories used (Supplementary Note 1.4 and Supplementary Figure S4).  
Only users who were observed more than 300 minutes each day were used for the analysis to remove users with substantial missing data. Stays (stops) longer than 10 minutes and shorter than 4 hours were then extracted from the dataset, and each stay was spatially matched with the closest place locations within 100 meters to infer stays at specific POIs. The results on income diversity were robust against the choice of data filtering parameters (Supplementary Note 1.5 and Supplementary Figure S5) and spatial threshold parameters for visit attribution (Supplementary Note 1.2 and Supplementary Figure S1). 

Given the estimated SES quantiles of individual users and the visited POIs, we measured the income diversity at each place $\alpha$ (denoted as $D_{\alpha}$) and experienced by each individual $i$ (denoted as $D_{i}$). $D_{\alpha}$ measures the evenness of the time spent by people from different income quantiles at each place, and $D_{i}$ measures the evenness of time spent with people from different income quantiles for each individual (see Methods and Supplementary Notes 3.1 and 3.2). For places, $D_{\alpha}=1$ when the place is fully diverse, with 25\% of time spent by people from each of the four income quantiles, and $D_{\alpha}=0$ when the place is visited by members of only a single income quantile. Similarly, to calculate the diversity of individuals $D_i$, we measure the exposure of the individual $i$ to each income quantile $q$ across all the places $\alpha$ the individual has visited. The robustness of the results to the choice of diversity metric was tested (Supplementary Note 3.3 and Figure S12). The diversity measures were computed for each 2-month moving window to ensure sufficient number of visits to POIs, and were de-seasonalized using monthly trends observed in 2019. The panels in Figure \ref{fig:fig1}A show how income diversity at places around the Boston and Cambridge area substantially decreased during the first wave of the pandemic. The diversity of encounters gradually recovers, however, not fully even after more than 1 and a half years from the lockdown, in October 2021. Similar patterns can be observed in all three other cities in the study (Supplementary Figure S7). The maps highlight the significant spatial heterogeneity of income diversity (e.g., Back Bay area is more diverse compared to the suburban areas), which is further investigated in the later sections.

\subsection*{Diversity of urban encounters have decreased during the pandemic}
The pandemic substantially changed people's mobility patterns in the early waves, as documented in previous studies using mobility data (e.g., \cite{chang2021mobility}). However, several individual mobility metrics indicate that individual based mobility patterns have returned back to pre-pandemic levels by late 2021. Figure \ref{fig:fig1}B shows monthly average values of several individual mobility metrics across the two years in 2020 and 2021. Mobility metrics, more specifically the daily number of visits per individual, daily amount of time spent at POIs per individual, average dwell time spent per visit, and number of visited unique POIs per individual, have all returned back to pre-pandemic levels (annotated by horizontal dashed lines) by late 2021 in all four CBSAs. The drop in the rate of visits to POIs as well as the duration of visits to POIs during the earlier stages of the pandemic agree with the findings in previous studies \cite{lucchini2021living}, however our analysis extends the analysis to two years into the pandemic and confirms how activity patterns have recovered back to pre-pandemic levels by October 2021. The mobility data confirms that people have resumed spending time outside their homes and visiting different POIs, similar to before the pandemic.

Given the recovery of aggregate mobility metrics, one could expect the income diversity of encounters to also return back to pre-pandemic levels by late 2021. However, as shown in Figure \ref{fig:fig1}C, income diversity experienced at places and by individuals are consistently lower than the pre-pandemic levels for all four cities even after 2 years into the pandemic. Absolute values of $D_{\alpha}$ and $D_{i}$ are shown in the Supplementary Figure S10. Cities experience the most decrease of diversity in April 2020, $30$\% lower than pre-pandemic levels during the lockdown. A second peak in the loss of diversity is observed in late 2020, which corresponds to the increase in cases due to the first SARS-CoV-2 variant. Despite the recovery of individual mobility metrics, income diversity of encounters is still around $10$\% less than pre-pandemic levels even by late 2021. 
The decrease in income diversity was robust to the choice of diversity metrics, such as the entropy of income quantiles for encounters at places and for individuals (Supplementary Note 3.3 and Supplementary Figures S11 and S12). 

Dissecting the place-based diversity results into POI categories, we further observe that diversity in places in Boston decreased in all POI categories both on the short-term (e.g., April 2020) and long-term (e.g., October 2021) in Figure \ref{fig:fig1}D and Supplementary Figure S9. Especially, `Museums', `Leisure', `Transportation', and `Coffee' places had the largest decrease in diversity, while `Grocery' places consistently experienced the least effects of the pandemic. This agrees with the fact that we observe the number of visits to follow similar patterns in Supplementary Figure S8, where we see a decrease during the early stages of the pandemic and gradual recovery by late 2021 for all POI categories, with the exception of grocery stores, which experienced no reduction in the number of visits even during the first waves. This suggests that the reduction in the number of visits indeed is one of the factors that cause the decrease in diversity of encounters. In the following section, we employ a counterfactual analysis approach to further understand why the diversity of encounters have consistently decreased during the pandemic.

\begin{figure}[!t]
\centering
\includegraphics[width=\linewidth]{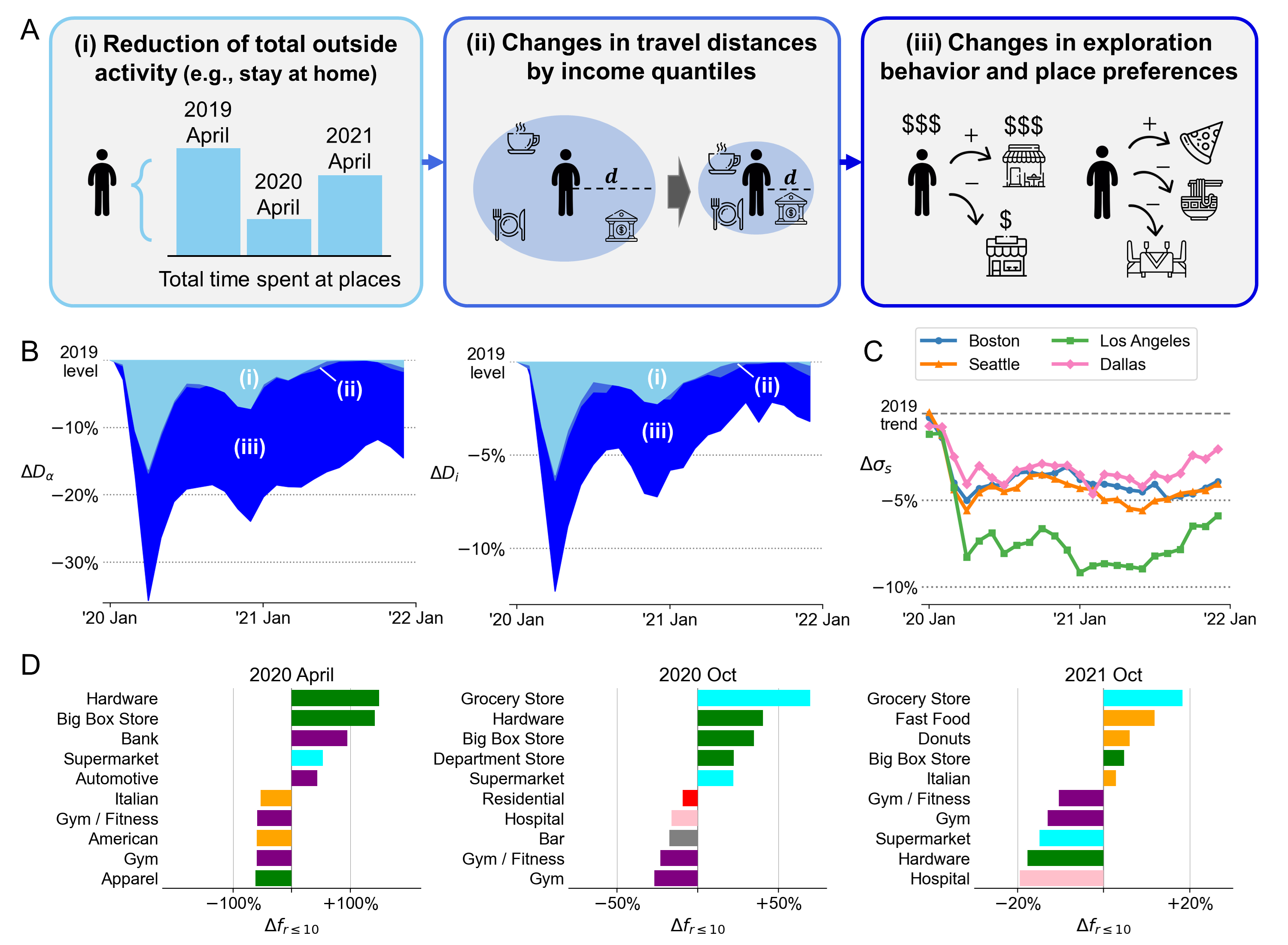}
\caption{\textbf{Behavioral changes worsened income diversity in cities.} A) Three hierarchical levels of behavioral changes were simulated to understand why experienced income diversity decreased: (i) reduction in total outside activity by income groups, (ii) changes in traveled distances by income groups, and (iii) microscopic changes in place preferences, including exploration behavior and place sub-categories. B) Decrease in diversity of encounters for places and individuals decomposed into the three behavioral factors for Boston. Counterfactual simulations show that reduction in total activities (i) in the short-term, and mostly changes in exploration and place preferences (iii) in the long-term, were the major factors that decreased diversity. 
C) Social exploration decreased during the pandemic compared to 2019 trends in all four cities. 
D) POI subcategories which were more (and less) visited in different periods during the pandemic.
}
\label{fig:fig2}
\end{figure}


\subsection*{Behavioral changes worsened income diversity in cities}

To investigate the behavioral factors that led to the consistent decrease in income diversity experienced at places and by individuals, we consider three possible hierarchical levels of changes in the behavior of individuals due to the pandemic. As illustrated in \ref{fig:fig2}A, the pandemic led, especially during the beginning of the pandemic, to a (i) reduction in the total amount of time spent at places outside homes and workplaces. Moreover, due to stay-at-home orders and also reluctance towards long-distance trips compared to before, we also consider (ii) changes in travel distances for each income quantile. In addition, since some type of activity categories were particularly affected by social-distancing policies, we also consider changes in visits to major activity categories and traveled distances for each income quantile, shown in the Supplementary Note 4.2 and Supplementary Figure S16. Finally, we also consider the possibility of (iii) microscopic changes in place preferences, including changes in exploration behavior and visitation patterns across place sub-categories. 

To disentangle the relative weights of these behavioral changes, we created different counterfactual mobility datasets. For example, to estimate the effects of reduction of total activity time on the loss of diversity, we created a counterfactual mobility dataset that contains the same total visit duration at places during the pandemic (e.g., 2020 April), by randomly down-sampling visits from pre-pandemic data observed on the same month (e.g., 2019 April) (see Methods and Supplementary Note 4.1). The resulting generated counterfactual data retains the behavioral mobility patterns observed in 2019, but includes the effects of reduced activity during the pandemic. By comparing the place and individual-based diversity measures computed from the actual and the counterfactual mobility datasets, we are able to delineate the effects of activity reduction on the decrease in diversity. Similarly to measure the effects of (ii) changes in traveled distances by income quantiles, we extended the previous counterfactual to have the same total visit duration by distance ranges for each income quantile (see Methods and Supplementary Note 4).

Figure \ref{fig:fig2}B shows the decreased diversity experienced at places and by individuals decomposed into the three behavioral factors (full results shown in Supplementary Figure S17). The counterfactual simulations show that (i) reduction in total activities caused around 50\% of the decrease in diversity during the first pandemic wave, however, decreases to almost 2\% by late 2021 when mobility metrics have recovered back to normal, as shown in Figure \ref{fig:fig1}B. 
Although we observe different rates of dwell time decrease and recovery across income quantiles where the richer populations disproportionately reduce dwell times at places than poorer populations (Supplementary Figure S14b), the overall diversity measures are not affected since the relative mixing of population groups across income groups are consistent (Supplementary Note 4.2 and Supplementary Figure S15). Changes in distance distributions, where people prefer trips to closer places during the pandemic (Supplementary Figure S14c) has slight negative effects on the income diversity of encounters (Supplementary Note 4.2). Surprisingly, changes in dwell time duration at major activity categories had no effects on the income diversity metrics (Supplementary Note 4.2 and Supplementary Figure S16). 


Heterogeneity in activity reduction rates across income quantiles and changes in traveled distances explain around $55$\% of the decreased diversity during the first wave of the pandemic, however, the remaining $45$\% is due to more microscopic, place-based preference changes.
These effects become the single dominant factor in the later stages of the pandemic. To identify the changes in the mobility behavior during the pandemic, we fit the social exploration and preferential return (Social-EPR) model \cite{moro2021mobility,song2010modelling} to the data for each time period and assess the model parameters (see Supplementary Note 4.3). 
Among the parameters of the social-EPR model, the parameter which changed the most between before and during the pandemic was the social exploration parameter $\sigma_s$, as shown in Figure \ref{fig:fig2}C and Supplementary Figure S18. Social exploration $\sigma_s$ measures the probability of an individual to visit a place where their income group is not the majority income quantile group when they decide to explore a new place.
During the pandemic, people's willingness to socially explore substantially decreased compared to the 2019 levels (horizontal dashed line) in all four cities, leading to less experienced diversity.

Furthermore, we observe changes in place level preferences across POI sub-categories. Sub-category popularity $f_k$ is measured by computing the probability that a POI sub-category is included in an individual's top $k$ most frequently visited places. Figure \ref{fig:fig2}D and Supplementary Figure S19 shows the POI sub-categories which were more (and less) visited in different periods during the pandemic compared to 2019 levels. Hardware stores, big box stores, grocery stores were POI sub-categories which gained popularity during the pandemic, and gyms, movie theaters, American food places were subcategories which were less visited frequently. 
Taken together with the results that controlling by major activity categories did not explain additional decreased diversity to scenario (ii) as shown in Supplementary Note 4.2 and Supplementary Figure S13, this result shows that people have not changed their proportion of time spent for major activity categories, but have changed which specific types of places they visit within each major activity (e.g., less time at American restaurants, but more time at fast food and donut stores). 
To summarize, not only reduction in activity, but also microscopic behavioral changes especially during the later stages of the pandemic, including less exploration and shift in preferences, led to decreased diversity in urban encounters. 

\begin{figure}[!t]
\centering
\includegraphics[width=\linewidth]{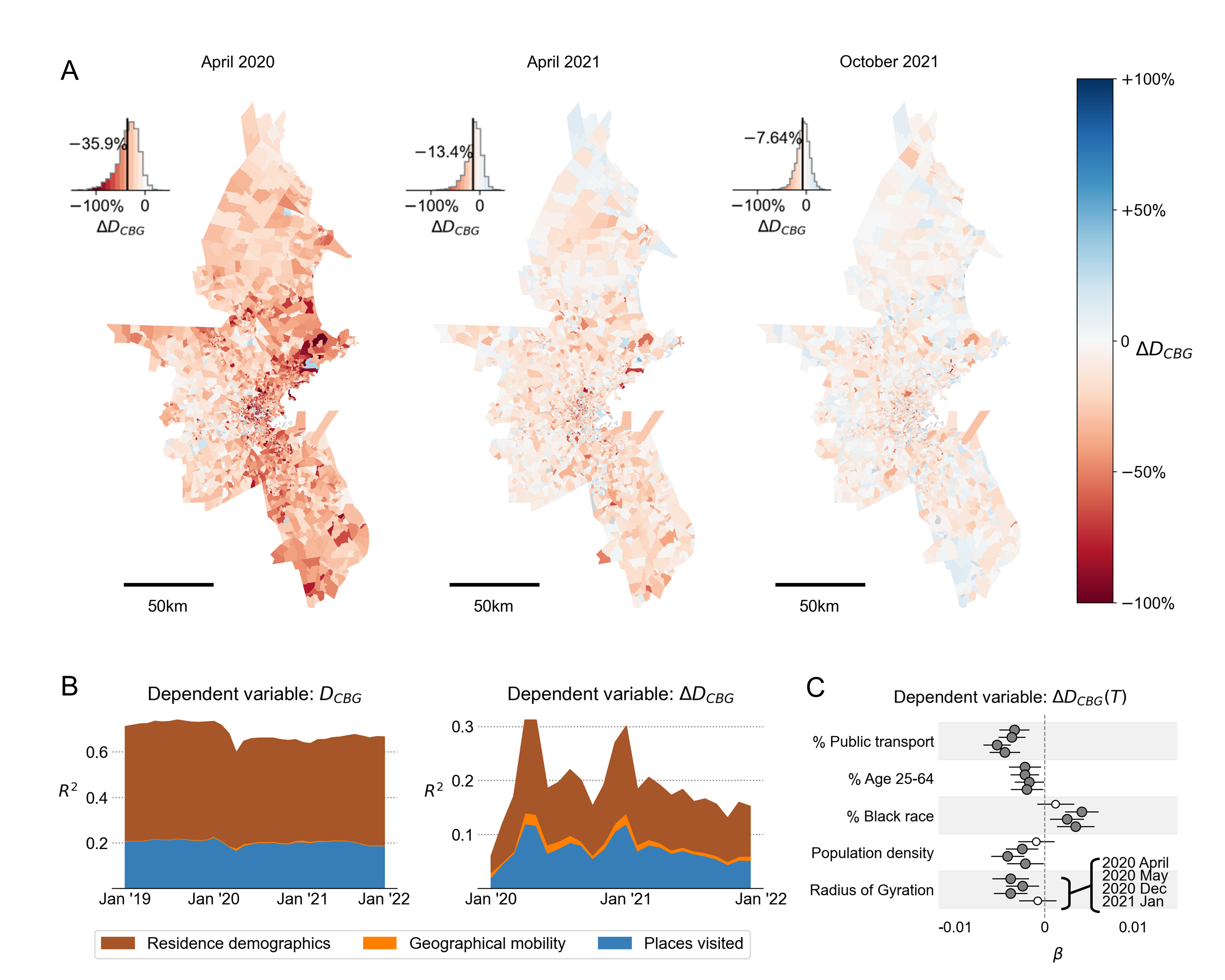}
\caption{\textbf{Spatial and socioeconomic heterogeneity in decreased diversity.} A) Change in mean income diversity on the census block group levels in the Boston CBSA for three different time periods (April 2020, April 2021, and October 2021), compared with the corresponding months in 2019. Insets show histogram of difference in income diversity $\Delta D_{CBG}$. 
B) Adjusted $R^2$ of regression models for $D_{CBG}$ and $\Delta D_{CBG}$, respectively, across different time periods. The three groups of variables (places visited, geographical mobility, residence and demographics) explain around 55\% to 70\% of the variance in income diversity. However, the same variables explain much lower variance of $\Delta D_{CBG}$, indicating that regions became less diverse homogeneously.
C) Regression coefficients that explain the heterogeneity in $\Delta D_{CBG}$ for the four different time periods where the $R^2$ was relatively higher. Filled variables are statistically significant at the $P<0.05$ threshold. 
}
\label{fig:fig3}
\end{figure}

\subsection*{Spatial and socioeconomic heterogeneity in decreased diversity}
Which sociodemographic groups and areas were more affected by the decrease in income diversity? To understand the heterogeneity in decreased diversity, the mean CBG-level income diversity of all individuals living in the CBG were computed  for each CBG in the four CBSAs, thus $D_{CBG} = \frac{1}{|N_{CBG}|} \sum_{i \in N_{CBG}} D_i$, where $N_{CBG}$ denotes the set of individuals living in the corresponding CBG. By visualizing $\Delta D_{CBG} = 100\% \times (D_{CBG} - D_{CBG}^{2019})/D_{CBG}^{2019}$ in the Boston-Cambridge-Newton CBSA in Figure \ref{fig:fig3}A (and other CBSAs in Supplementary Figure S20), we observe spatial heterogeneity in the changes in diversity in the early stages of the pandemic, however more homogeneity in the long-term. The insets also show the magnitude of $\Delta D_{CBG}$ decreasing as cities recovery from the pandemic. 
The correlation between $D_{CBG}$ in April 2020 and $D_{CBG}$ in April 2019 is much smaller ($R^2=0.37$) than for October 2021 and October 2019 ($R^2=0.71$), indicating the larger heterogeneity in $\Delta D_{CBG}$ during the earlier stages of the pandemic (Supplementary Figure S21). 

To understand the spatial and sociodemographic heterogeneity in the decreased diversity of encounters during the pandemic compared to 2019, we model $D_{CBG}$ and its difference $\Delta D_{CBG}$, using a simple regression model (see Methods and Supplementary Note 5). We include variables describing the places visited by the residents in the CBG (in 2019), mobility metrics including the average total traveled distance and radius of gyration (in 2019), and sociodemographic and economic characteristics of the CBG, including its population density, median income, age and race composition, and transportation behavior (e.g., public transportation usage), all of which were standardized (Supplementary Table S3). Regression analysis was conducted for each month, including all four cities. To control for the difference between areas across and within the metropolitan areas, we include geographical fixed effects at the level of Public Use Microdata Areas (PUMAs), which typically span around 20km and contain a residential population of 150 thousand people. Detailed summary statistics (Supplementary Table S3), collinearity and correlations between variables (Supplementary Figure S22), variance inflation factor analysis, and full regression results can be found in Supplementary Note 5. 

Figure \ref{fig:fig3}B shows the adjusted $R^2$ of regression models for $D_{CBG}$ and $\Delta D_{CBG}$, respectively, across different time periods. The three groups of variables (places visited, geographical mobility, residence and demographics) explain around $60$\% to $70$\% of the variance of income diversity ($D_{CBG}$), which agrees with previous findings \cite{moro2021mobility} (Supplementary Tables S4 -- S6). However, the difference in diversity with respect to 2019 levels ($\Delta D_{CBG}$) has lower explained variance (at most $R^2 = 0.31$), and also decreases where there is no pandemic outbreak. In the long-term (October 2021), the regression model has low explained variance ($R^2=0.11$), indicating that regions homogeneously became less diverse, irrespective of sociodemographic or behavioral characteristics of the areas. Figure \ref{fig:fig3}C shows the factors that were most important in explaining the variance of $\Delta D_{CBG}$ in the months where $R^2$ was relatively high (April, May, December 2020 and January 2021) (Supplementary Tables S7 -- S10). The highlighted regression coefficients suggest that whenever there is an outbreak, areas with higher population density and higher proportion of working age populations (age 25 -- 64), higher reliance on public transport, and larger movement range (radius of gyration) experience the largest decrease in income diversity of encounters.  


\begin{figure}[!t]
\centering
\includegraphics[width=\linewidth]{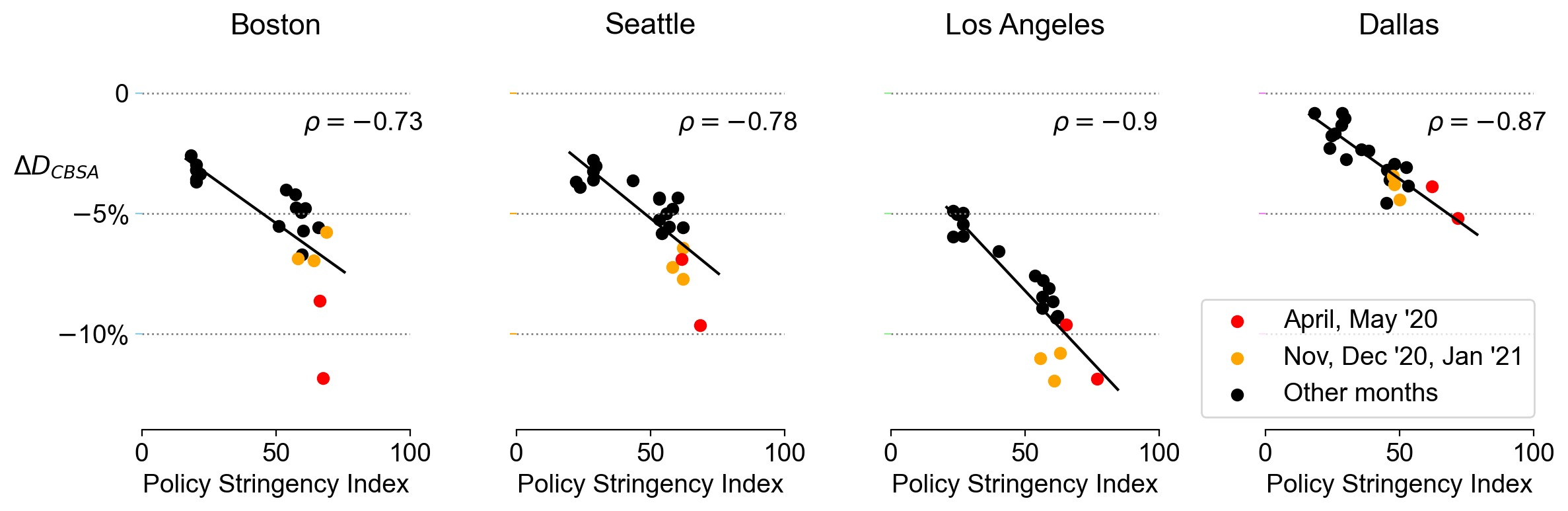}
\caption{\textbf{Trade-off between decreased income diversity of encounters and stringency of COVID-19 policies.} Decrease in income diversity of encounters $\Delta D_{CBSA}$ has strong and significant correlation with the stringency of COVID-19 measures in all four CBSAs, with outliers during the pandemic waves especially in Boston, Seattle (first wave; in red) and Los Angeles (second wave; in orange).}
\label{fig:fig4}
\end{figure}

\subsection*{Trade-off between income diversity of encounters and stringency of policy measures}

From a public policy perspective, an important and interesting question is to understand how COVID-19 containment measures, including lockdowns, school and workplace closures, and restrictions on public gatherings, have affected resulted in the loss of diversity in urban encounters. To measure the relationship between the stringency of COVID-19 measures and experienced income diversity, we utilize the COVID-19 Stringency Index \cite{hale2021global} (Supplementary Figure S23), which is a composite measure of nine response metrics, including school and workplace closures, restrictions and cancellation of public events and gatherings, and restrictions on movement and travel (See Supplementary Note 6). 

Figure \ref{fig:fig4} shows the relationship between the stringency of COVID-19 policies and the decrease in diversity of urban encounters. In all four cities we observe statistically significant ($p<0.01$) and strong negative correlation ($\rho(SI_{CBSA},\Delta D_{CBSA}) \in [-0.9,-0.73]$). 
The robust negative correlations suggest a strong trade-off relationship between income diversity and COVID-19 policy and outbreak intensity in all cities. 
The decrease in diversity become pronounced during COVID-19 outbreaks, especially during the first pandemic wave (red plots) in Boston and Seattle, and during the second pandemic wave (orange plots) in Los Angeles, where the number of cases and deaths were substantial in the respective cities. 
Moreover, for Boston, Seattle, and Los Angeles, even though the Stringency Index has decreased to around 20 in late 2021 (which indicate already less strict policies in place), the decrease in income diversity is positive, suggesting that the COVID-19 pandemic may have had a long-lasting decreasing effect on the income diversity of urban encounters. 
Regression results using additional exogenous variables such as the number of COVID-19 cases and deaths on the federal and local (CBSA) levels are shown in Supplementary Note 6.1 and Supplementary Figure S24. 
Since $\Delta D_{CBSA}(t)$ is a temporal data with autocorrelation, we tested ARIMA type models as well, however the regression results and especially the estimated coefficients were found to be robust (see Supplementary Note 6.2, Supplementary Tables S13 and S14, and Supplementary Figure S25).

\section*{Discussion}
Cities around the world currently face a wide array of challenges, ranging from combating inequality in wealth and economic opportunities \cite{chong2020economic}, to avoiding catastrophic outcomes caused by climate change induced disasters \cite{aldrich2012building}. Improving the inherent social capital of local communities and neighborhood networks, which are the fundamental units of collective decision making and support, is crucial for tackling these complex and global-scale societal challenges. With many cities expanding and urban inhabitants increasing at an unprecedented pace, the importance of promoting diverse encounters has never been higher \cite{sampson2017urban}. Previous literature show that physical co-location and encounters are known to be significant factors \cite{small2019role} and predictors \cite{cho2011friendship} for real world friendship formation, accounting for around 30\% of new friendship additions \cite{scellato2011exploiting}. Therefore, decrease in income diversity over the long-term could have substantial cumulative effects on the number and diversity of friendship ties, leading to more income segregation and polarization.

In this study, we make three important contributions towards understanding the dynamics of urban income diversity during and beyond the COVID-19 pandemic. First, we empirically revealed that physical encounters in US cities have indeed become less diverse than pre-pandemic levels even two years after the first case in the US, despite almost full recovery in aggregate mobility statics (e.g., number of visits per day). Second, we identified key behavioral changes that resulted in lower income diversity of encounters during the pandemic, including the consistent decrease in the exploration of socially diverse places and shifts in visitation preferences. Third, comparative analysis with COVID-19 policies suggested a strong trade-off relationship between COVID-19 policy stringency and income diversity. Thus, although social-distancing policies helped to mitigate the propagation of the epidemic, they also had negative effects on the social fabric of our cities. These insights, which are extremely difficult to quantify using traditional residence-based measures, collectively allow us to understand how and why urban encounters have become less diverse due to the pandemic. 

Studies have suggested that while the development of effective vaccines have successfully suppressed the mortality rates of COVID-19, the new behavioral habits and social norms that we have acquired during the pandemic, such as higher rates of work from home, and dramatic changes in physical activity, sleep, time use, and mental health \cite{giuntella2021lifestyle}, could have long-lasting impact on society \cite{pewwork}. Behavioral changes that were observed in this study, such as less social exploration when visiting new places and changes in place preferences, may also remain for a long period due to persistent fear of infections. Our results suggest that policy interventions on urban mobility, such as the introduction of fare-free transit systems and development of public spaces, should target and evaluate the recovery of social exploration to potentially improve income diversity after the pandemic. Increasing the quantity and diversity of our social encounters \cite{small2019role} could help communities to acquire social capital, which could improve the resilience to natural hazards \cite{aldrich2015social} and foster economic growth \cite{whiteley2000economic}. 

The results of our study should be interpreted in light of its limitations. Regarding the limitations of the mobility data, we are not able to identify the purpose of visits or the types of the encounters, for example, whether it is a co-visitation at a cafe where no conversations take place, or a cocktail party where strangers meet and have a conversation over a common topic. Therefore, the metrics computed in our study should be interpreted as a proxy for all meaningful encounters, and as a bound for urban income diversity. Regarding the study design, we focus on income diversity and not other socioeconomic and demographic dimensions, including racial diversity \cite{athey2021estimating,wang2018urban}. The methods and approaches may be applied to other sociodemographic data available in the American Community Survey to understand the dynamics of these other types of social diversity experienced in cities.

\section*{Methods}
\subsection*{Mobility data}
We utilize an anonymized location dataset of mobile phones and smartphone devices provided by Spectus Inc., a location data intelligence company which collects anonymous, privacy-compliant location data of mobile devices using their software development kit (SDK) technology in mobile applications and ironclad privacy framework. Spectus processes data collected from mobile devices whose owners have actively opted in to share their location, and require all application partners to disclose their relationship with Spectus, directly or by category, in the privacy policy. With this commitment to privacy, the dataset contains location data for roughly 15 million daily active users in the United States. Through Spectus’ Social Impact program, Spectus provides mobility insights for academic research and humanitarian initiatives. All data analyzed in this study are aggregated to preserve privacy. 
The home locations of individual users are estimated at the CBG level using different variables including the number of days spent in a given location in the last month, the daily average number of hours spent in that location, and the time of the day spent in the location during nighttime. See Supplementary Note 1.1 for more details. 
The representativeness of this data has been tested and corrected in Supplementary Note 2 using post-stratification techniques.

\subsection*{Estimation of stays at places}
Stops, which are location clusters where individual users stay for a given duration, are estimated using the Sequence Oriented Clustering approach \cite{xiang2016extracting}. The stops are attributed to places (or points-of interest; POIs) by simply searching for the closest place from the stops within a 100 meter radius. The robustness of the estimated income diversity to this spatial parameter was tested in Supplementary Note 1.2. Stays between 10 minutes and four hours, of individuals who were observed more than 300 minutes each day were used for the analysis. The results were shown to be robust against the choice of these temporal parameters in Supplementary Note 1.5. Moreover, the robustness of the results on income diversity against the choice of place datasets were tested using the ReferenceUSA dataset \cite{GW2P3G2014} in Supplementary Note 1.3.

\subsection*{Income diversity of encounters}
To measure the income diversity of encounters experienced at each place $\alpha$ in each city, we compute the proportion of total time spent at place $\alpha$ by each income quantile $q$, $\tau_{q\alpha}$. Income thresholds for the quantiles are chosen based on the income distributions in each city. We checked that the results on income diversity are independent of the choice of the number of income quantiles in Supplementary Note 1.4. We define full diversity of encounters at a place when people from all income quantiles spend the same amount of time, $\tau_{q\alpha}=\frac{1}{4}$ for all $q$. Using the metric used to compute income segregation in urban encounters in previous studies \cite{moro2021mobility}, we define the income diversity experienced at each place $\alpha$, $D_{\alpha}$ as a measure of evenness of time spend by different income quantiles $D_{\alpha} = 1 - \frac{2}{3} \sum_{q} |\tau_{q\alpha} - \frac{1}{4}|$. 
The diversity measure is bounded between 0 and 1, where $D_{\alpha}=0$ means there is no diversity (the place is visited by people from only one income quantile), and $D_{\alpha}=1$ indicates that all income quantiles spent equal amount of time at the place. 
Similarly for individuals, given the proportion of time individual $i$ spent at place $\alpha$, $\tau_{i\alpha}$, the individual's relative exposure to income quantile $q$, $\tau_{iq}$ can be computed by $\tau_{iq} = \sum_\alpha \tau_{i\alpha}\tau_{q\alpha}$.
Then, the income diversity experienced by individual $i$ can be measured using the same equation used for places $D_{i} = 1 - \frac{2}{3} \sum_{q} |\tau_{iq} - \frac{1}{4}|$. 
Most of the results in the main manuscript are shown by percentage differences, which is computed by $\Delta D_i(t) = \frac{D_i(t) - D_i(2019)}{D_i(2019)} \times 100(\%)$, where $D_i(2019)$ is the income diversity of encounters observed on the same month as $t$ in 2019, before the pandemic.
Results in Supplementary Note 3.3 show that using different popular measures of diversity such as entropy does not affect the results on income diversity of encounters.

\subsection*{Counterfactual simulation of mobility}

To understand the underlying behavioral changes that contributed to the decrease of income diversity in urban encounters, we design a simulation framework that leverages the pre-pandemic data to create synthetic, counterfactual mobility patterns. 
The synthetic, counterfactual mobility dataset is designed so that while the fundamental behavioral patterns observed in 2019 are kept consistent, the number of users and stays at different place categories by different income quantiles are reduced to post-pandemic levels. This way, we are able to delineate the effects of different levels of behavioral changes to the total decrease in income diversity. 

The following steps are performed to simulate the synthetic mobility datasets. To create the synthetic counterfactual data for year $y$ and month $m$, denoted as $\mathcal{S}_{y,m}$, we use the mobility data observed in the year 2019 on the same month $m$ as input data $\mathcal{D}_{2019,m}$, for example, to create a synthetic mobility dataset for April 2020, we use the mobility data observed in April 2019. Several different synthetic datasets, $\mathcal{S}^{(i)}_{y,m}$ and $\mathcal{S}^{(ii)}_{y,m}$ (and their variants), are created based on different levels of detail (see Supplementary Note 4). 
More specifically, the first synthetic dataset $\mathcal{S}^{(i)}_{y,m}$ is created by randomly removing visits from $\mathcal{D}_{2019,m}$ to adjust the total amount of dwell time spent at visits to places to match $\mathcal{D}_{y,m}$. The second synthetic dataset $\mathcal{S}^{(ii)}_{y,m}$ employs a more granular removal process, where we randomly remove visits to places from $\mathcal{D}_{2019,m}$ by income quantiles $q$ and traveled distance $d$ (binned into 7 distance ranges: $[0km,1km)$, $[1km,3km)$, $[3km,5km)$, $[5km,10km)$, $[10km,20km)$, $[20km,40km)$, $[40km,\infty]$) to adjust the amount of dwell time spent at visits to places. We also tested removing visits by income quantiles $q$, traveled distance $d$, and place taxonomy $c$, however the results were similar to scenario (ii), as shown in Supplementary Note 4.2. More details on creating the counterfactual synthetic datasets can be found in Supplementary Note 4. After creating the synthetic counterfactual datasets, we compute the income diversity of encounters and compare with the income diversity measured using the actual observed data $\mathcal{D}_{y,m}$ to delineate the effects of reduction in active users and visits to place categories on the decrease in income diversity.

\subsection*{Modeling the heterogeneity in income diversity}
To further understand how the income diversity of encounters decreased heterogeneously across sociodemographic groups during throughout the pandemic, we build simple linear regression models of the form:
\begin{equation}
    D_{CBG}(t), \Delta D_{CBG}(t) \sim \{R_{CBG}\}+\{P_{CBG}\}+\{M_{CBG}\}
\end{equation}
where $D_{CBG}(t)$ and $\Delta D_{CBG}(t)$ denote the differences in diversity at time $t$ compared to the same month in the year 2019. $\{R_{CBG}\}$ is the set of all residential variables from the census that describe the demographic, transportation, education, race, employment, wealth, etc. of the Census Block Group. $\{P_{CBG}\}$ is a vector of variables that indicate the places where individuals living in the CBG spent most of their time in 2019, out of the place subcategories which have at least 100 venues. For each individual, we identify the subcategories where the individual stays more than 0.3\% of their time and obtain a binary vector with the length of 564, which is the number of place subcategories. 
$\{M_{CBG}\}$ is a set of variables that describe the geographical mobility behavior of people living in the corresponding CBG. We use two variables: (i) the radius of gyration of all the places visited by each user, and (ii) the average distance traveled to all places from each individual's home. 
Details of the regression covariates, including their summary statistics and correlations, are studied in Supplementary Note 5.

To further understand the differences in decreased income diversity across CBSAs, the correlation between the stringency of COVID-19 policies and the decrease in diversity was analyzed.
The stringency index $SI_{CBSA}(t)$ is a composite metric that measures the strictness of COVID-19 policies calculated using data collected in OxCGRT \cite{hale2021global}, and are provided at the state levels for the United States. The stringency index takes into account policies including the closings of schools and universities, closings of workplaces, cancelling of public events and gatherings, closing of public transport, orders to shelter-in-place, restrictions on internal movement between cities/regions and international travel, and presence of public info campaigns.
More details are provided in the codebook in the github webpage \footnote{\url{https://github.com/OxCGRT/covid-policy-tracker/blob/master/documentation/codebook.md}}.

\section*{Acknowledgements}
We would like to thank Spectus who kindly provided us with the mobility dataset for this research through their Data for Good program.

\section*{Author contributions statement}
T.Y. designed the algorithms, performed the analysis, developed models and simulations. B.G.B.B. and E.M. performed part of the analysis, partially developed models and simulations. A.P., X.D. and E.M. supervised the research. All authors wrote the paper. Company data were processed by T.Y. and partially by B.G.B.B. and E. M. All authors had access to aggregated (nonindividual) processed data. All authors reviewed the manuscript.

\section*{Data Availability}
The data that support the findings of this study are available from Spectus through their Social Impact program, but restrictions apply to the availability of these data, which were used under the licence for the current study and are therefore not publicly available. Information about how to request access to the data and its conditions and limitations can be found in \url{https://spectus.ai/social-impact/}. 

\section*{Code Availability}
The analysis was conducted using Python. Code to reproduce the main results in the figures from the aggregated data is public available on github \url{https://github.com/takayabe0505/IncomeDiversity}.

\section*{Competing Interests}
The authors declare no competing interests.

\bibliographystyle{unsrt}  
\bibliography{ms}

\end{document}


\maketitle

\tableofcontents
\newpage

\listoffigures

\listoftables

\newpage

\setcounter{figure}{0}
\setcounter{table}{0}


\section{Mobility data}

\subsection{Home estimation and stop detection}
In this study we utilize an anonymized location dataset of mobile phones and smartphone devices provided by Spectus Inc., a location data intelligence company which collects anonymous, privacy-compliant location data of mobile devices using their software development kit (SDK) technology in mobile applications and ironclad privacy framework. Spectus processes data collected from mobile devices whose owners have actively opted in to share their location, and require all application partners to disclose their relationship with Spectus, directly or by category, in the privacy policy. With this commitment to privacy, the data set contains location data for roughly 15 million daily active users in the United States. Through Spectus’ Data for Good program, Spectus provides mobility insights for academic research and humanitarian initiatives. All data analyzed in this study are aggregated to preserve privacy\footnote{\url{https://spectus.ai/privacy/privacy-policy/}}. Each entry in the data table comprises anonymized device ID, location coordinates, start time, and dwell time of the stop for the device. 

To define the type of location (Home or Work), different variables are used, including the number of days spent in a given location in the last month, the daily average number of hours spent in that location, and the time of the day spent in the location (nighttime/daytime). To estimate the home position of a user, the algorithm combines the three variables and creates a score that represents the probability that the position points to the home. The more days and the average number of hours spent in the position, the higher the score is. Higher scores will also be assigned to the most common places during the night. The location that maximizes this score is defined as the home of the device.

Once the location of the home location is identified, the algorithm looks for the work position. Note that the algorithm requires the work location to be located at least 100 meters apart from the home location. The same variables used for the detection of the home location are used, but a higher score is given to daytime locations for the work location rather than nighttime locations.
Spectus runs the algorithm every week in order to confirm or update the inferred home and work locations as we observe new data. We will only consider devices that have been present in Spectus’ dataset for at least 15 days. 
Spectus tightly restricts access to the inferred precise home and work locations of devices. Furthermore, it is used as input into various downstream processes to create more privacy-protected versions of Spectus datasets.
For example, we only expose home and work datasets in Spectus Workbench associated with standard Census Block Groups, created by the U.S. Census Bureau, rather than the precise locations. This offers a good balance between utility and privacy: according to the U.S. Census Bureau, there are between 600 and 3000 people living in each block group. Each block group is an aggregate of contiguous U.S. blocks sharing similar socio-demographic characteristics.
The representativeness of this data has been tested and corrected in Section \ref{repre} in the Supplementary Material.
The stops, which are location clusters where individual users stay for a given duration, are estimated using the Sequence Oriented Clustering approach \cite{xiang2016extracting}.


\begin{figure}
\centering
\includegraphics[width=\linewidth]{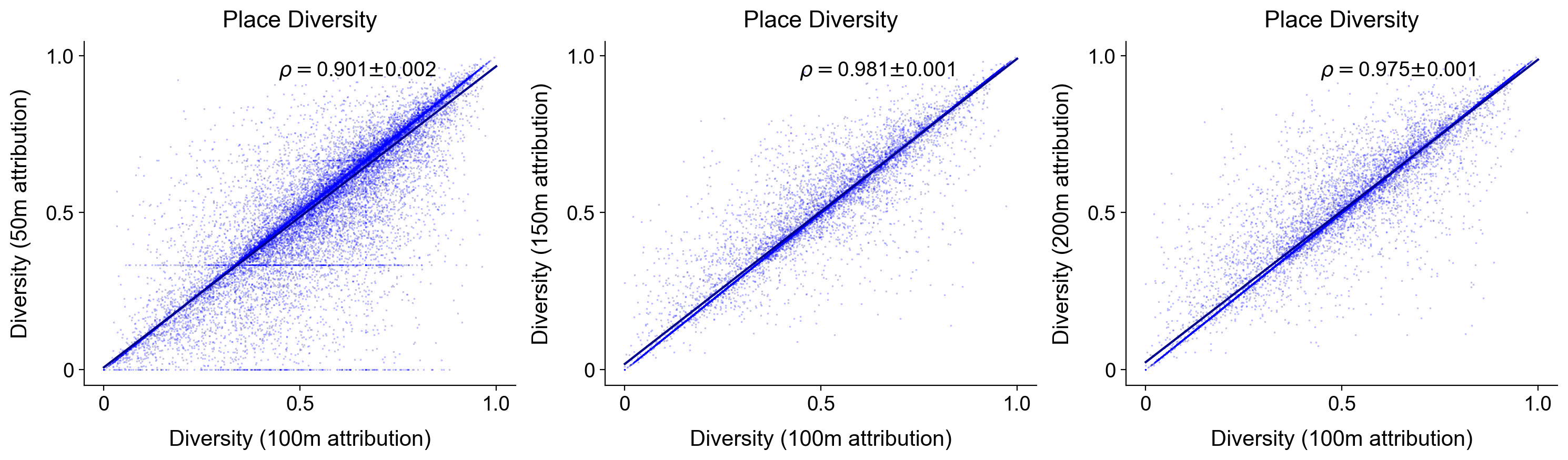}
\caption{Sensitivity of place based diversity of encounters with respect to spatial parameters for the visit attribution algorithm.}
\label{fig:visitattribution}
\end{figure}

\subsection{Robustness to threshold distance for attribution of stays to places}
To measure the diversity of physical encounters in urban environments, we attribute the stops of individual users to specific places in the city. To study the stops at different places, we use stops that are longer than 10 minutes but shorter than four hours. In our study, we use location data of places collected via the Foursquare API \footnote{\url{https://developer.foursquare.com/}}. To protect the users' privacy, we have removed various privacy-sensitive places from our places database. Sensitive places include health-related places, places where the vulnerable population are located, military-related, religious facilities, places that are related to sexual-orientation, and adult-oriented places \footnote{\url{https://spectus.ai/privacy/spoi-policy/}}. As a result, we have a total of 71K places in Boston, 57K places in Seattle, 206K places in Los Angeles, and 97K places in Dallas. The breakdown of the number of places by the place category is shown in Table \ref{table:pois}.
To attribute a stop to a place, we simply attribute each stop the closest place in our dataset. o avoid attributing a stop to place far away, we attribute the stop to a place within $d_{max}=100$ meters from the observed location of the stop. If the stop is further away than 100 meters from any place in the dataset, the stop is discarded from our dataset and not used for computing the diversity of encounters. 

\begin{table}
\centering
\caption{Number of places in the Foursquare dataset in the four core-based statistical areas (CBSAs) analyzed in this study.}
\begin{tabular}{lrrrr}
\toprule
& \multicolumn{4}{c}{CBSAs} \\
\cmidrule{2-5}
Place category & Boston & Seattle & Los Angeles & Dallas  \\
\midrule
Arts and Museums  & 3,346  & 2,797  & 12,019 & 4,340  \\
City and Outdoors & 9,370  & 6,794  & 22,364 & 9,719  \\
Coffee and Tea    & 872   & 1,968  & 3,284  & 1,064  \\
Entertainment     & 5,548  & 3,991  & 18,533 & 7,065  \\
Food              & 14,791 & 10,936 & 46,411 & 22,812  \\
Grocery           & 2,017  & 1,166  & 4,602  & 1,808  \\
Health            & 318   & 209   & 979   & 554   \\
Service           & 20,500 & 15,692 & 53,972 & 30,044  \\
Shopping          & 8,612  & 6,134  & 26,538 & 14,269  \\
Transportation    & 6,615  & 7,460  & 18,165 & 5,538  \\
\midrule
All places        & 71,989 & 57,147 & 206,867 & 97,213 \\
\bottomrule
\end{tabular}
\label{table:pois}
\end{table}

The robustness of our results on the diversity of encounters have been tested using different spatial thresholds of $d_{max}$. Different levels of $d_{max}$ could change how individuals' stays are attributed to the places, and thus could affect our estimates of the income diversity of physical encounters. Figure \ref{fig:visitattribution} compares the diversity of encounters for places in Boston when we use different levels of $d_{max}$ (y-axis) with our default parameter $d_{max}=100m$ (x-axis). For all values $d_{max}=\{50,150,200\}$, the Pearson correlation of place based diversity metrics are extremely high, where $\rho [D_{\alpha}^{d_{max}=100m},D_{\alpha}^{d_{max}=50m}]=0.901 \pm 0.002$, $\rho [D_{\alpha}^{d_{max}=100m},D_{\alpha}^{d_{max}=150m}]=0.981 \pm 0.001$, and $\rho [D_{\alpha}^{d_{max}=100m},D_{\alpha}^{d_{max}=200m}]=0.975 \pm 0.001$. This robustness check shows that the estimated diversity values do not depend on the choice of the spatial threshold parameter for visit attribution.

\subsection{Robustness against choice of POI dataset}
\begin{figure}
\centering
\includegraphics[width=.9\linewidth]{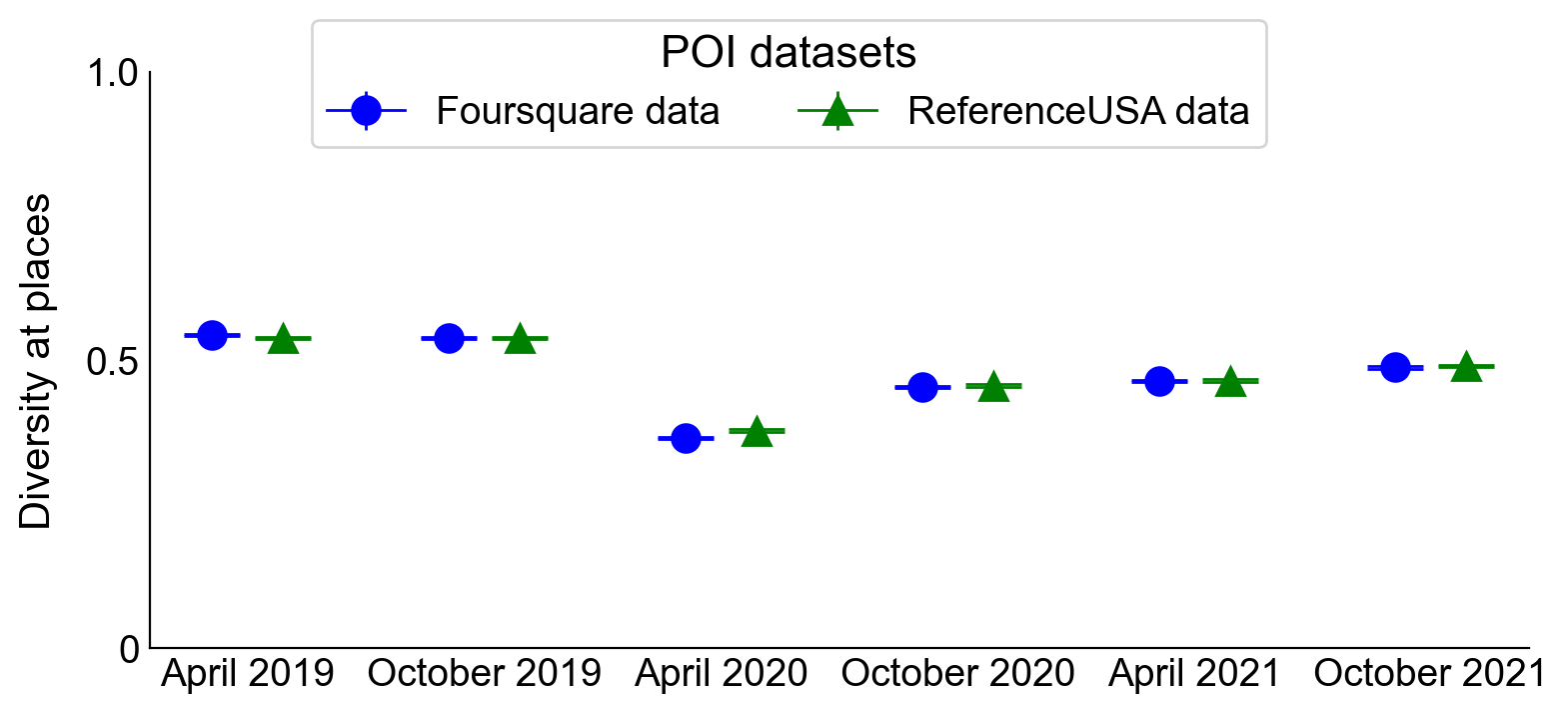}
\caption{Robustness of income diversity at food and grocery places to using different POI datasets. Horizontal bars show the standard errors, which are very small due to the large number of places.}
\label{fig:altpois}
\end{figure}

Although we may assume that our dataset of places (name, location coordinates, business category) collected via the Foursquare API is relatively comprehensive, there could be places that are missing from the dataset, which could affect our results on income diversity. To check whether our findings in our study are independent on the selection of the dataset of places, we used the ``ReferenceUSA Business Historical Data'', which is a record of companies across the US. The dataset is created annually from Infogroup’s U.S. Business Database, and a snapshot of the data is saved each December (we used the 2020 version). The data, similar to the Foursquare data, contains the company name, mailing address, SIC and NAICS codes, employee size, sales volume, latitude/longitude, and other variables about each company \cite{GW2P3G_2014}. In Boston, there were 12641 food and restaurant places (NAICS code starts with 722) and 3886 grocery stores (NAICS code starts with 445) in the ReferenceUSA dataset, compared to the 14,791 and 2,017 places in the Foursquare data, respectively. The income diversity experienced at food, restaurant, and grocery places were calculated using the two different datasets for several time periods (April and October in 2019, 2020, 2021). Figure \ref{fig:altpois} shows the mean $\pm$ standard errors of income diversity of encounters at places. Despite the differences in the number of places and the minor differences in category labels between the Foursquare data and the ReferenceUSA datasets, similar levels of decrease in income diversity are observed between the two datasets, suggesting the results we obtain are robust against the choice of place datasets.

\subsection{Robustness against definition of income quantiles} \label{sec:incomeqs}
\begin{figure}
\centering
\subfloat[Income quantile range for the four CBSAs.]{\includegraphics[width=.85\linewidth]{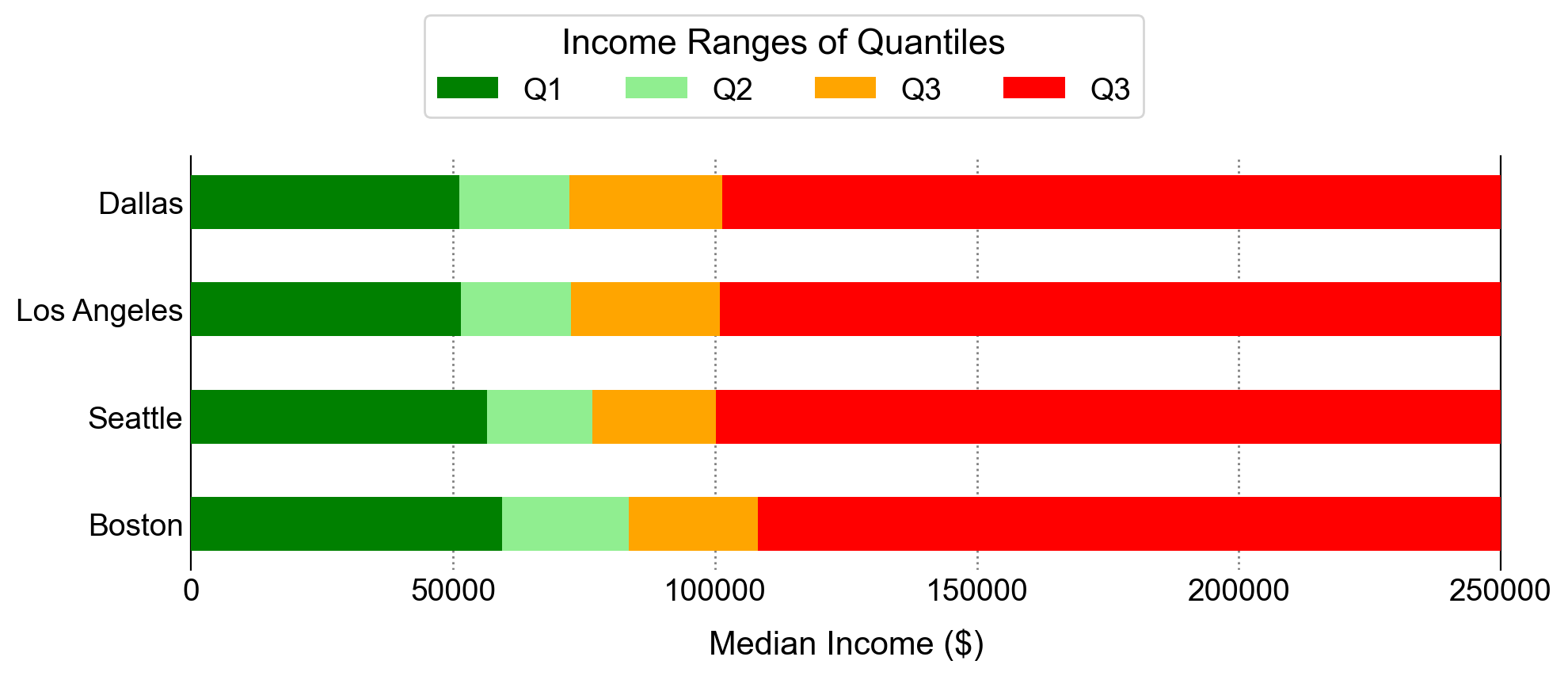}}\\
\subfloat[Income quantile ranges when $n=\{2,3,4,5,6\}$ in the Boston CBSA.]{\includegraphics[width=.85\linewidth]{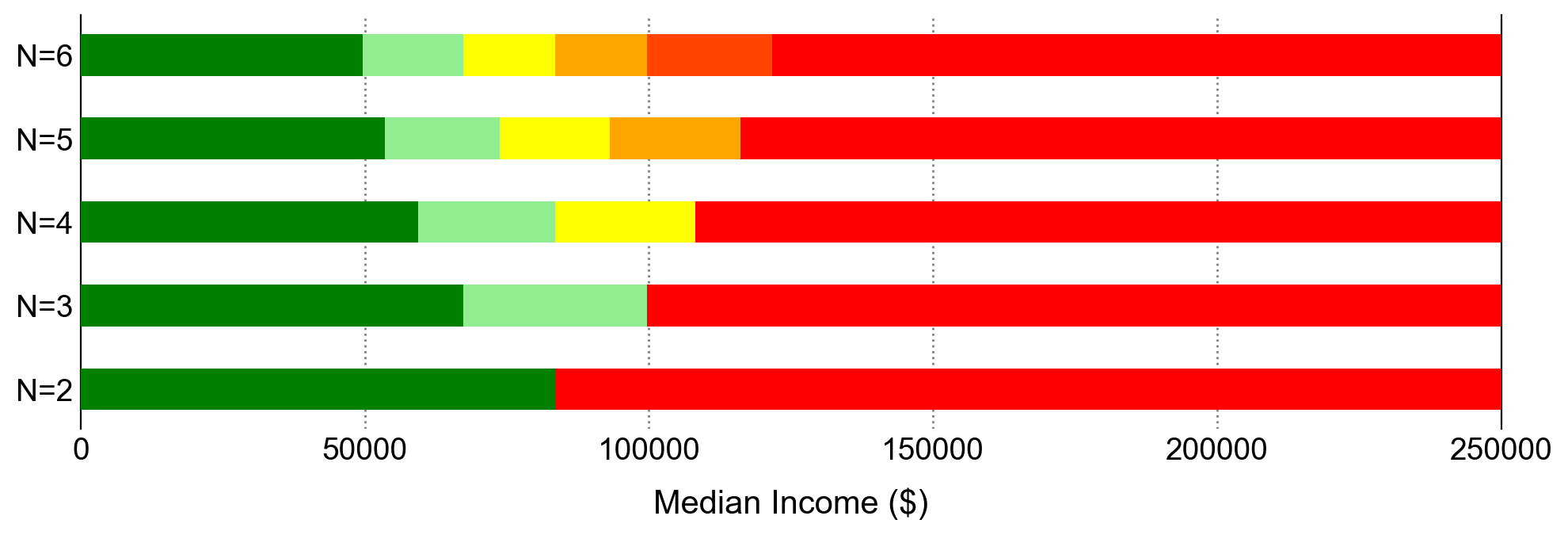}}
\caption{Income quantile ranges when using different $n$.}
\label{fig:nquant_4cities}
\end{figure}

To estimate the socioeconomic status of each individual, we use the median income of the census block group (CBG) where their estimated homes are located in as a proxy for their income. Individuals in our dataset are then grouped into four equal-size quantiles of economic status within each city. The diversity of encounters at places and for individuals are hereon calculated using these assigned quantile values.
For Boston, the median income thresholds for the quantile classification are: [\$0,\$59K] for quantile 1 (low income), [\$59K,\$84K] for quantile 2 (medium-low income), [\$84K, \$108K] for  quantile 3 (medium-high income), and [\$108K, \$250K] for quantile 4 (high income). The four income quantile ranges for the four cities are shown in Figure \ref{fig:nquant_4cities}a. While Boston has the highest quantile thresholds, Los Angeles and Dallas have slightly lower income quantile ranges. 

\begin{figure}
\centering
\includegraphics[width=\linewidth]{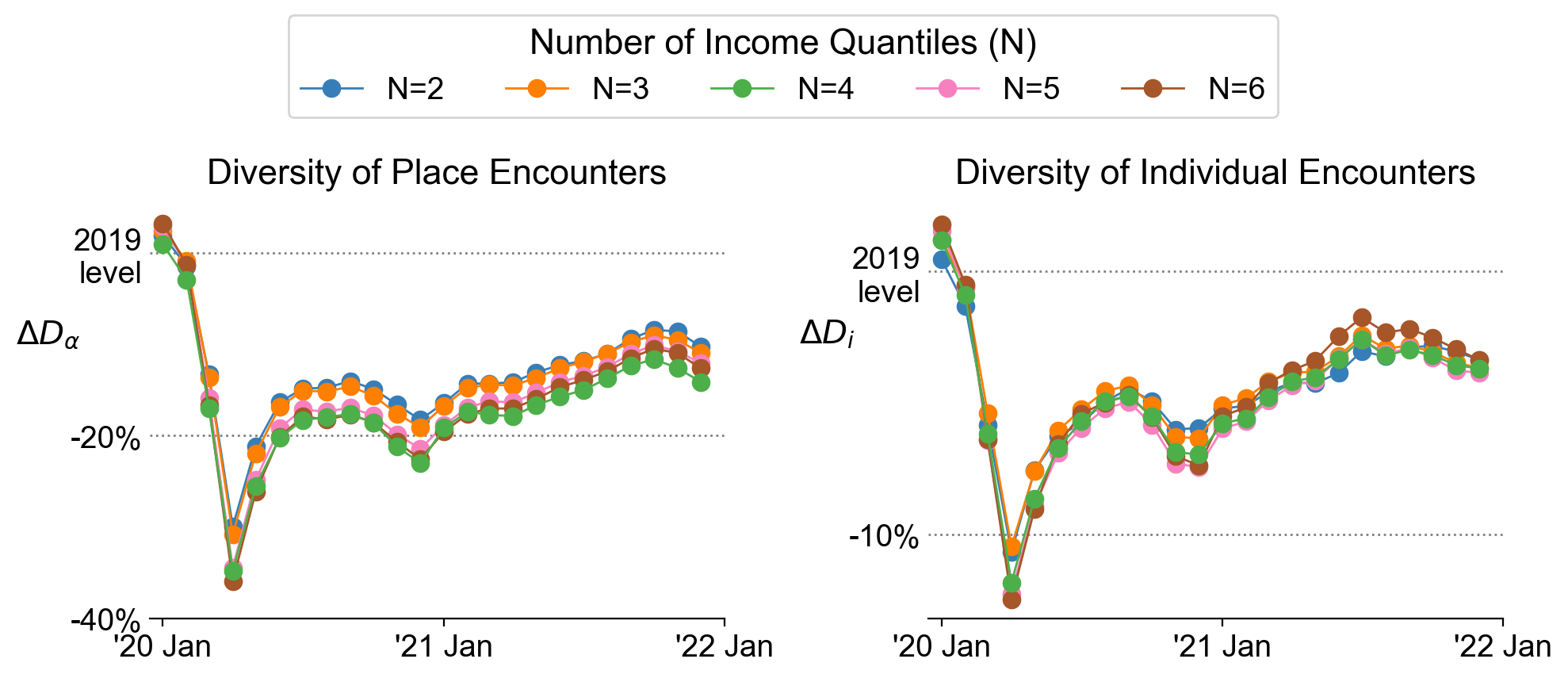}
\caption[Sensitivity of income diversity of encounters with respect to number of income quantiles used.]{Sensitivity of income diversity of encounters with respect to number of income quantiles used. The results on income diversity are robust and independent of the choice of $n$.}
\label{fig:nquant}
\end{figure}

Since our estimation of income diversity is conducted by grouping the encountered individuals into income quantile groups and measuring the unevenness of the group sizes, it is important to check whether our income diversity estimates are affected by the number of income quantile groups we use. 
To check the robustness of our income diversity measures against the selection of the number of income quantiles $n$, we compute the place-based and individual-based diversity measures when using different number of income quantiles ($n=2,3,4,5,6$). The income diversity metric under a given $n$ is computed as the following:
\begin{equation*}
    D_{\alpha}^{\{n\}} = 1 - \frac{n}{2n-2} \sum_{q=1}^n |\tau_{q\alpha} - \frac{1}{n}|,
\end{equation*}
where $n$ is the number of quantiles used for income quantile classification. 
For Boston's case, the income ranges of quantiles under different number of quantiles are shown in Figure \ref{fig:nquant_4cities}b.
Figure \ref{fig:nquant} shows the estimated decrease in diversity in Boston when using different number of income quantiles. We observe that both the dynamics of the diversity of encounters experienced at places and by individuals are consistent across time, showing high agreement with the result obtained using $n=4$ (green color). Therefore, we conclude that our findings related to the loss of diversity in the short- and long-term during the pandemic is independent of the choice of the number of quantiles. 

\begin{figure}
\centering
\subfloat[Number of smartphone users selected under different thresholds for minimum minutes observed per day. ]{\includegraphics[width=.5\linewidth]{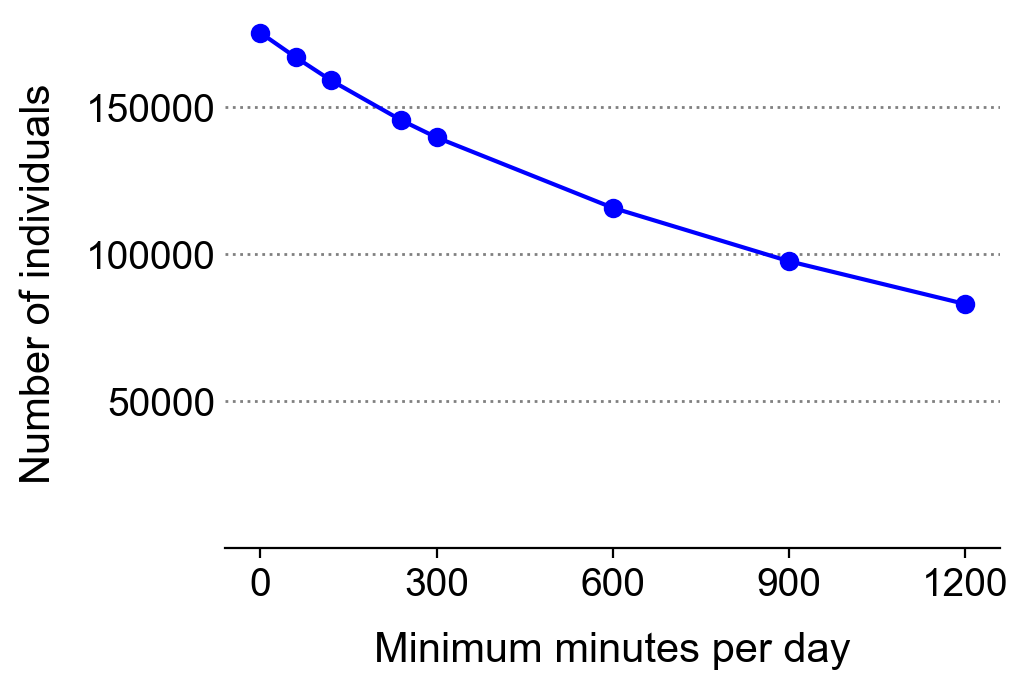}}\\
\subfloat[Sensitivity of income diversity of encounters with respect to the minimum observation threshold for user selection. The decrease in diversity becomes amplified when selecting a smaller set of users with longer observed duration.]{\includegraphics[width=\linewidth]{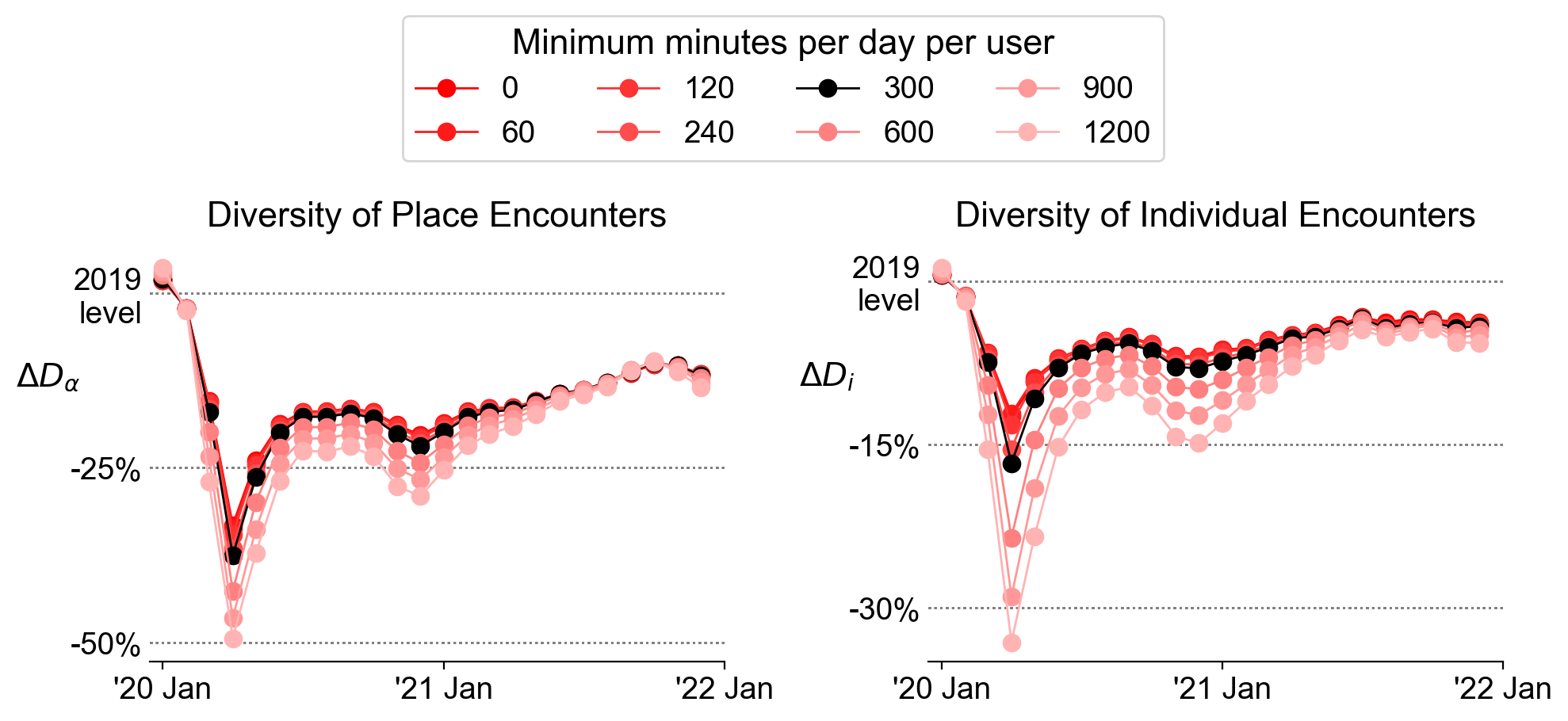}}
\caption{Sensitivity of income diversity of encounters with respect to the minimum observation threshold for user selection.}
\label{fig:minminutes}
\end{figure}

\subsection{Robustness against choice of data filtering parameters}
Since mobile phone location pings are collected via various smartphone apps at asynchronous timings and frequencies, some users are observed for a long duration during the day while others could be observed for just a very short period of time. Using a group of individuals with very short observation times could skew the results of the income diversity of encounters. Therefore, we limit the group of individual users analyzed in this study to those who are observed a substantial amount of time each day. In this study, we use users who are observed more than $t_{min}=300$ minutes across all visited places (including their homes) to select the users used in our analysis. 

Since 300 minutes is an arbitrary temporal threshold, we tested whether income diversity experienced at places and by individuals are affected by the selection of the $t_{min}$ parameter. There is an obvious trade-off between the number of available users in the dataset and the temporal coverage of the users' mobility patterns, as shown in Figure \ref{fig:minminutes}a for the Boston CBSA. Out of all the 175K users in the dataset, 140K users were observed more than 300 minutes. 

Figure \ref{fig:minminutes}b shows how the income diversity dynamics experienced at places (left panel) and by individuals (right panel) vary when using different $t_{min}$ parameters. The losses in diversity in encounters are amplified for both places and individuals when we employ a stricter threshold for selecting the users, mainly due to the lack of individuals visiting each place, which increases the likelihood of lower diversity. However, the main takeaways of the dynamics in income diversity are consistent -- the income diversity in urban encounters have become decreased in both the long and short term, both from the places' and individuals' perspectives. 




To summarize the mobility data filtering process, we 1) estimate home and stop locations for each individual, 2) attribute the stays to specific places, 3) estimate each individual user's socioeconomic status using census-block group level data, and 4) select users who are observed more than 300 minutes per day. 
After pre-processing the mobility datasets for each of the four urban areas, the entire dataset contains a total of 1.16 million unique users and 23.4 million stays across a total of 97K places. Table \ref{table:stats} shows the summary statistics for the four CBSAs. 

\begin{table}
\centering
\caption{Description of the four core-based statistical areas (CBSAs) analyzed in this study.}
\resizebox{\textwidth}{!}{\begin{tabular}{lrrrr}
\toprule
CBSA & Population & \# users (monthly) & \# stays (monthly) & \# places  \\
\midrule
Boston-Cambridge-Newton & 4.64M & 144K & 2.34M & 71,989  \\
Seattle-Tacoma-Bellevue & 3.55M & 141K & 2.23M & 57,147  \\
Los Angeles-Long Beach-Anaheim & 13.05M & 452K & 10.00M & 206,867  \\
Dallas-Fort Worth-Arlington & 6.70M & 425K & 8.83M & 97,213  \\
\midrule
Total & 27.94M & 1.16M & 23.4M & 433,216 \\
\bottomrule
\end{tabular}}
\label{table:stats}
\end{table}


\section{Data representativeness}\label{repre}

The location data used in our study is collected from smartphones via various apps and services. Although a significant portion (85\% according to 2021 data\footnote{\url{https://www.statista.com/topics/2711/us-smartphone-market/#topicHeader__wrapper}}) of the US population owns a smartphone, one could question the representativeness of the 1.16 million user samples across geographical regions and income quantiles. Studies have reported digital divide and smartphone usage gaps across sociodemographic groups in the US \cite{tsetsi2017smartphone}. In this section, we test whether our group of users in the mobility data are representative of the total population, and further employ post-stratification techniques to correct for any potential biases in the sampling rates across places and socioeconomic status and to test whether the results on income diversity dynamics are robust to such uncertainties concerning data representativeness.

\subsection{Population and income representativeness}
\begin{figure}
\centering
\subfloat[(left) Comparison of number of smartphone users with the census population. (right) Proportion of smartphone users in the four income quantiles.]{\includegraphics[width=\linewidth]{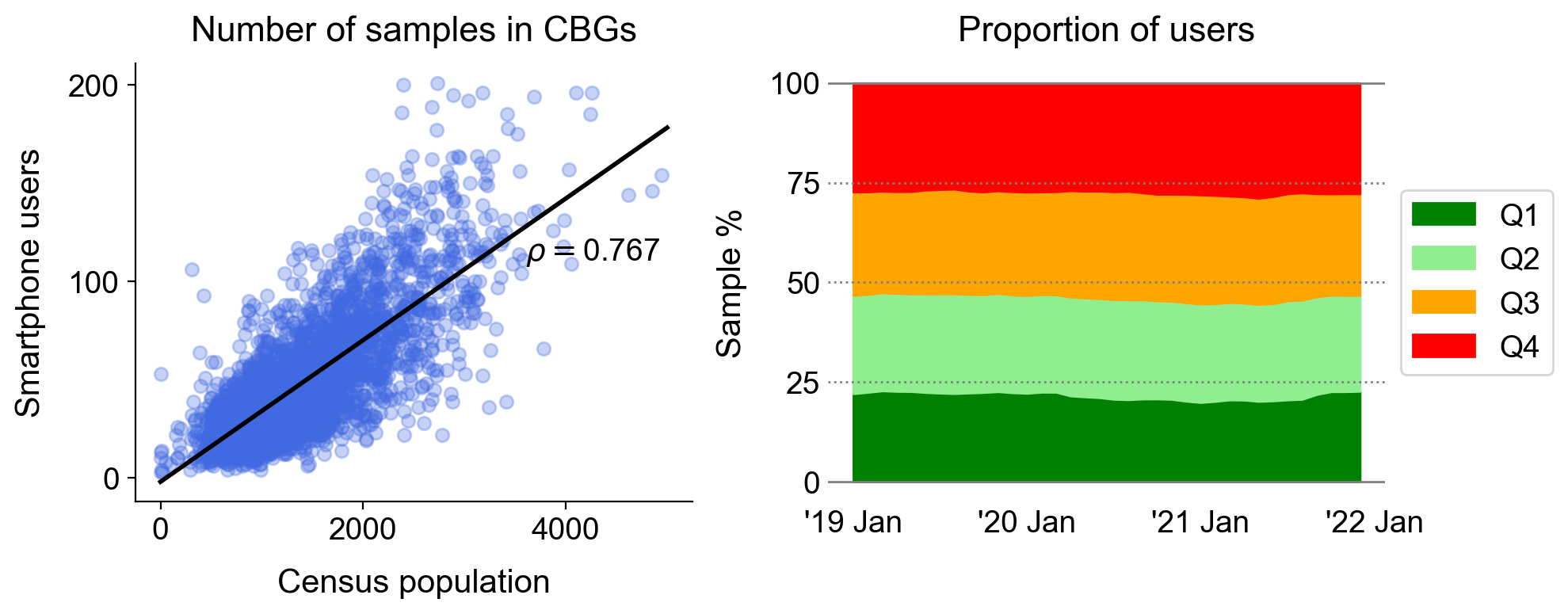}}\\
\subfloat[Sensitivity of income diversity of encounters with respect to the representativeness of smartphone location data via post-stratification.]{\includegraphics[width=\linewidth]{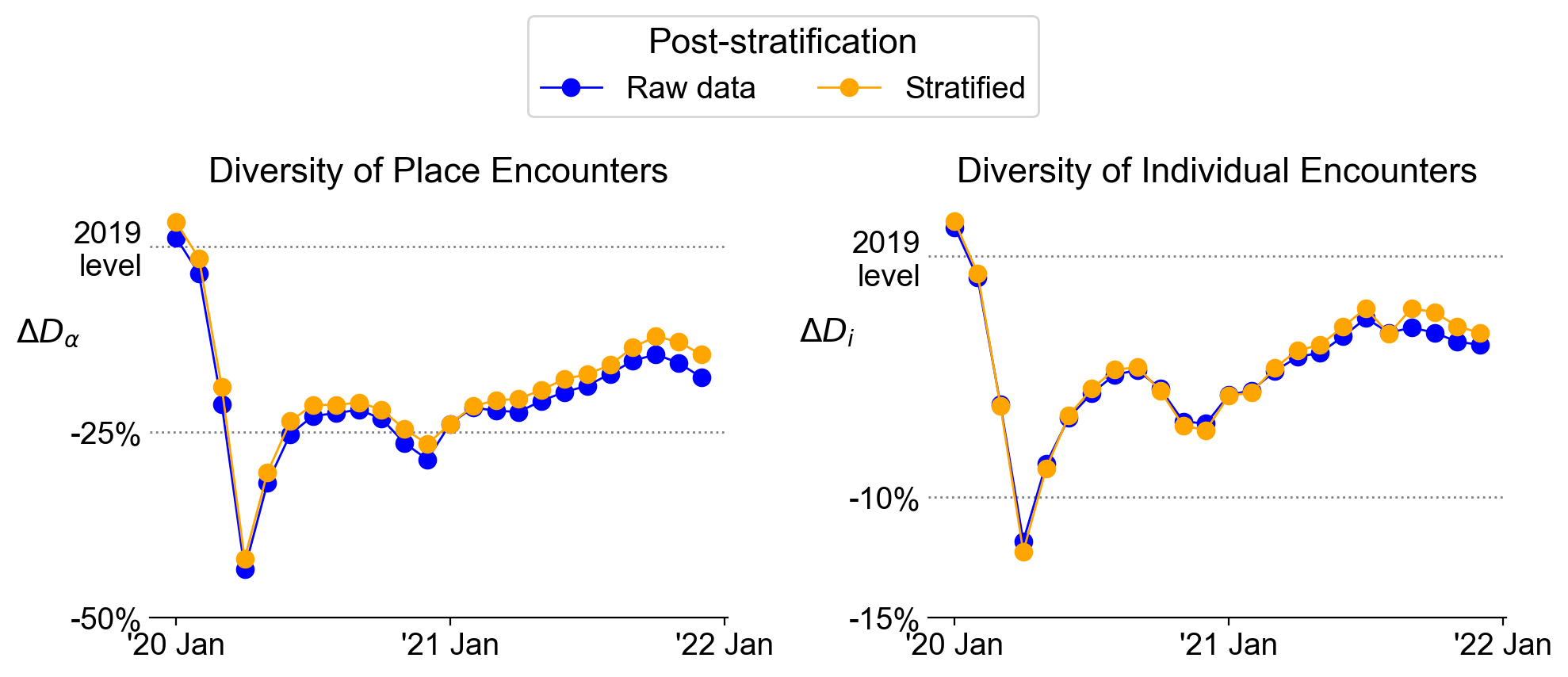}}
\caption{Sensitivity of income diversity of encounters with respect to the representativeness of mobile phone users across CBGs and income groups.}
\label{fig:representativeness}
\end{figure}

The sampling percentage of the mobility data (100\% $\times$ number of observed mobile phone users divided by the total population from the census data) is around 3\% across all metropolitan regions. 
To test whether the users in the location data are representative of the entire population, first we compare the population detection in our mobility data and the 2019 ACS data for each of the CBGs in the cities. The left panel in Figure \ref{fig:representativeness}a shows the comparison between the census population (x-axis) and the number of observed smartphone users (y-axis) on the CBG scale in the month of January 2020 in the Boston CBSA. The correlation is moderately high ($\rho=0.767$) showing that despite the use of such small census areas and potential bias in the smartphone usage patterns, we are able to obtain a good representation of the population. 
This correlation is relatively stable before and during the pandemic at around $\rho=0.75$, which is moderately high. In Section \ref{poststrat} we use post-stratification techniques to correct for any differences in the sample percentages across CBGs and assess whether our estimates on income diversity of urban encounters are affected by the representativeness of the data.  

In addition to the differences in sampling rates across CBGs, differences in representativeness across income quantiles are important for our study. To study the representativeness across income quantiles, we compute the proportion of users in the four income quantiles across time, which is shown in the right panel of Figure \ref{fig:representativeness}a. A completely balanced dataset would have all income quantiles each represent 25\% of the proportion of the users. However, we can observe that the highest income quantile (Q4) is over-represented in the dataset throughout the 3 years period, while the lowest income quantile (Q1) is under-represented. In Section \ref{poststrat}, we investigate whether this bias in income representativeness affects our estimates on income diversity using post-stratification techniques. 


\subsection{Robustness check via post-stratification} \label{poststrat}
To understand the effects of the varying sampling rates across CBGs and income groups on our estimation on income diversity of urban encounters experienced at places and by individuals, we apply a post-stratification technique, which is used in a previous study \cite{moro2021mobility}. 
Post-stratification is a well know sampling tool \cite{salganik2019bit} and is typically used to study the impact of sampling biases in mobile phone location data \cite{jiang2016timegeo} or (geolocated) social media data \cite{wang2018urban} on various downstream tasks and analyses. 
Following the methods employed in Moro et al. \cite{moro2021mobility}, we denote $w_g$ the expansion factor, which is the ratio of the population of census block $g$ to the population detected in our mobility data. 
We then weight the time people from census block group $g$ spends at place $\alpha$ by 
\begin{equation*}
    \hat{\tau}_{g\alpha} = w_g \tau_{g\alpha}
\end{equation*}
where the assumption is that $\tau_{g\alpha}$ is proportional to the number of people visiting the place. Using this method, we could increase (decrease) the time spent at places by people coming from census block groups that are under-estimated (over-estimated). 

Recomputing the income diversity of urban encounters using the corrected duration of stays $\hat{\tau}_{g\alpha}$, as shown in Figure \ref{fig:representativeness}b we observe that the dynamics of the income diversity decrease between the raw mobility data and the post-stratified  data are very similar. These results show the robustness of the insights on income diversity, and that even though the representativeness of mobile phone users are not perfect, the effect on our estimations are very limited.

\section{Income diversity of encounters}

\begin{figure}
\centering
\subfloat[Seattle]{\includegraphics[width=\linewidth]{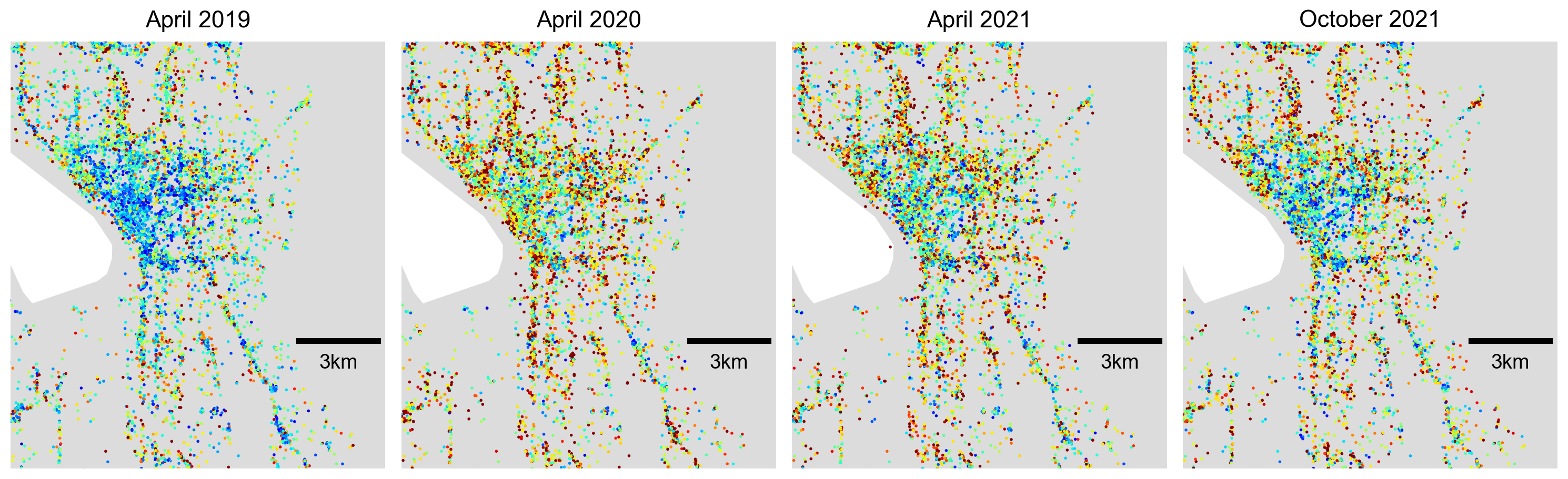}}\\
\subfloat[Los Angeles]{\includegraphics[width=\linewidth]{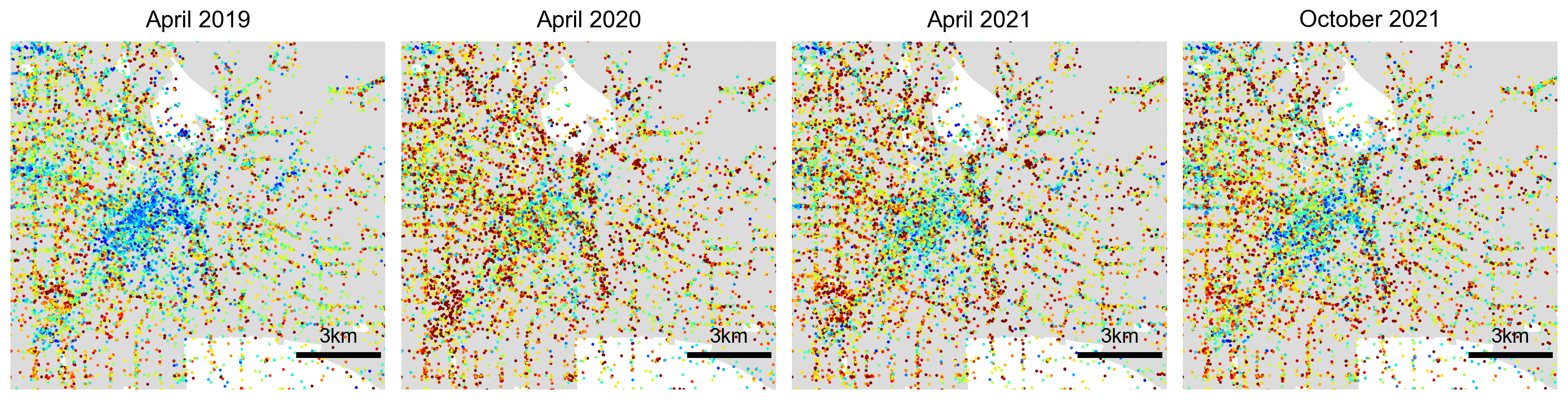}}\\
\subfloat[Dallas]{\includegraphics[width=\linewidth]{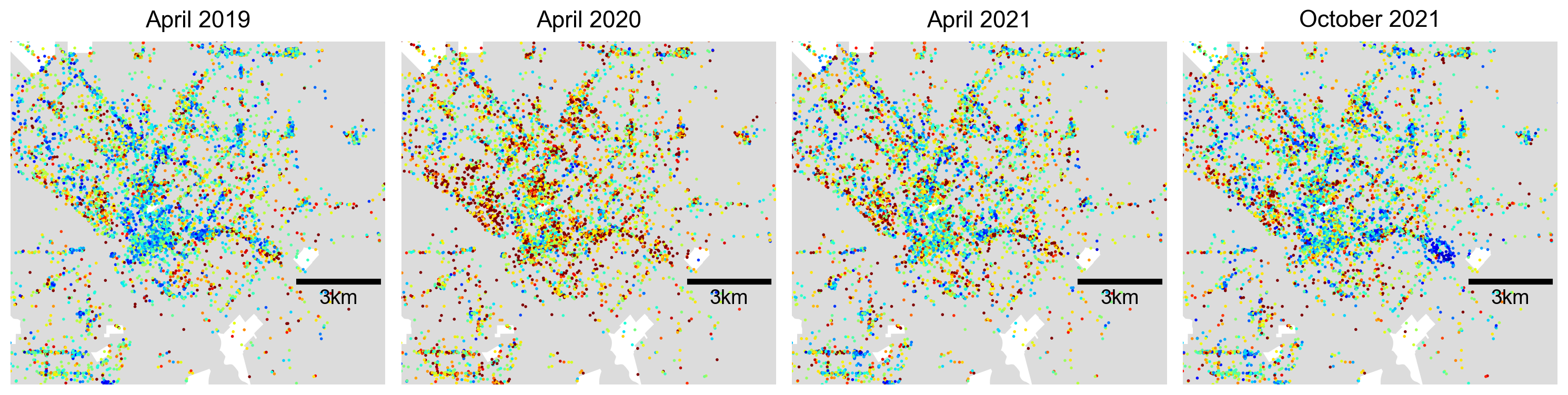}}
\caption{Income diversity of encounters in places in the three CBSAs (Boston is shown in main manuscript).}
\label{fig:placeseg_maps}
\end{figure}

\begin{figure}
\centering
\includegraphics[width=.9\linewidth]{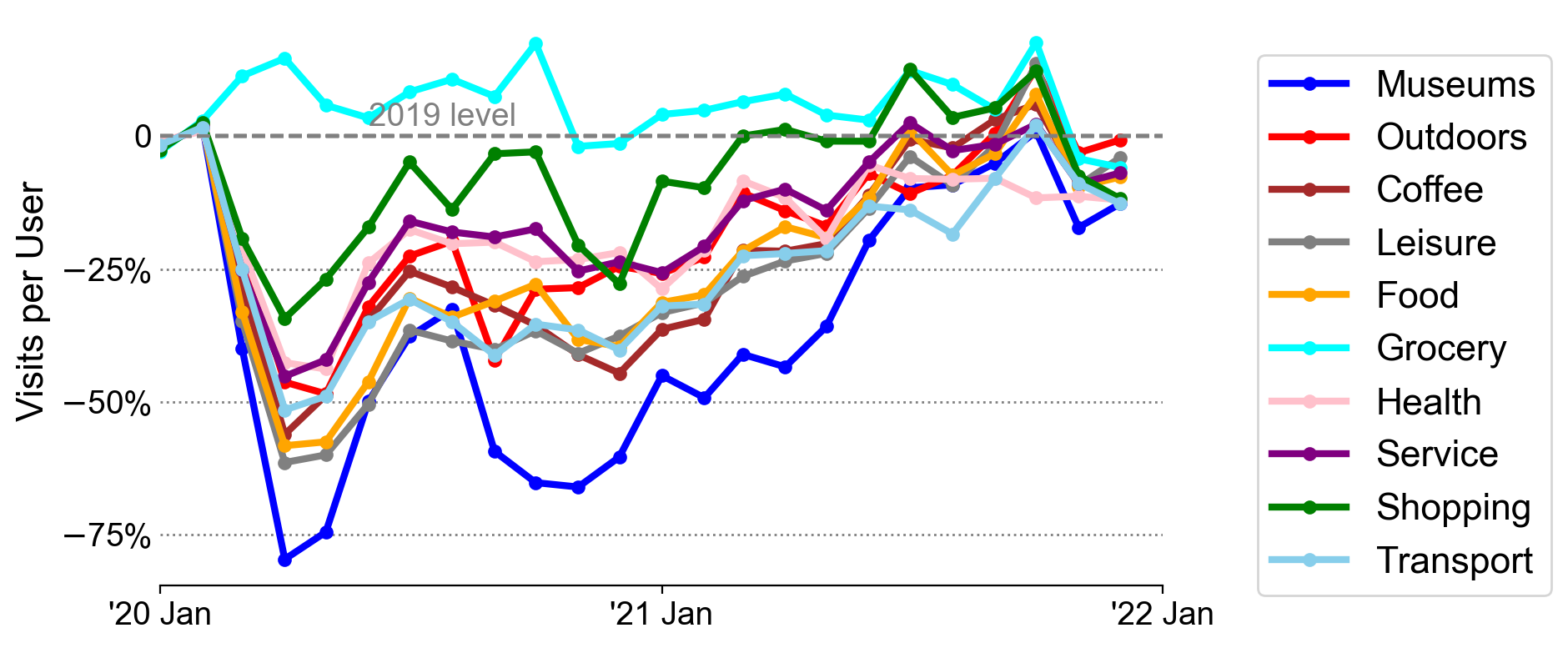}
\caption{Normalized visits per user to different place categories.}
\label{fig:poi_visits}
\end{figure}

\begin{figure}
\centering
\subfloat[Seattle]{\includegraphics[width=\linewidth]{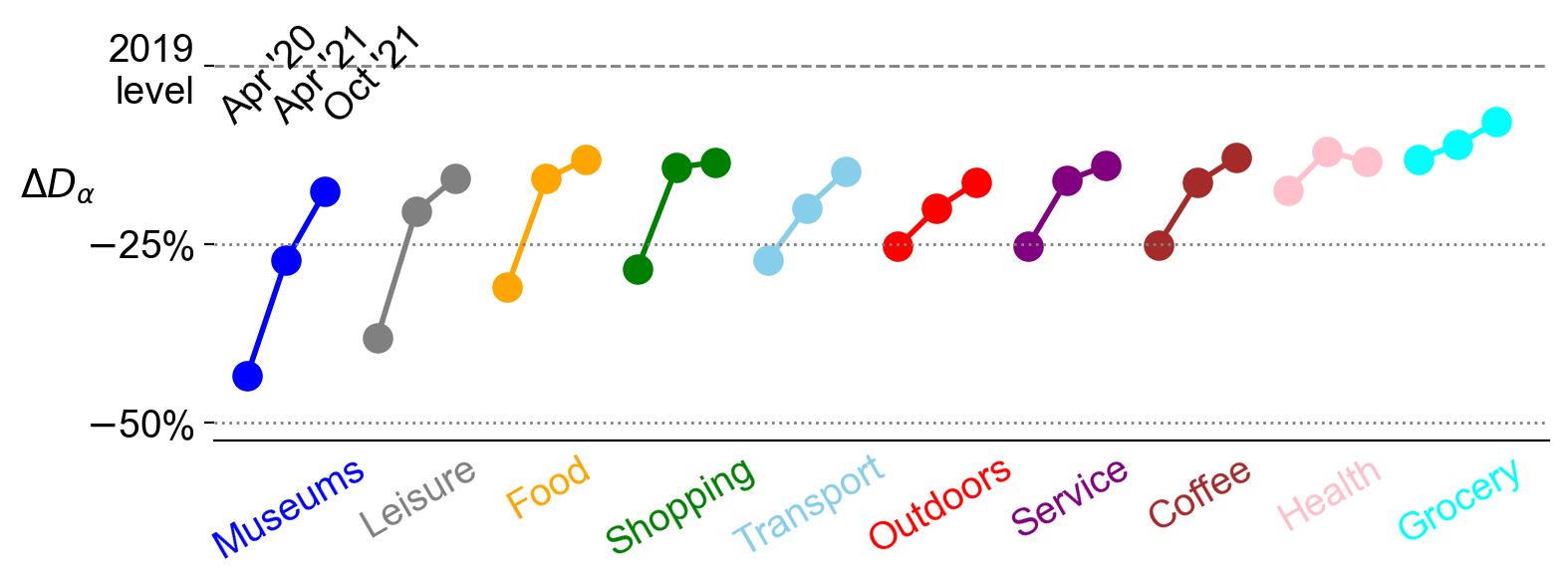}}\\
\subfloat[Los Angeles]{\includegraphics[width=\linewidth]{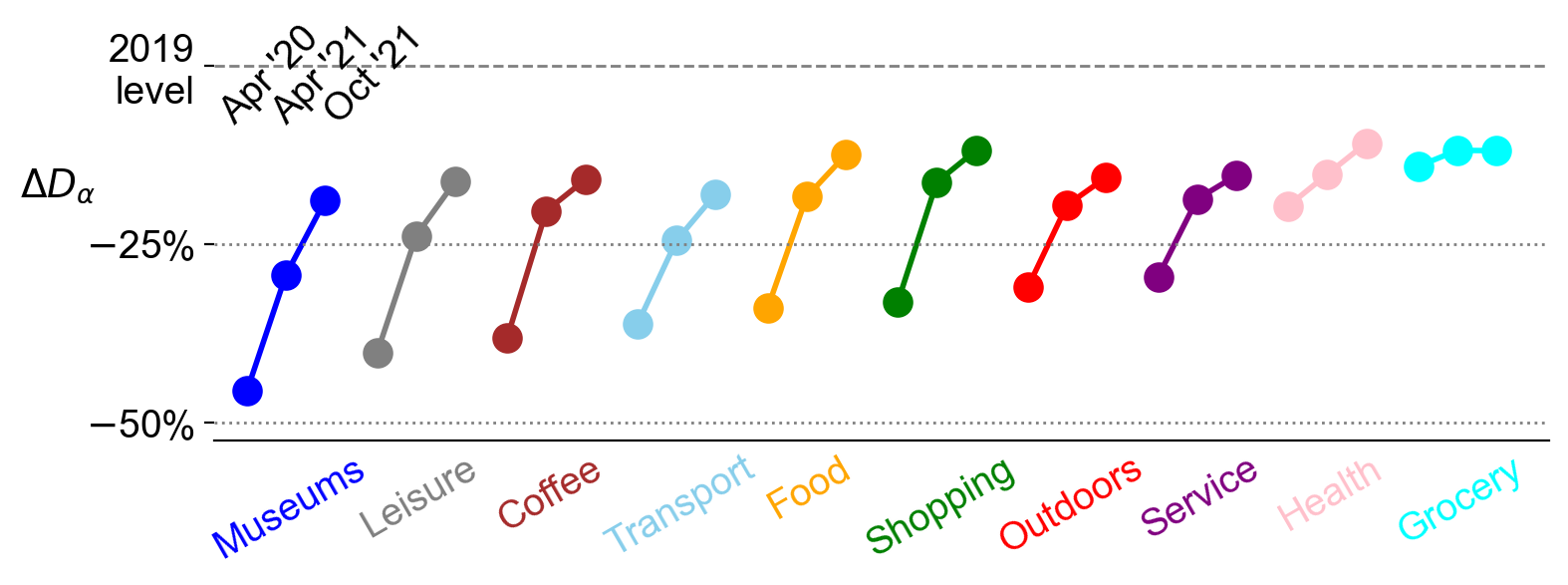}}\\
\subfloat[Dallas]{\includegraphics[width=\linewidth]{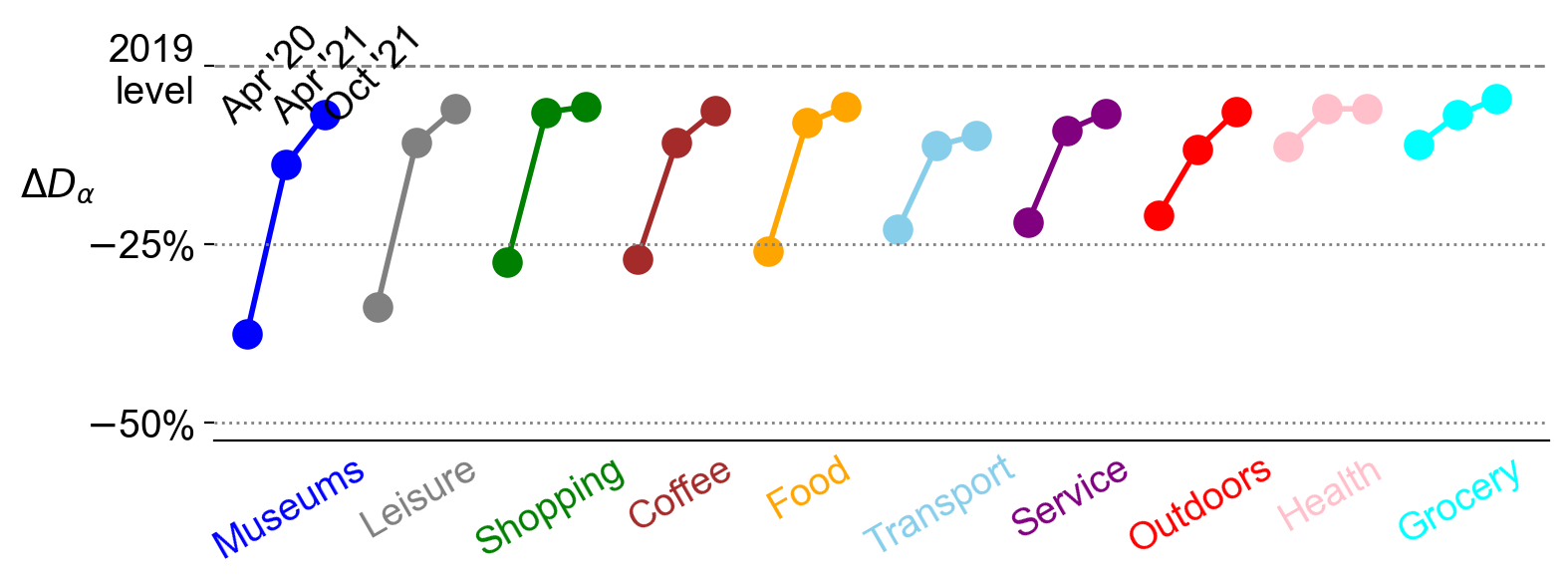}}
\caption{Average income diversity of encounters at different place categories in the three CBSAs (excluding Boston, which was in main manuscript) across four time periods.}
\label{fig:poi_diversity_fourcities}
\end{figure}

\subsection{Income diversity at places}
To measure the income diversity of encounters experienced at each place $\alpha$ in each city, we compute the proportion of total time spent at place $\alpha$ by each income quantile $q$, $\tau_{q\alpha}$. Income thresholds for the quantiles are chosen based on the income distributions in each city, as described in Section \ref{sec:incomeqs}. We also checked that the results for income diversity are independent of the choice of the number of income quantiles in Section \ref{sec:incomeqs}. We define full diversity of encounters at a place when people from all income quantiles spend the same amount of time, $\tau_{q\alpha}=\frac{1}{4}$ for all $q$. Using the metric used to compute income segregation in urban encounters in previous studies \cite{moro2021mobility}, we define the income diversity experienced at each place $\alpha$, $D_{\alpha}$ as a measure of evenness of time spend by different income quantiles:
\begin{equation*}
    D_{\alpha} = 1 - \frac{2}{3} \sum_{q} |\tau_{q\alpha} - \frac{1}{4}|.
\end{equation*}
The diversity measure is bounded between 0 and 1, where $D_{\alpha}=0$ means there is no diversity (the place is visited by people from only one income quantile), and $D_{\alpha}=1$ indicates that all income quantiles spent equal amount of time at the place. Results in Section \ref{sec:diversity} show that using different popular measures of diversity such as entropy does not affect the results on income diversity of encounters.

Figure \ref{fig:placeseg_maps} shows the changes in income diversity at places across four time periods: April 2019 (before the pandemic), April 2020, April 2021, and October 2021 in the four CBSAs. 
Figure \ref{fig:poi_diversity_fourcities} shows the income diversity experienced at different types of places across the four cities, across four time periods. Similar to the results for Boston in Figure 1D in the main manuscript, museums and leisure places had the largest decrease in diversity while health and grocery related places had the smallest decrease in diversity. 
This result agrees with the large decrease in visits to places such as museums, food places, and leisure places, as shown in Figure \ref{fig:poi_visits}, indicating that the decrease in number of visits per user is correlated to the decrease in income diversity experienced at places. We further investigate how much of income diversity reduction is due to the decrease in the number of visits in Supplementary Note 4. 

\begin{figure}
\centering
\includegraphics[width=.9\linewidth]{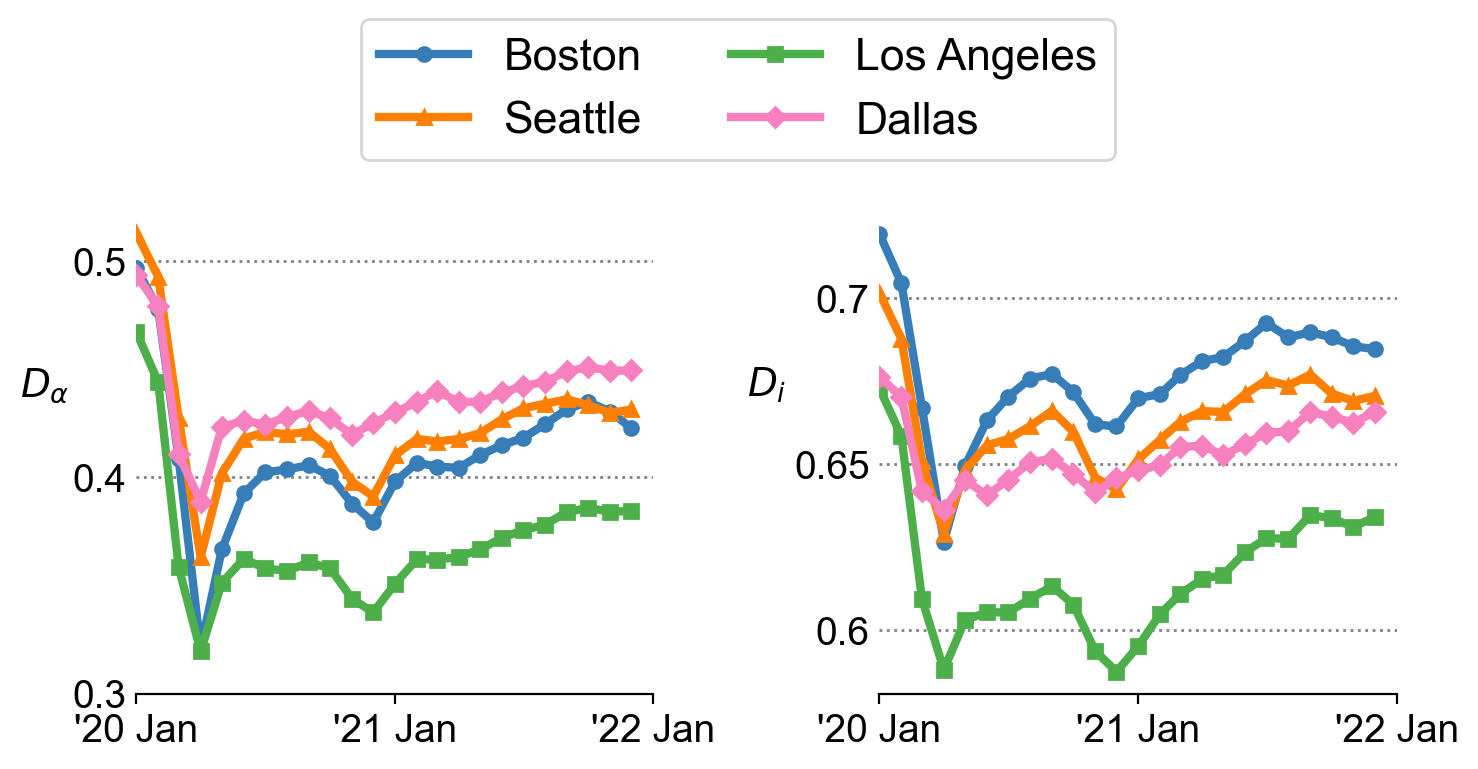}
\caption{Income diversity of encounters experienced at places and by individuals across time for the four CBSAs.}
\label{fig:diversity_absolute}
\end{figure}

\subsection{Income diversity experienced by individuals}
In addition to the income diversity experienced at places, we are interested in measuring the income diversity that each individual experiences across all places they visit. Given the proportion of time individual $i$ spent at place $\alpha$, $\tau_{i\alpha}$, the individual's relative exposure to income quantile $q$, $\tau_{iq}$ can be computed by:
\begin{equation*}
    \tau_{iq} = \sum_\alpha \tau_{i\alpha}\tau_{q\alpha}.
\end{equation*}
Then, the income diversity experienced by individual $i$ can be measured using the same equation used for places: 
\begin{equation*}
    D_{i} = 1 - \frac{2}{3} \sum_{q} |\tau_{i\alpha} - \frac{1}{4}|.
\end{equation*}
Note that the exposure to income quantiles are calculated in a probabilistic manner across a two month time horizon to overcome the sparsity in actual encounters observed in the mobility data. Figure \ref{fig:diversity_absolute} shows the average income diversity at places and experienced by individuals for the four CBSAs. Los Angeles has the lowest income diversity both at places and by individuals out of the four cities. Different cities, which are located in different states, were restricted with COVID-19 lockdown polices of different levels of strictness. We investigate the regional differences from this perspective in Figure 4 in the main manuscript and in Section \ref{sec:stringency} in the Supplementary material. All monthly time series data, including the mean place diversity and individual diversity data are de-seasonalized by removing the monthly fluctuations (simply the deviations from the annual mean) observed in 2019. Most of the results in the main manuscript are shown by percentage differences, which is computed by $\Delta D_i(t) = \frac{D_i(t) - D_i(2019)}{D_i(2019)} \times 100(\%)$, where $D_i(2019)$ is the income diversity of encounters observed on the same month as $t$ in 2019, before the pandemic.







\subsection{Other measure of diversity: entropy} \label{sec:diversity}
The metric for diversity used in our study captures the (un)evenness of exposure between different income quantile groups adopted in previous studies \cite{moro2021mobility}. Another popular metric used to measure the (un)evenness of distribution groups is the entropy metric, which has been used in previous studies related to the diversity of communication networks across cities \cite{eagle2010network}. In our scenario, the entropy of the physical encounters at places are computed as the following: 
\begin{equation*}
    H_{\alpha} = \frac{1}{\log 4} \sum_{q=1}^4 \tau_{q\alpha} \log \tau_{q\alpha}.
\end{equation*}

\begin{figure}
\centering
\includegraphics[width=\linewidth]{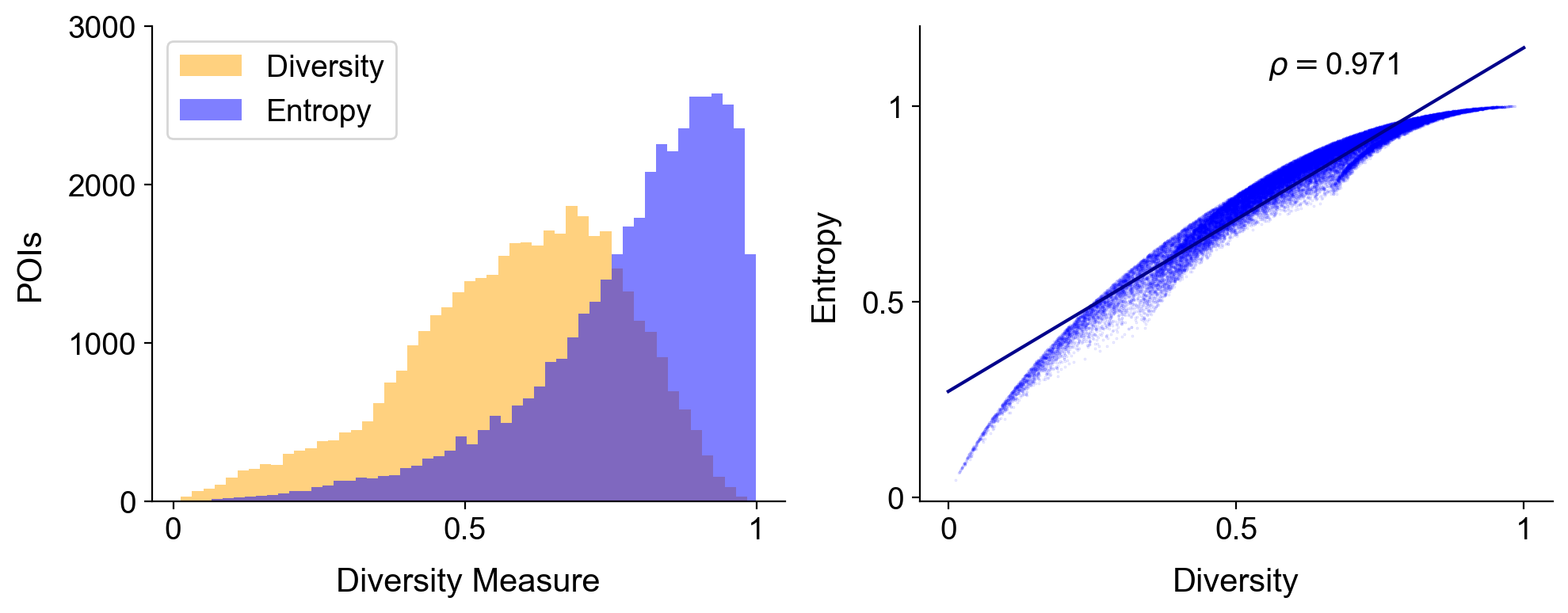}
\caption{Comparison of the diversity and entropy metrics.}
\label{fig:entropy_comparison}
\end{figure}

\begin{figure}
\centering
\includegraphics[width=\linewidth]{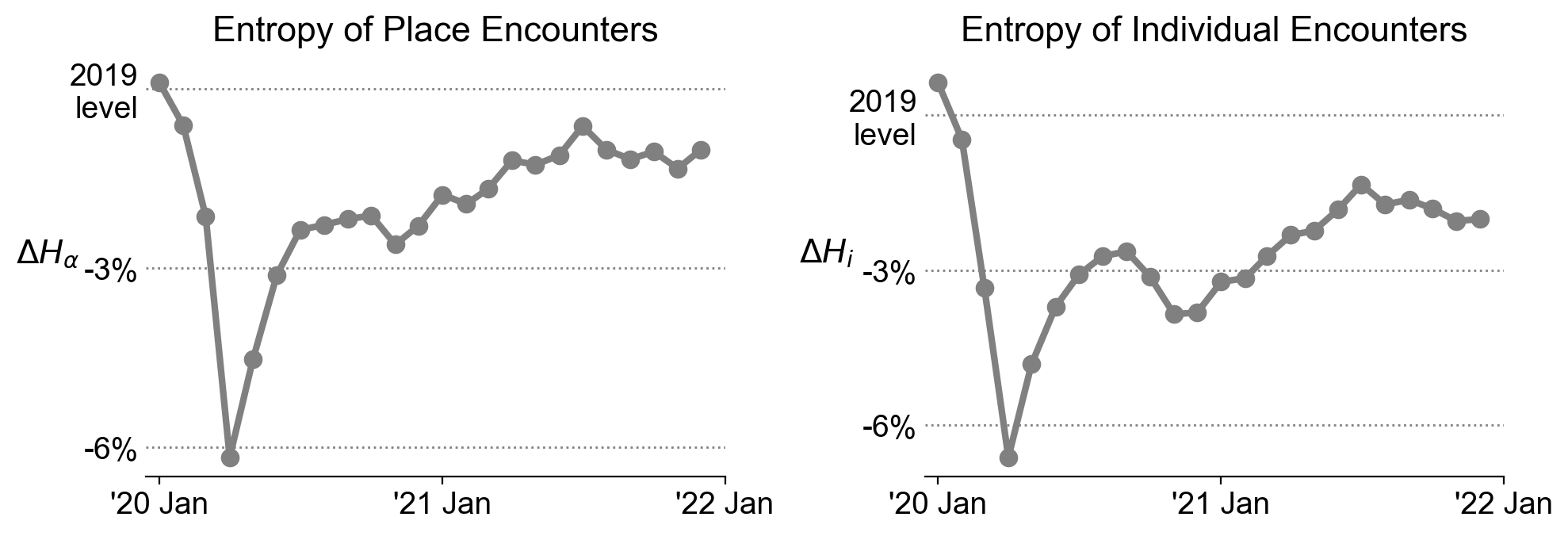}
\caption{Income diversity of encounters measured using the entropy metric.}
\label{fig:entropy_timeseries}
\end{figure}

The left panel in Figure \ref{fig:entropy_comparison} shows the histogram of the diversity (used in our study) and entropy of the encounters taken place at places. The histograms shows how the entropy metric is heavily skewed to high values between 0.8 and 1.0, whereas the diversity metric has relatively larger variability, spanning from 0 to 1. Despite these different characteristics, the right panel in Figure \ref{fig:entropy_comparison} plots the correlation between the diversity (x-axis) and entropy (y-axis) metrics. The Pearson's correlation between these two metrics is very high ($\rho=0.971$), indicating that these two different metrics are both able to capture the income diversity of encounters. 

Indeed, when using the entropy metric to measure the changes in diversity of encounters experienced at places and by individuals, we obtain similar results to when we use the diversity metric. 
Figure \ref{fig:entropy_timeseries} shows how similar to Figure 1C in the main manuscript, we observe a decrease in income diversity of encounters during the first and second waves (April 2020 and December 2020). Moreover, the long-term decrease in diversity in late 2021 is consistent with the results using the diversity metric. Because of the consistency in the key insights between the two metrics, both these metrics are suitable for measuring the income diversity in encounters. Given the wider variability in the range between 0 and 1, we employ the diversity metric as our main metric for measuring income diversity.

\section{Counterfactual simulations} \label{sec:counterfac}

To understand the underlying behavioral changes that contributed to the decrease of income diversity in urban encounters, we design a simulation framework that leverages the pre-pandemic data to create synthetic, counterfactual mobility patterns. 
The synthetic, counterfactual mobility patterns dataset is designed so that while the fundamental behavioral patterns observed in 2019 are kept consistent, the number of visits to different place categories, in different distance ranges, by different income quantiles are reduced to post-pandemic levels. This way, we are able to delineate the effects of different levels of behavioral changes to the total decrease in income diversity. 

\subsection{Synthetic data generation procedure}
The following steps are performed to simulate the synthetic mobility dataset. To create the synthetic counterfactual data for year $y$ and month $m$, denoted as $\mathcal{S}_{y,m}$, we use the mobility data observed in the year 2019 on the same month $m$ as input data $\mathcal{D}_{2019,m}$, for example, to create a synthetic mobility dataset for April 2020, we use the mobility data observed in April 2019. Three different synthetic data, $\mathcal{S}^{(i)}_{y,m}$, $\mathcal{S}^{(ii-1)}_{y,m}$, $\mathcal{S}^{(ii-2)}_{y,m}$, and $\mathcal{S}^{(ii-3)}_{y,m}$, are created based on different levels of detail. The steps for creating the synthetic datasets are as follows: 

\begin{itemize}
    \item $\mathcal{S}^{(i)}_{y,m}$: Randomly remove visits from $\mathcal{D}_{2019,m}$ to adjust the total amount of time spent at places outside home or workplaces to match $\mathcal{D}_{y,m}$
    \begin{itemize}
        \item Visits are randomly retained by rate $r(y,m) = \min \big( 1, \frac{\sum_{i\in {\mathcal{D}_{y,m}}} \tau_i}{\sum_{i\in {\mathcal{D}_{2019,m}}} \tau_i} \big)$, where $\sum_{i\in x} \tau_i$ is the total amount of dwell time duration spent by all users in dataset $x$. As a result, we obtain $\mathcal{S}^{(i)}_{y,m}$ which is a modified version of $\mathcal{D}_{2019,m}$ with adjusted total activity based on observations in the target year $y$ and month $m$.
    \end{itemize}
    \item $\mathcal{S}^{(ii-1)}_{y,m}$: Randomly remove visits from $\mathcal{D}_{2019,m}$ by income quantiles $q$ to adjust the total dwell time at places
        \begin{itemize}
            \item Visits are randomly retained by rate $r(y,m,q)= \min \big( 1,\frac{\sum_{i\in {\mathcal{D}_{y,m}(q)}} \tau_i}{\sum_{i\in {\mathcal{D}_{2019,m}(q)}} \tau_i}\big)$, where $\sum_{i\in x(q)} \tau_i$ is the total amount of dwell time spent by all users in dataset $x$ by users from income quantile $q$. As a result, we obtain $\mathcal{S}^{(ii-1)}_{y,m}$ which is a modified dataset of $\mathcal{D}_{2019,m}$ with adjusted number of visits based on observations in the target year $y$ and month $m$.
        \end{itemize}
    \item $\mathcal{S}^{(ii-2)}_{y,m}$: Randomly remove visits from $\mathcal{D}_{2019,m}$ by income quantiles $q$ and traveled distance $d$ to adjust the total dwell time at places
        \begin{itemize}
            \item Visits are randomly retained by rate $r(y,m,q,d)= \min \big( 1,\frac{\sum_{i\in {\mathcal{D}_{y,m}(q,d)}} \tau_i}{\sum_{i\in {\mathcal{D}_{2019,m}(q,d)}} \tau_i}\big)$, where $\sum_{i\in x(q,d)} \tau_i$ is the total amount of dwell time spent by all users in dataset $x$ by users from income quantile $q$ within distance $d$ from the user's home location. $d$ was binned into 7 distance ranges: $[0km,1km)$, $[1km,3km)$, $[3km,5km)$, $[5km,10km)$, $[10km,20km)$, $[20km,40km)$, $[40km,\infty]$ to obtain rates for each category. As a result, we obtain $\mathcal{S}^{(ii-2)}_{y,m}$ which is a modified dataset of $\mathcal{D}_{2019,m}$ with adjusted number of visits based on observations in the target year $y$ and month $m$.
        \end{itemize}
    \item $\mathcal{S}^{(ii-3)}_{y,m}$: Randomly remove visits from $\mathcal{D}_{2019,m}$ by income quantiles $q$, place taxonomy $c$, and traveled distance $d$ to adjust the total dwell time spent at places
    \begin{itemize}
        \item Visits are randomly retained by rate
        $r(y,m,q,d,c)= \min \big(1, \frac{\sum_{i\in {\mathcal{D}_{y,m}(q,d,c)}} \tau_i}{\sum_{i\in {\mathcal{D}_{2019,m}(q,d,c)}} \tau_i} \big)$,
        where $\sum_{i\in x(q,d,c)} \tau_i$ is the total amount of dwell time spent by all users in dataset $x$ by users from income quantile $q$, to places in major taxonomy $c$, within distance $d$ from the user's home location. Similar to the previous counterfactual, $d$ was binned into the same 7 distance ranges to obtain rates for each category. The 10 taxonomies shown in Table \ref{table:pois} are used. As a result, we obtain $\mathcal{S}^{(iii-3)}_{y,m}$ which is a modified version of $\mathcal{D}_{2019,m}$ with adjusted number of visits based on observations in the target year $y$ and month $m$.
        \end{itemize}
\end{itemize}

\begin{figure}
\centering
\includegraphics[width=\linewidth]{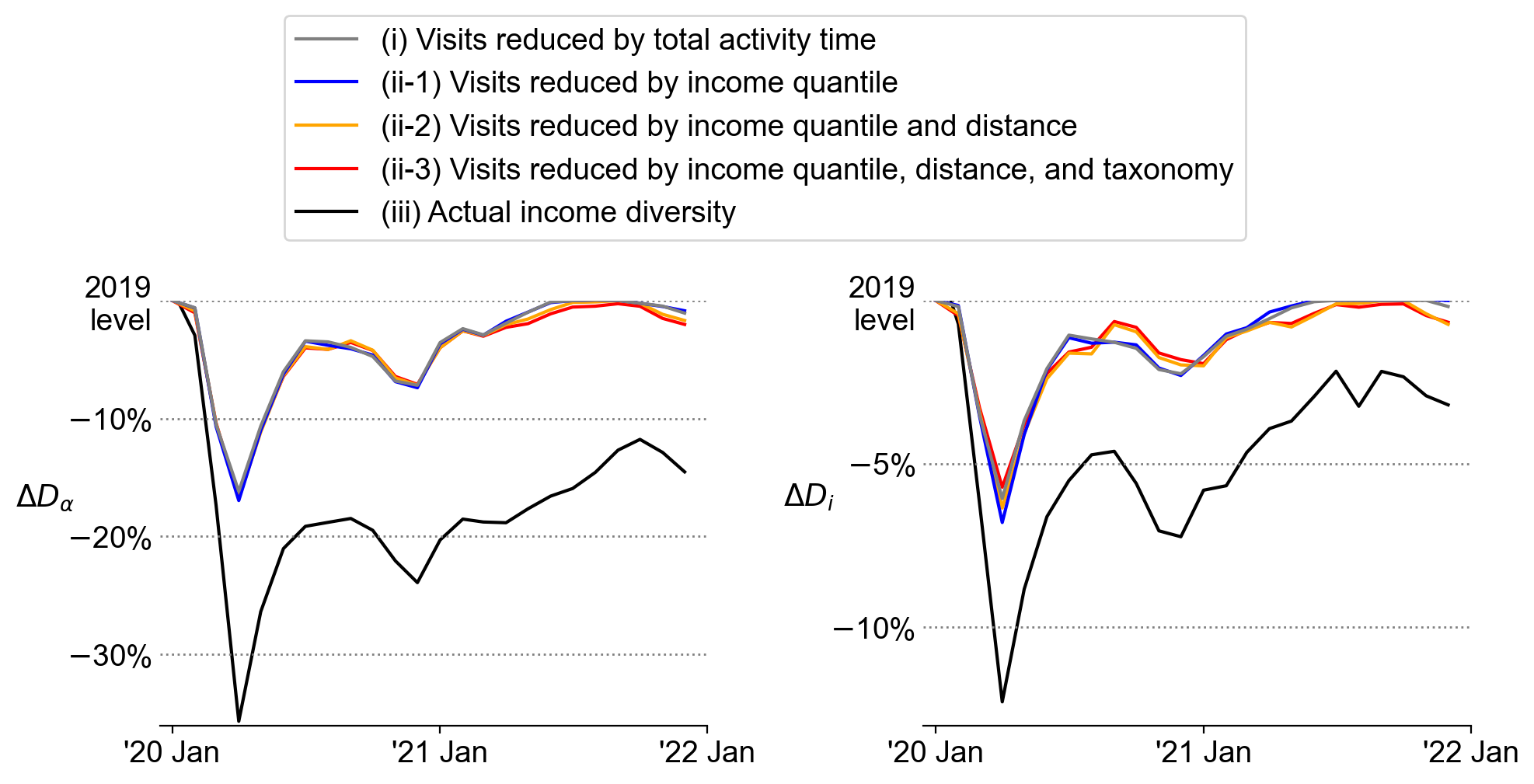}
\caption[Comparison of counterfactual scenarios]{Comparison of counterfactual scenarios for Boston where the visits are reduced based on (i) total activity time, (ii-1) activity time categorized by income quantiles, (ii-2) activity time categorized by income quantiles and distance distributions, and (ii-3) activity time categorized by income quantiles, distance distributions, and POI taxonomy, and (iii) actual income diversity. Scenarios (i) and (ii-2) was employed in the main manuscript since there was little difference between scenarios (i) and (ii-1), and (ii-2) and (ii-3), respectively.}
\label{fig:counterfac_all}
\end{figure}

\begin{figure}
\centering
\subfloat[Retain rate using total dwell time.]{\includegraphics[width=.5\linewidth]{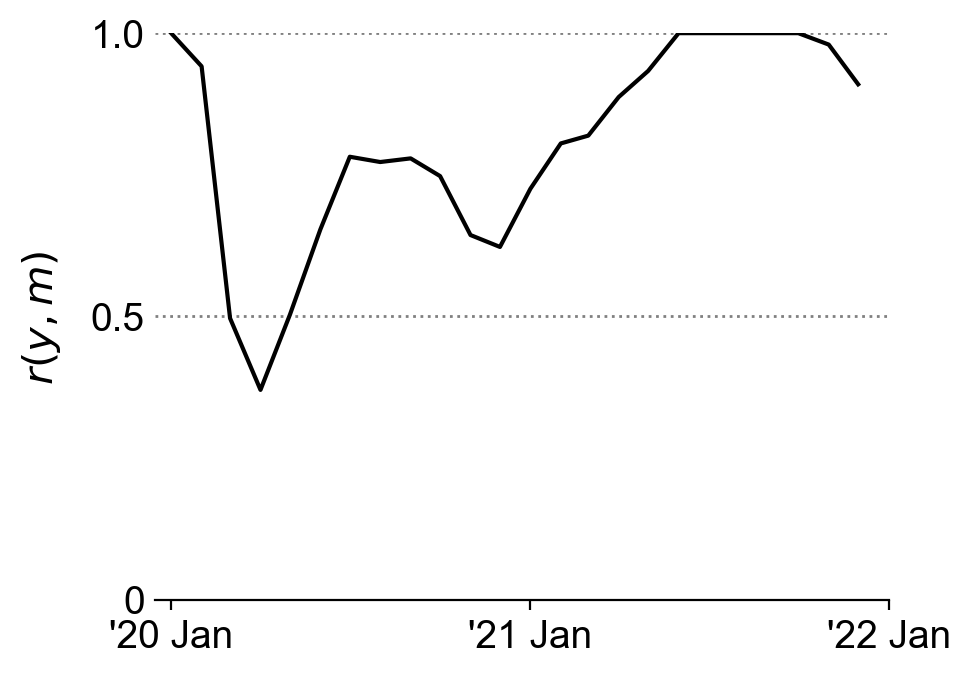}} \\
\subfloat[Retain rate by income quantiles.]{\includegraphics[width=.5\linewidth]{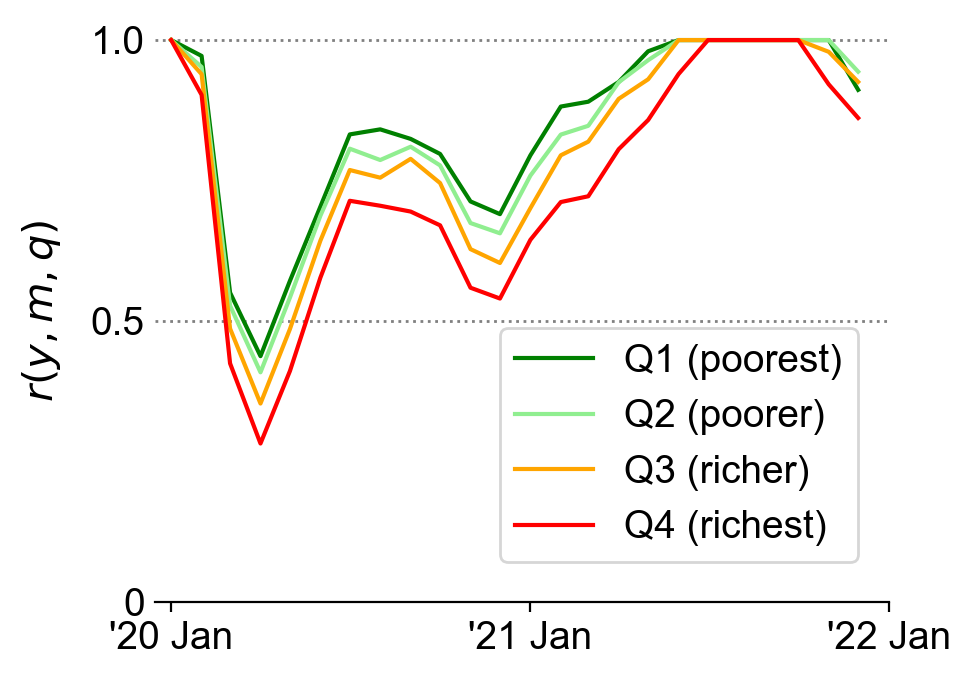}} \\
\subfloat[Retain rate by travel distance.]{\includegraphics[width=.5\linewidth]{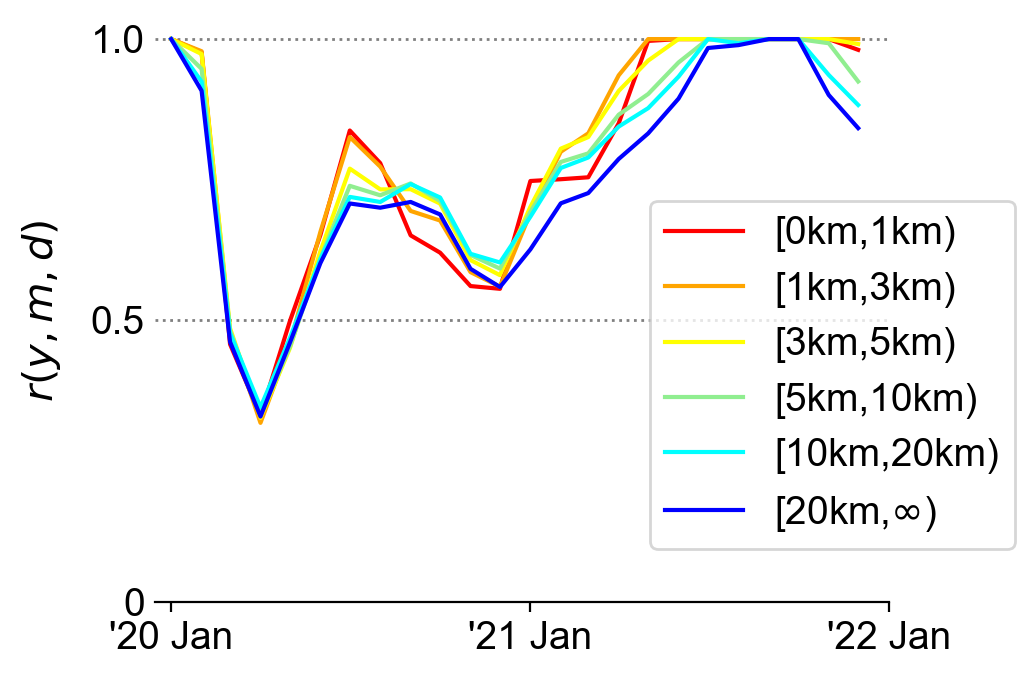}}
\caption{Retain rates used to generate mobility datasets under different counterfactual scenarios.}
\label{fig:retainrates}
\end{figure}

After creating the synthetic counterfactual datasets $\mathcal{S}^{(i)}_{y,m}$, $\mathcal{S}^{(ii-1)}_{y,m}$, $\mathcal{S}^{(ii-2)}_{y,m}$, and $\mathcal{S}^{(ii-3)}_{y,m}$ from the observed changes in aggregate behavior metrics, we compute the income diversity of encounters and compare with the income diversity measured using the actual observed data $\mathcal{D}_{y,m}$. Figure \ref{fig:counterfac_all} shows the percentage changes in income diversity at places $\Delta D_{\alpha}$ and by individuals $\Delta D_{i}$ computed using the different counterfactual datasets. 
The results indicate that the counterfactual scenarios using $\mathcal{S}^{(i)}_{y,m}$, $\mathcal{S}^{(ii-1)}_{y,m}$, $\mathcal{S}^{(ii-2)}_{y,m}$, and $\mathcal{S}^{(ii-3)}_{y,m}$ yield similar results. In particular, results from $\mathcal{S}^{(i)}_{y,m}$ and $\mathcal{S}^{(ii-1)}_{y,m}$, and  $\mathcal{S}^{(ii-2)}_{y,m}$ and $\mathcal{S}^{(ii-3)}_{y,m}$, generate similar patterns, indicating that the effects of controlling by income quantiles and place taxonomies are negligible. 

To summarize the findings, counterfactual simulations show that:
\begin{enumerate}
    \item Using different retain rates across income quantiles have no effect on income diversity measures (no difference between $\mathcal{S}^{(i)}_{y,m}$ and $\mathcal{S}^{(ii-1)}_{y,m}$);
    \item Using different retain rates across distance distributions have slight effects on income diversity measures (slight difference between $\mathcal{S}^{(ii-1)}_{y,m}$ and $\mathcal{S}^{(ii-2)}_{y,m}$), and;
    \item Using different retain rates across place taxonomies (major categories) have no effect on income diversity measures (no difference between $\mathcal{S}^{(ii-2)}_{y,m}$ and $\mathcal{S}^{(ii-3)}_{y,m}$),
\end{enumerate}
which will be further investigated in the following sections.

\begin{figure}
\centering
\subfloat[Histograms of $\tau_q$ for $q_1$,$q_2$,$q_3$,$q_4$, respectively, for counterfactual scenarios (i) and (ii-1).]{\includegraphics[width=\linewidth]{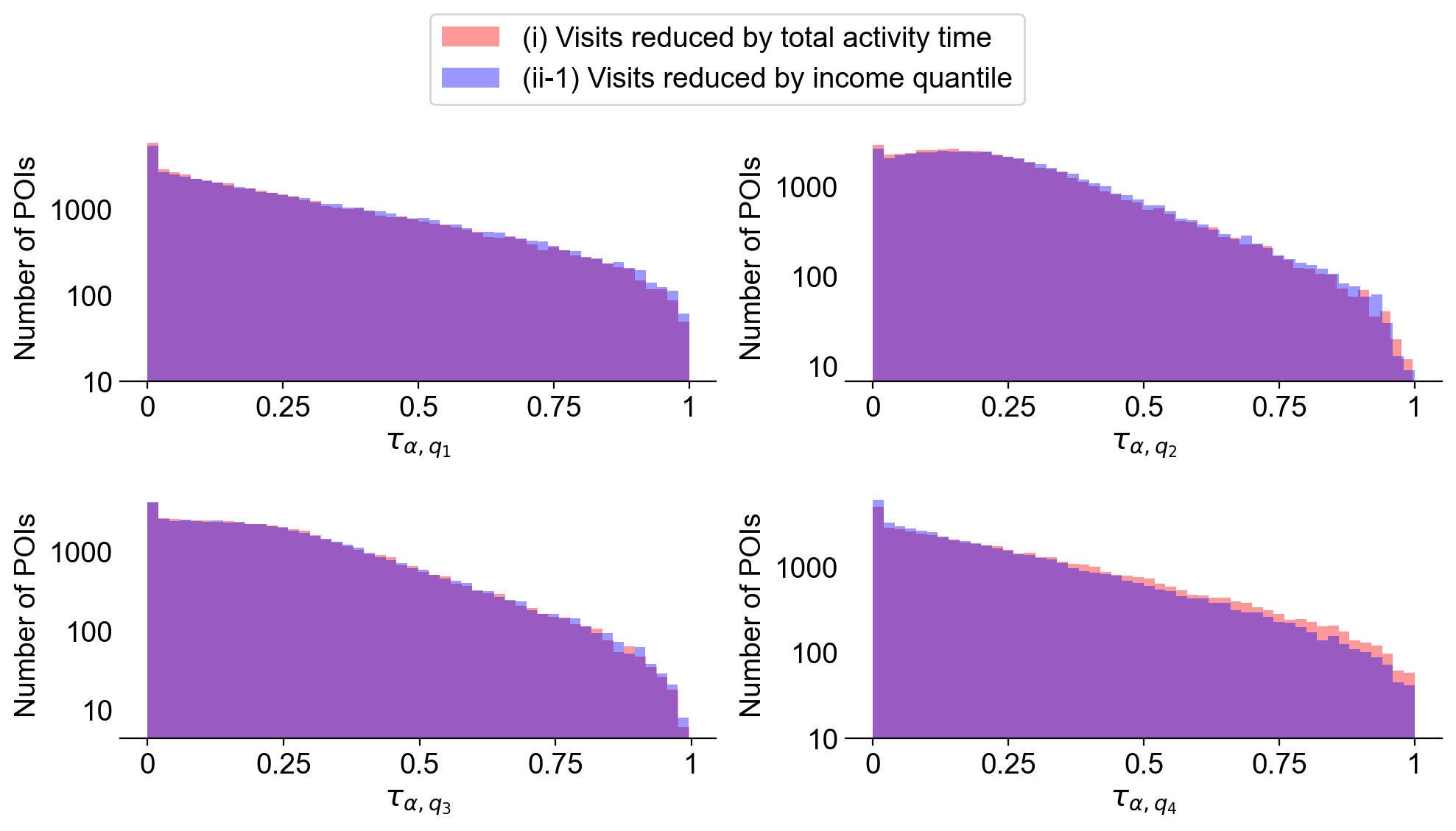}}\\
\subfloat[Histograms of $\tau_{q \in \{q_1, q_2, q_3, q_4\}}$ for counterfactual scenarios (i) and (ii-1).]{
\includegraphics[width=.6\linewidth]{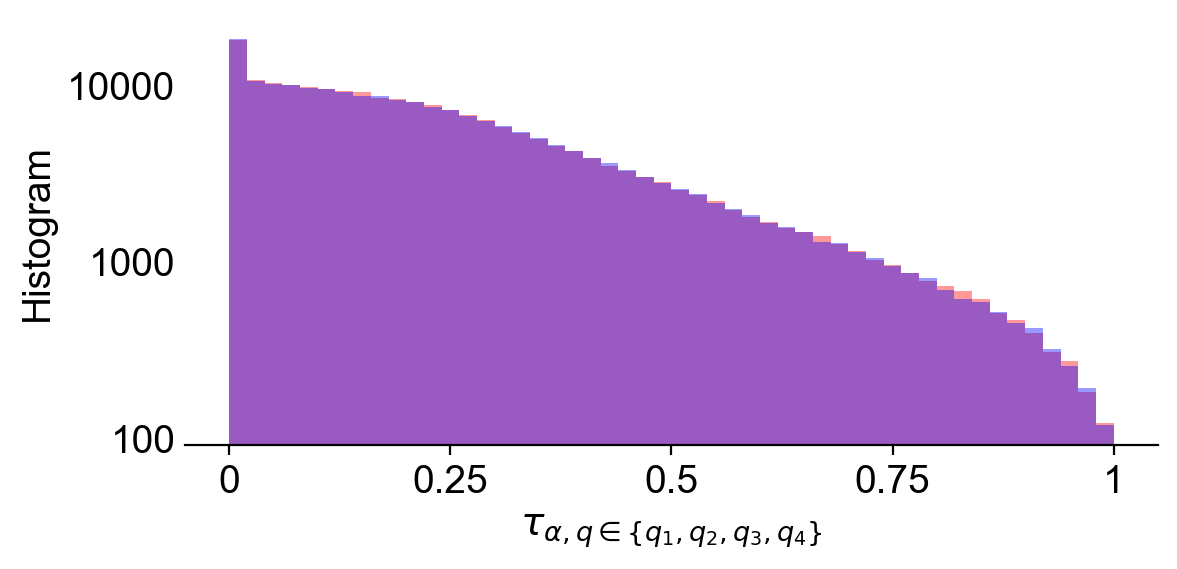}}
\caption{Differences in the distributions of $\tau_q$ between counterfactual scenarios (i) and (ii-1) are significant for each income quantile, but are nearly identical when aggregated across all income quantiles, yielding similar income diversity measures.}
\label{fig:taus}
\end{figure}

\begin{figure}
\centering
\subfloat[Retain rates for different place taxonomies (major categories).]{\includegraphics[width=\linewidth]{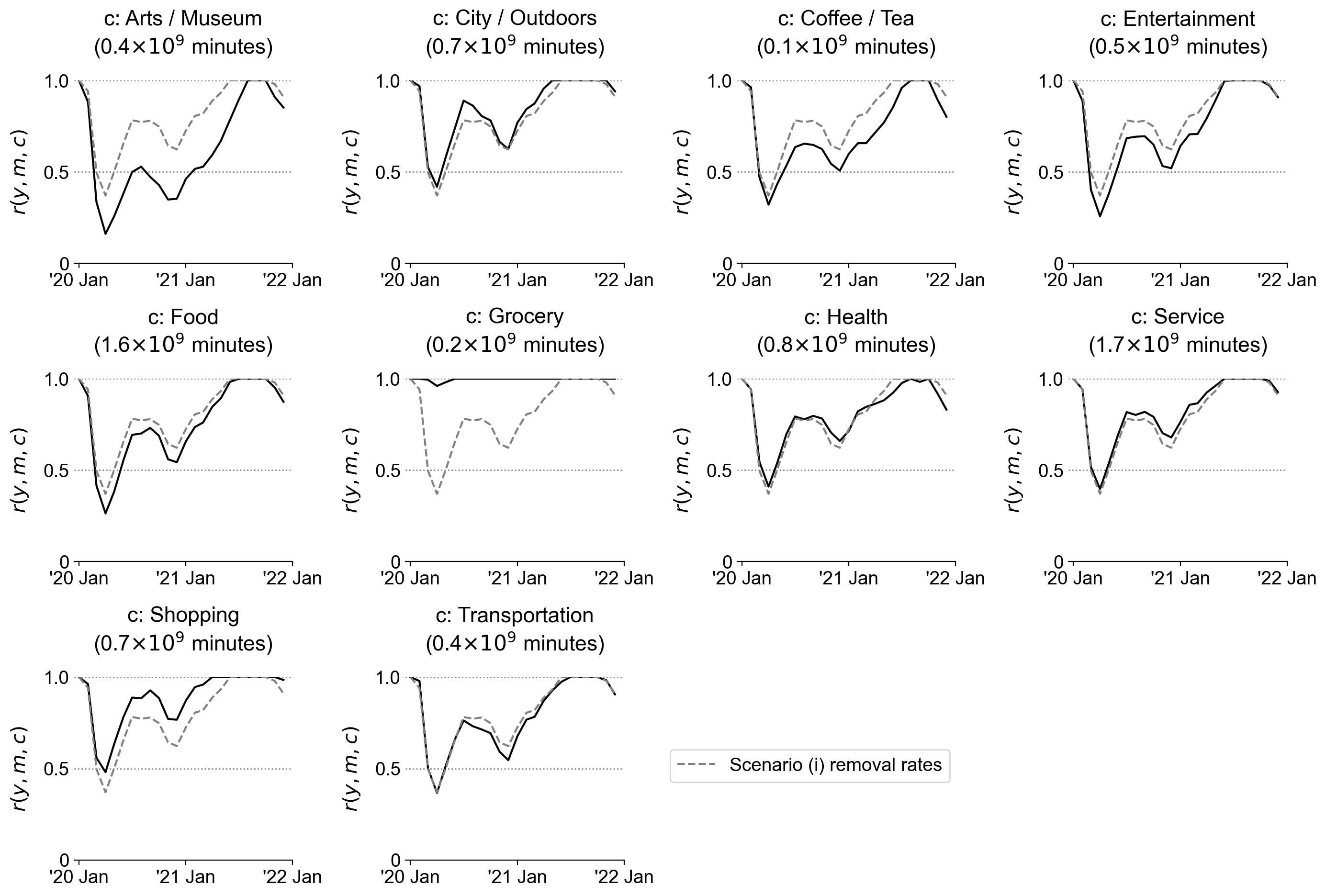}}\\
\subfloat[Average baseline diversity metric for each place taxonomy.]{
\includegraphics[width=.45\linewidth]{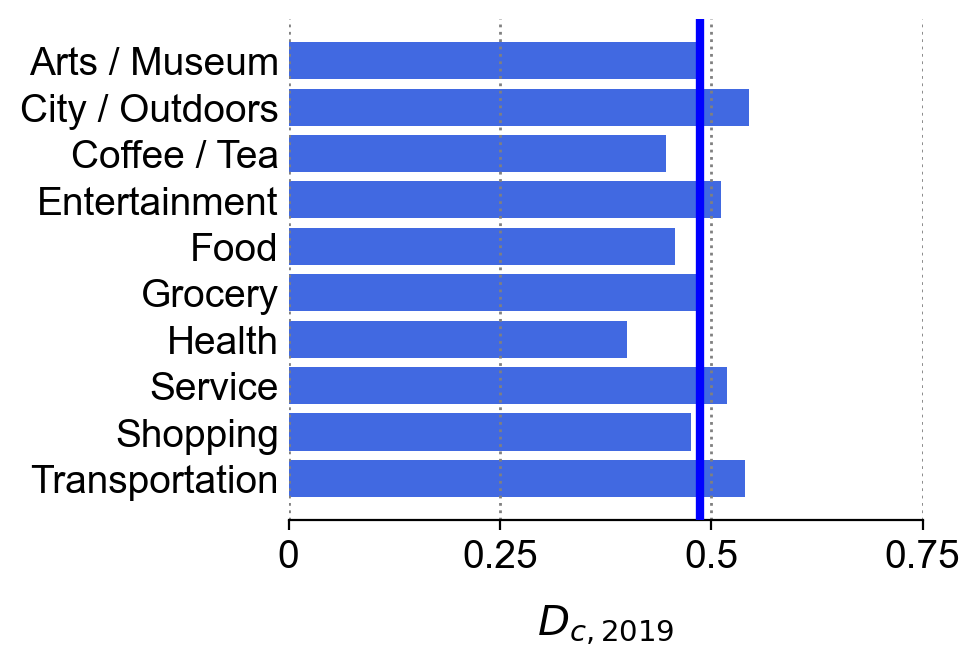}}
\subfloat[Average diversity weighted by category-level retain rates.]{
\includegraphics[width=.45\linewidth]{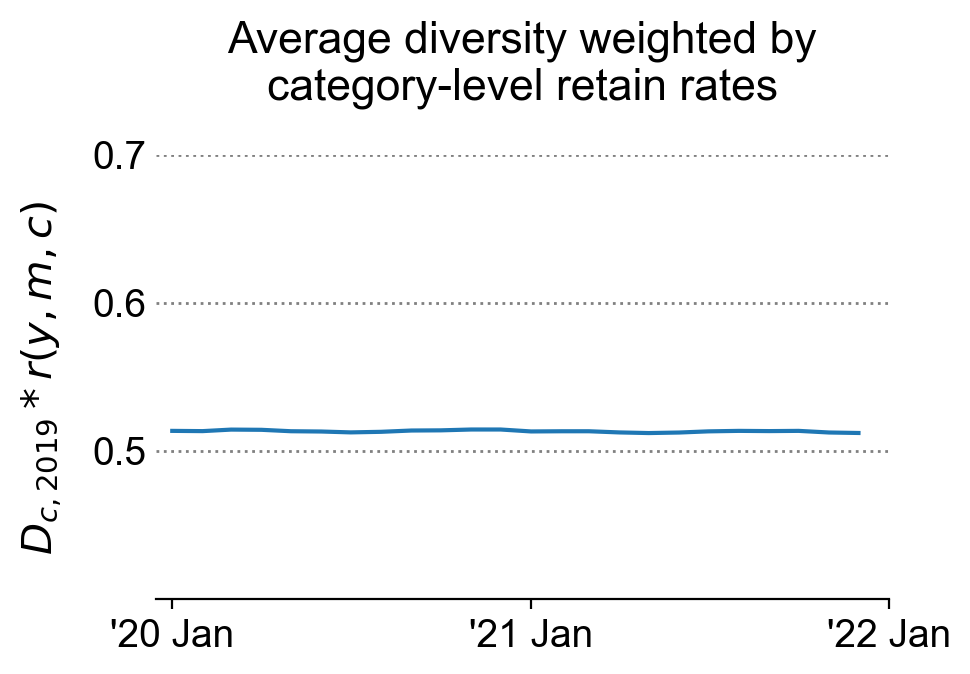}}
\caption{Heterogeneous retain rates across place taxonomies (major categories) suggest income diversity measures to be affected by adding place taxonomies as a constraint for creating counterfactual datasets in (ii-3). However, the effects are close to zero, since place categories which have substantially different retain rates (i.e., grocery and arts/museums) have average level diversity measures.}
\label{fig:activitymix}
\end{figure}

\subsection{Analysis of the impacts of removal rates under different scenarios}

\subsubsection*{1. Effects of different retain rates across income quantiles}
To understand why the impacts of using different retain rates across income quantiles (as shown in Figure \ref{fig:retainrates}b) yield no difference in diversity decrease, we plot the histograms of $\tau_\alpha,q$ of each place $\alpha$ for each income quantile $q$ and in aggregate, in Figures \ref{fig:taus}a and \ref{fig:taus}b, respectively, for the two counterfactual scenarios (i) and (ii-1). We observe that, in agreement with Figure \ref{fig:retainrates}b, during the pandemic $\tau_{q_1}$ and $\tau_{q_2}$ increased and $\tau_{q_4}$ decreased due to poorer populations disproportionately visiting places than richer people. 
However, when we aggregate and plot the $\tau_{q}$ values for all $q\in \{q_1,q_2,q_3,q_4\}$, there is no significant difference across the two counterfactual scenarios, consequentially yielding similar values of diversity, since the diversity measure does not differentiate whether $q_1$ or $q_4$ had disproportionate dwell time spent at places. 

\subsubsection*{2. Effects of different retain rates across distance distributions}
The retain rates across distance ranges shown in Figure \ref{fig:retainrates}c show that during most of the periods in the pandemic, shorter distance trips (e.g., $[0,1km)$, $[1km,3km)$) have higher retain rates compared to longer distance trips (e.g., $[20km,\infty)$), indicating that people preferred shorter distance trips than longer ones during the pandemic. As shown in previous studies, longer distance trips tend to result in higher diversity, whereas shorter distance trips are less diverse due to stronger effects of residential segregation \cite{moro2021mobility}. Indeed, when we compare results (ii-1) and (ii-2) in Figure \ref{fig:counterfac_all}, especially $\Delta D_i$, scenario (ii-2) has lower diversity during periods when $r(y,m,d)$ for shorter distances are higher than longer distances (i.e., June -- September 2020, January 2021 -- June 2021). On the other hand, scenario (ii-2) has higher diversity during periods when $r(y,m,d)$ for shorter distances are lower than longer distances (i.e., September -- December 2020). These observations show that changes in distance distributions does play a role in the income diversity of urban encounters, despite the small magnitude of the effects as shown in Figure \ref{fig:counterfac_all}. 
 
\subsubsection*{3. Effects of different retain rates across place taxonomies (major categories)}
The retain rates across place taxonomies (major categories) shown in Figure \ref{fig:activitymix}a indicate that different major categories had varying rates during the pandemic. While most categories follow similar patterns as the overall average retain rates shown in Figure \ref{fig:retainrates}a, places such as grocery stores had significantly higher (almost full) retain rates, indicating that dwell times at grocery stores had very small decreases. On the other hand, arts and museums had the largest decrease in retain rates. 
These heterogeneous rates suggest that using different retain rates across place categories when producing the counterfactual dataset (ii-3) could significantly affect the income diversity of $\mathcal{S}^{(ii-3)}_{y,m}$. 

However, as shown in Figure \ref{fig:counterfac_all}, this additional constraint of controlling by place taxonomies yield negligible effects. 
We test this by computing the average diversity weighted by category-level retain rates across time. More specifically, we take the 2019 level diversity of each place taxonomy, $D_{c,2019}$, and re-weight them by the time-varying retain rates $r(y,m,c)$. The results in Figure \ref{fig:activitymix}c show almost a flat trend across time, indicating that the heterogeneity in the time-varying retain rates have no effects on the overall income diversity. 
This can be explained by looking at place taxonomies that had the largest deviations in retain rates -- grocery stores and arts/museum places had close to the average diversity measures, as shown in Figure \ref{fig:activitymix}b. 

\subsection{Summary of counterfactual simulation results}
Since the effects of heterogeneous retain rates across place taxonomies was insignificant, results for counterfactual diversity decrease using the $\mathcal{S}^{(ii-1)}_{y,m}$ and $\mathcal{S}^{(ii-3)}_{y,m}$ datasets were omitted from Figure 2B in the main manuscript.
Figure \ref{fig:counterfac_all} shows the comparison of counterfactual scenarios for Boston where the visits are reduced based on (i) total activity time ($\mathcal{S}^{(i)}_{y,m}$), (ii-1) activity time categorized by distance and income quantile ($\mathcal{S}^{(ii)}_{y,m}$), and (ii-2) activity time categorized by distance, income quantile, and POI taxonomy ($\mathcal{S}^{(iii)}_{y,m}$), and (iii) actual income diversity. Scenario (ii-1) was employed in the main manuscript since there was little difference between scenarios (ii-1) and (ii-2).
For all cities, the decrease in income diversity when we consider the reduction in users and visits by quantile accounts ($\mathcal{S}^{(i)}_{y,m}$) for around 50\% of the reduction in diversity in the initial stages of the pandemic in the early stages of the pandemic. The marginal decrease in the diversity due to the reduction in visits based on place categories and travel distances ($\mathcal{S}^{(ii)}_{y,m}$) is relatively small compared to the reduction in visits. 

\begin{figure}
\centering
\subfloat[Seattle]{\includegraphics[width=\linewidth]{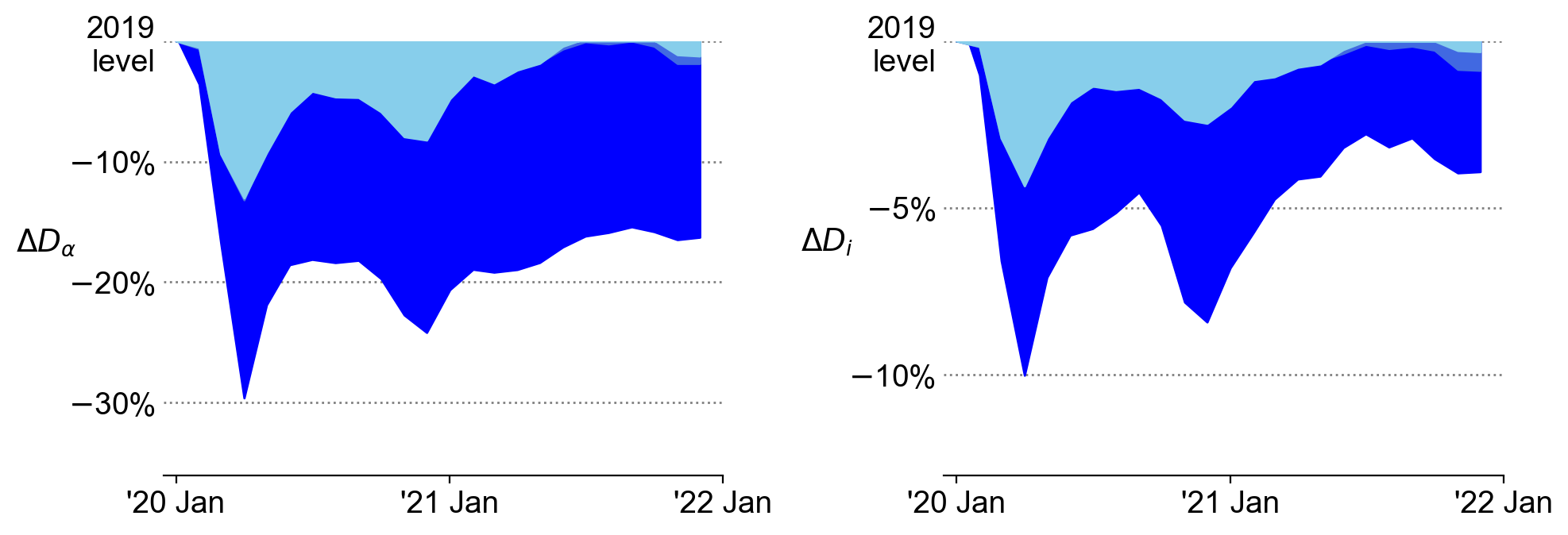}}\\
\subfloat[Los Angeles]{\includegraphics[width=\linewidth]{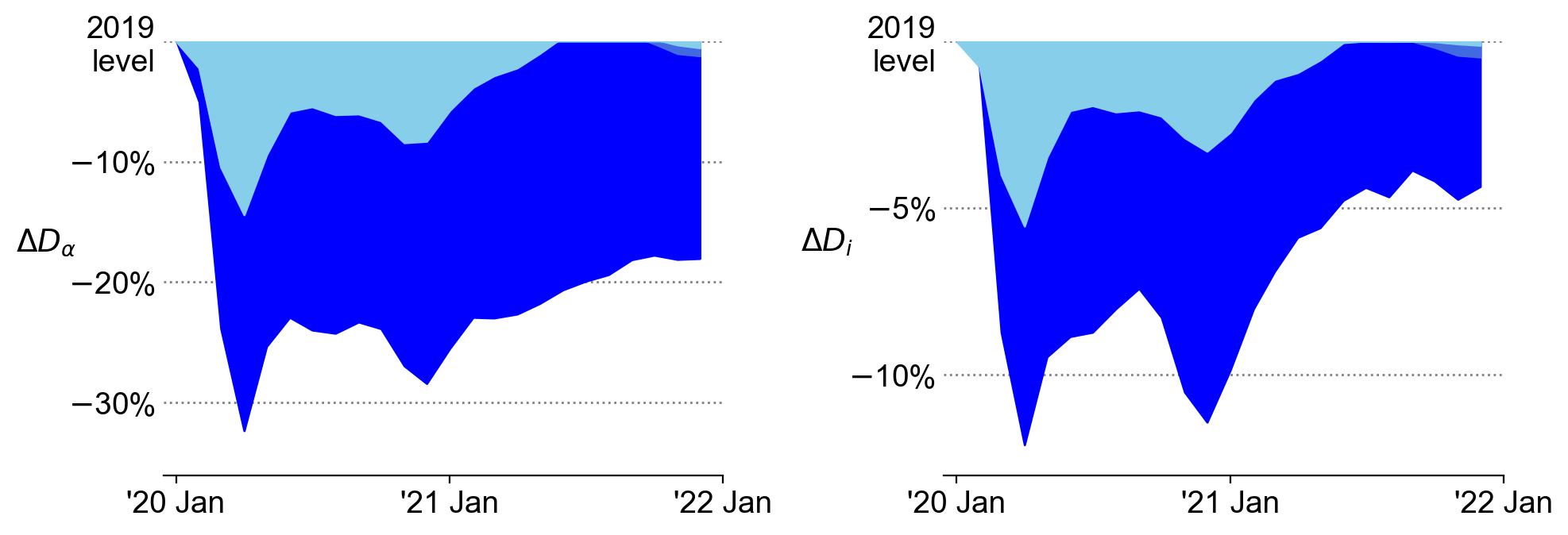}}\\
\subfloat[Dallas]{\includegraphics[width=\linewidth]{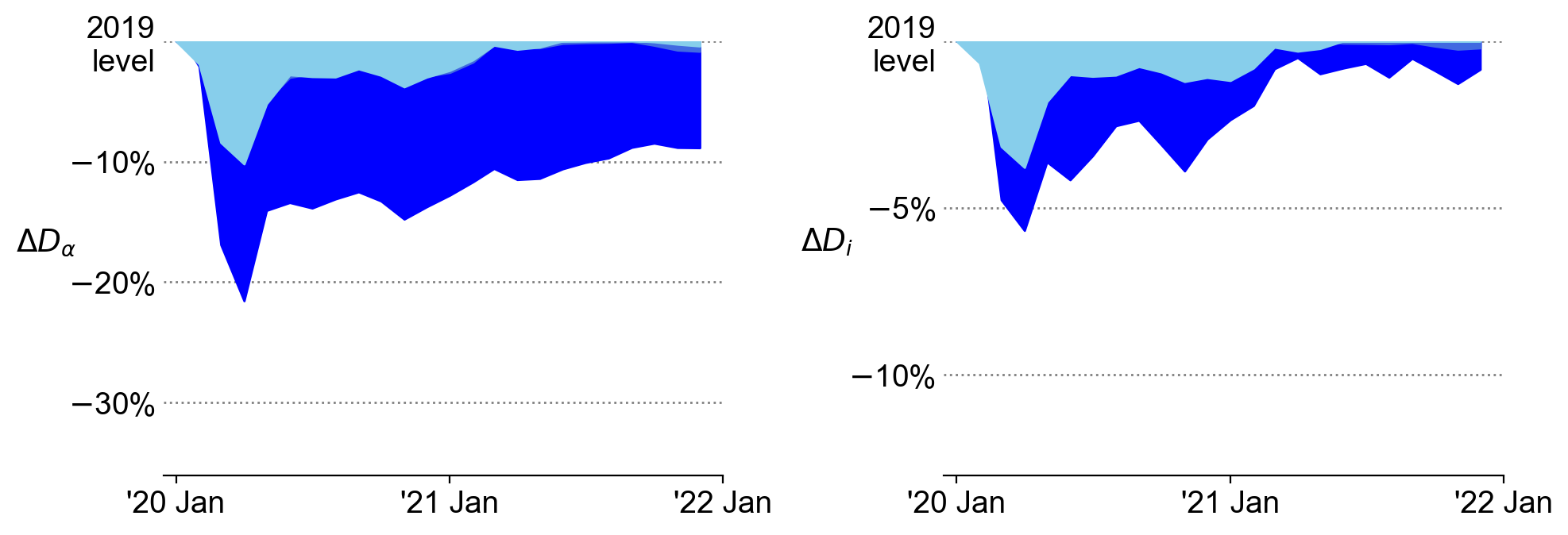}}
\caption{Percentage changes in income diversity of encounters in places and by individuals in the four CBSAs under different synthetic counterfactual scenarios.}
\label{fig:counterfac_cities}
\end{figure}

However, as shown in Figures \ref{fig:counterfac_cities} and \ref{fig:counterfac_all}, these reductions in active users and visits to different categories do not account for all of the reduction in income diversity, and indicates that more microscopic changes in human behavior have contributed to a further decrease in income diversity in cities during the pandemic. 
To investigate what behavioral changes during the pandemic contributed to the decrease in income diversity, we seek to find any microscopic, individual level behavior that changed during the later stages of the pandemic. To do that, we analyze the behavioral parameters of the Social-EPR model (proposed in \cite{moro2021mobility}, which extended the EPR model proposed in \cite{song2010modelling}).

\subsection{Parameters of the Social-EPR model}

The social exploration and preferential return (Social-EPR) model \cite{moro2021mobility,song2010modelling} characterizes visitation patterns of individuals using two mechanisms: exploration (visiting a new place) or preferential return (visiting an already visited place). The probability of exploration when an individual has already visited $S_T$ places is modeled as $P_{new}= \rho S_T^{-\gamma}$, where $\rho$ and $\gamma$ are model parameters. If an individual decides to explore, they then decide whether to socially explore (visit a new place where their income group is not the majority income quantile group) with probability $\sigma_s$. In the case that the individual decides to return, the individual selects the destination $\alpha$ with probability $\Pi_\alpha \sim \tau_{\alpha,i}$, where $\tau_{\alpha,i}$ is the proportion of time already spent at place $\alpha$ by individual $i$. 

\begin{figure}
\centering
\subfloat[Place exploration parameters ]{\includegraphics[width=.48\linewidth]{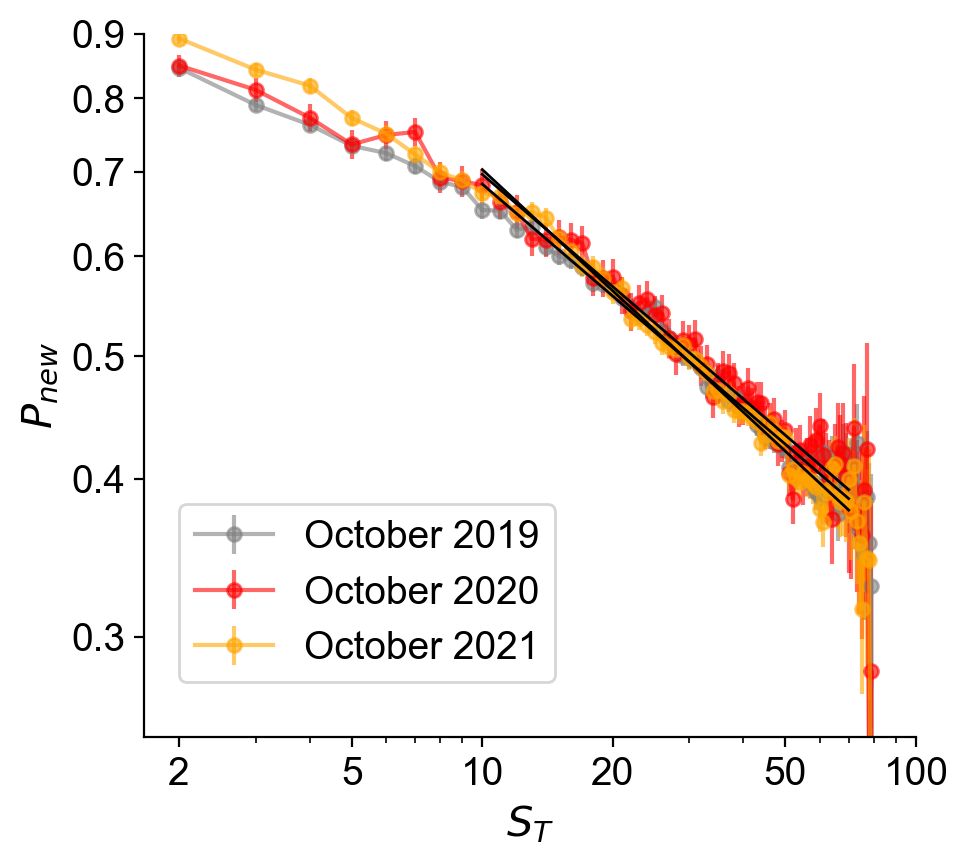}}
\subfloat[Preferential return parameters]{\includegraphics[width=.48\linewidth]{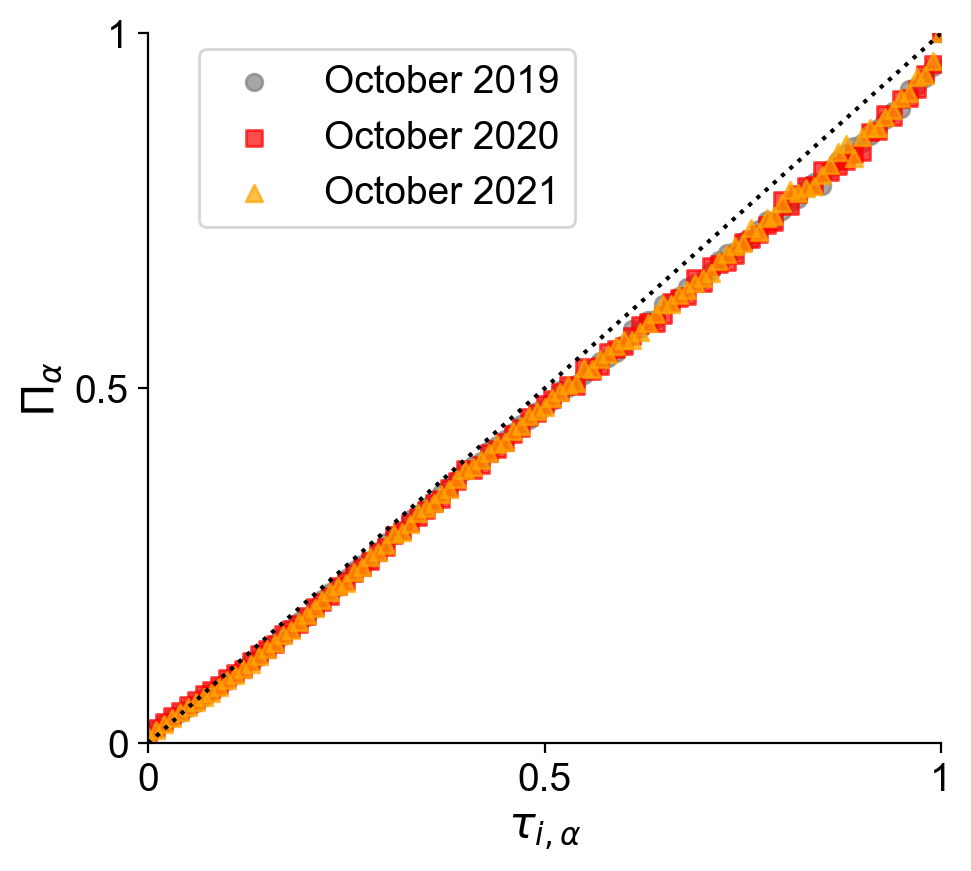}}
\caption{Key parameters of the Social-EPR model, $\rho$, $\gamma$, and $\pi$, are fairly consistent during the pandemic. The social exploration parameter $\sigma_s$ (shown in main manuscript Figure 2D) was the only parameter with significant changes.}
\label{fig:socialexp}
\end{figure}

To investigate whether any of the fundamental behavioral characteristics have changed due to the pandemic, we fitted the Social-EPR model to the observed mobility data patterns and estimated the model parameters across different periods of time. 
The fitted parameters are shown in Figure \ref{fig:socialexp}. 
Surprisingly, we find that the key parameters of the Social-EPR model, including $\rho$, $\gamma$, and linear relationship between $\Pi_\alpha$ and $\tau_{\alpha,i}$, are mostly consistent across time (with the exception of April and May 2020 due to the initial lockdown). This indicates that the fundamental characteristics of individual mobility, including exploration and preferential return, were consistent during the pandemic, when controlled by the number of visits an individual makes. The model parameter with the most significant change during the pandemic was the social exploration parameter $\sigma_s$, as shown in Figure 2D in the main manuscript. 

\begin{figure}
\centering
\subfloat[Seattle]{\includegraphics[width=\linewidth]{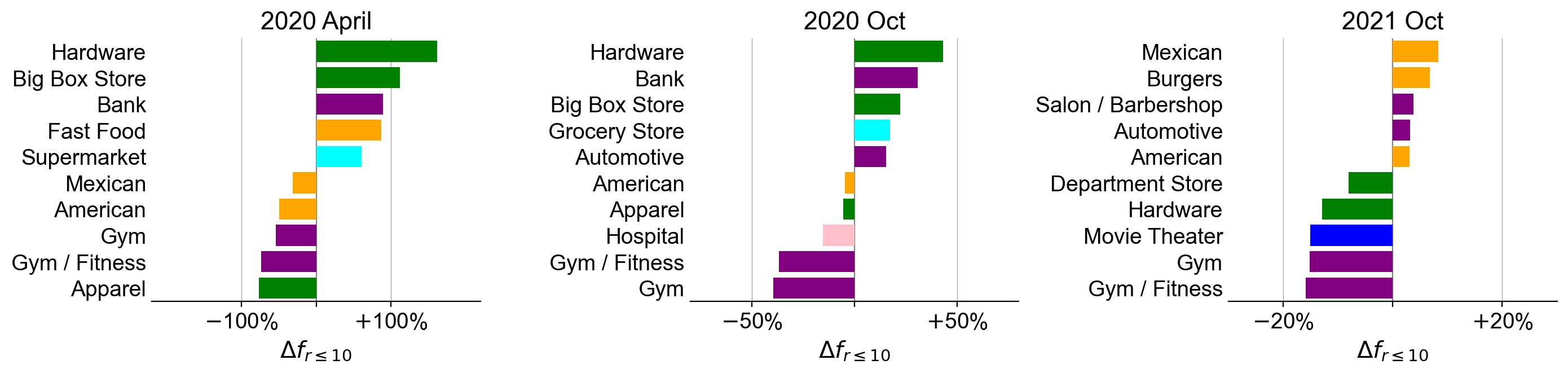}}\\
\subfloat[Los Angeles]{\includegraphics[width=\linewidth]{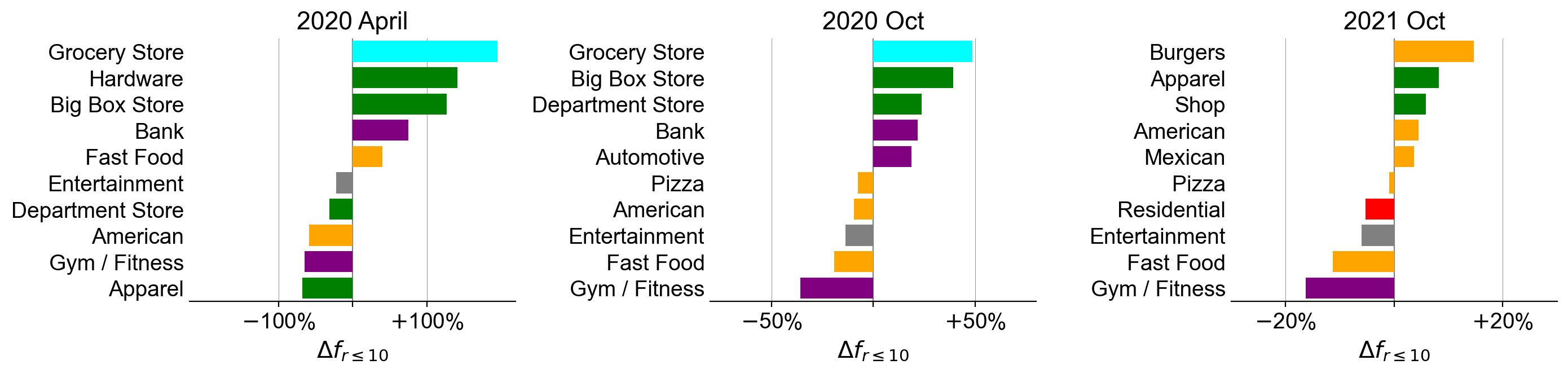}}\\
\subfloat[Dallas]{\includegraphics[width=\linewidth]{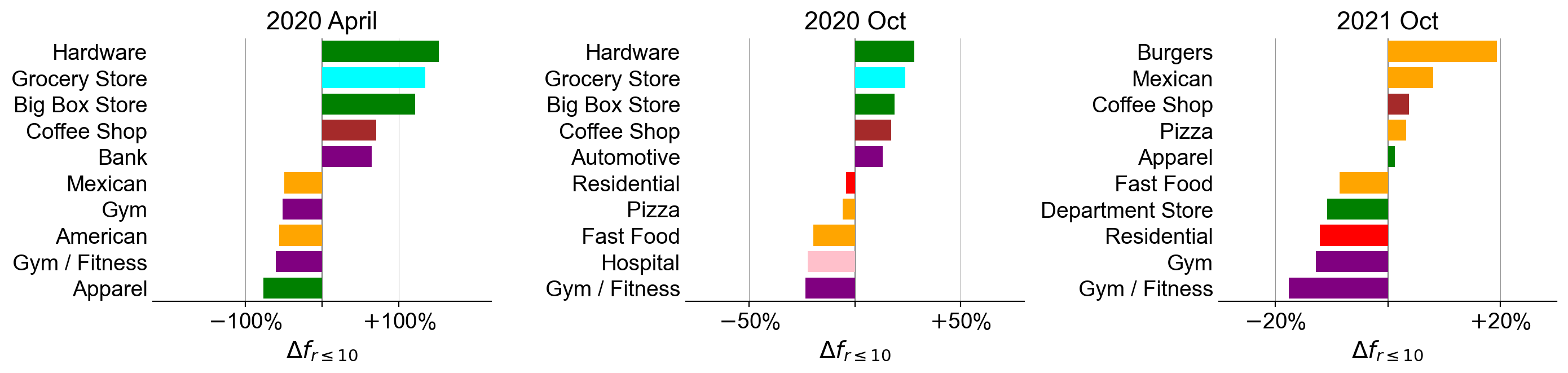}}
\caption{Changes in proportion of high-frequency visitation to place subcategories across different periods of the pandemic in Seattle, Los Angeles, and Dallas (Boston is shown in main manuscript).}
\label{fig:placeprefs}
\end{figure}

From the counterfactual experiment, we found that there is an excess level of decrease in diversity in urban encounters even when controlled for the number of visits to different place categories by different income groups, by travelled distance. The Social-EPR model revealed that such decrease was not due to changes in exploration and preferential return behavior, but because of decrease in social exploration behavior and microscopic changes in where people prefer to visit (sub-category level changes), which is shown in Figure \ref{fig:placeprefs}. Across all four CBSAs, we observe that places such as hardware, big box stores, banks, and grocery stores were the places with the highest increase in the proportion of individuals who visited them with a top-10 frequency, while gyms, food places (pizza, fast food), apparel, movie theaters were places with the largest decrease.


\section{Explaining spatial heterogeneity in diversity}

\begin{figure}
\centering
\subfloat[Seattle]{\includegraphics[width=\linewidth]{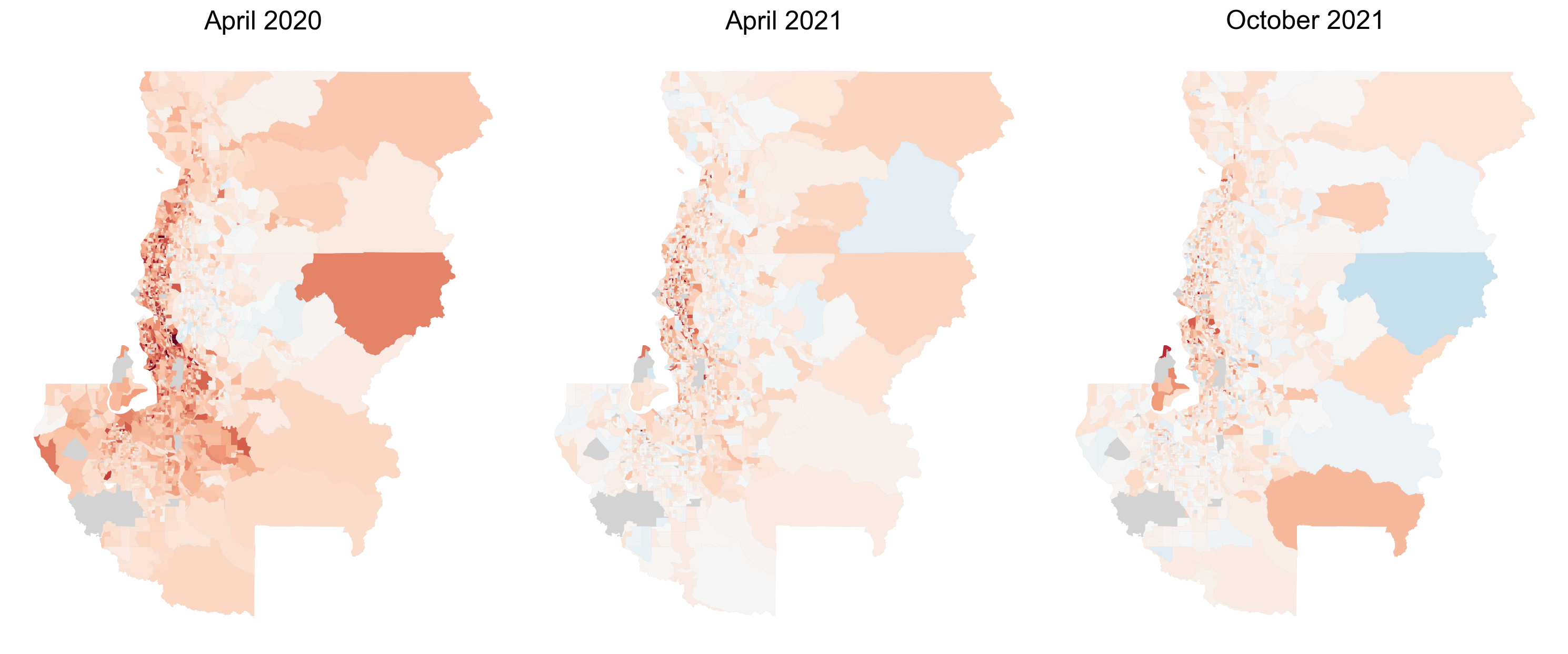}}\\
\subfloat[Los Angeles]{\includegraphics[width=\linewidth]{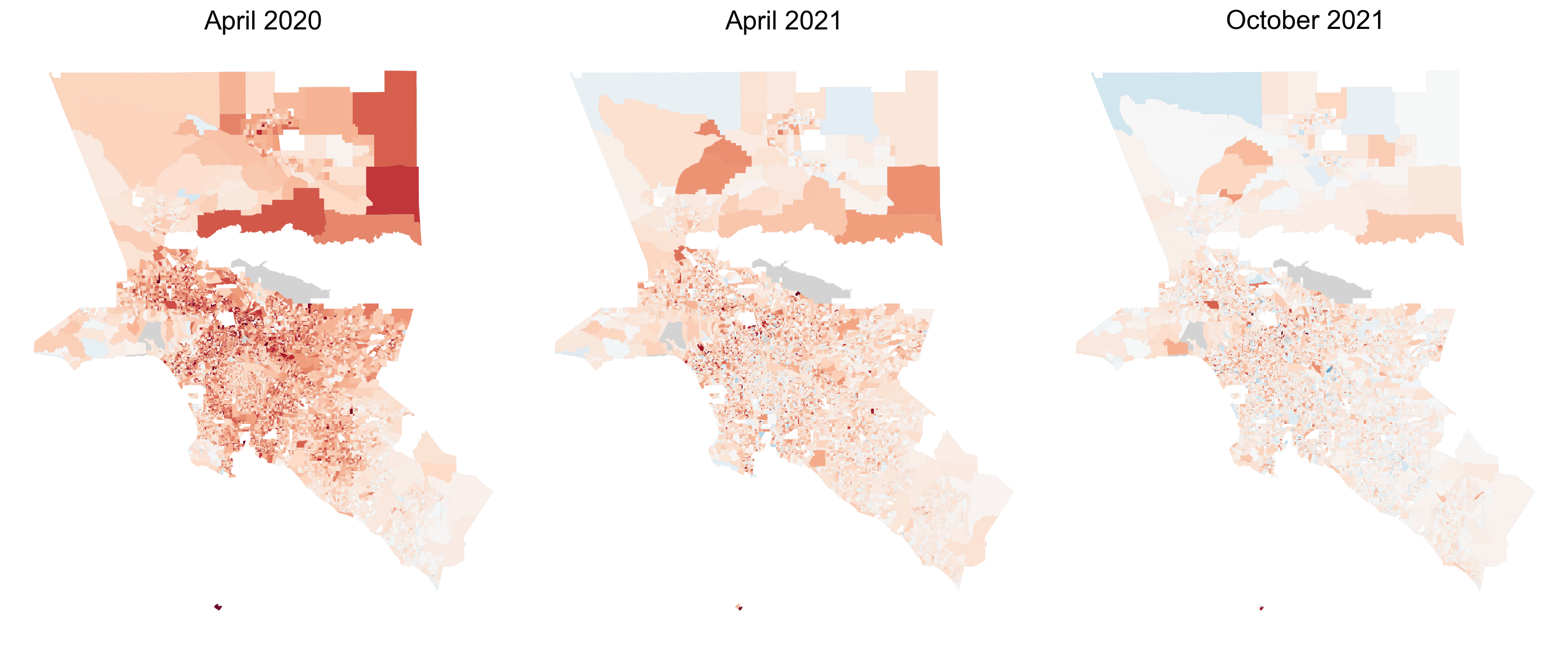}}\\
\subfloat[Dallas]{\includegraphics[width=\linewidth]{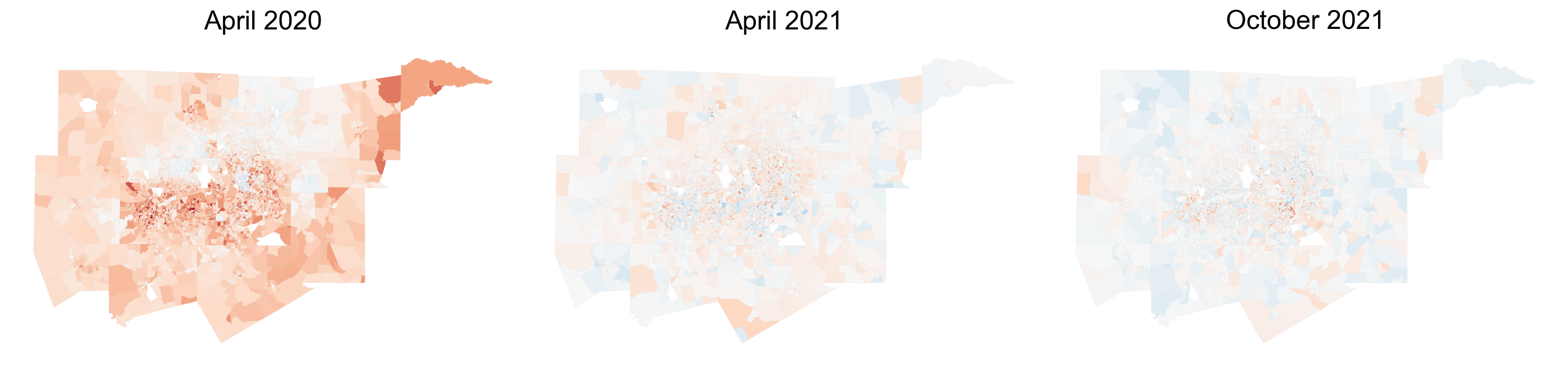}}
\caption{$\Delta D_{CBG}$ for different time periods in Seattle, Los Angeles, and Dallas.}
\label{fig:cbg_div_diff}
\end{figure}

\begin{figure}
\centering
\includegraphics[width=\linewidth]{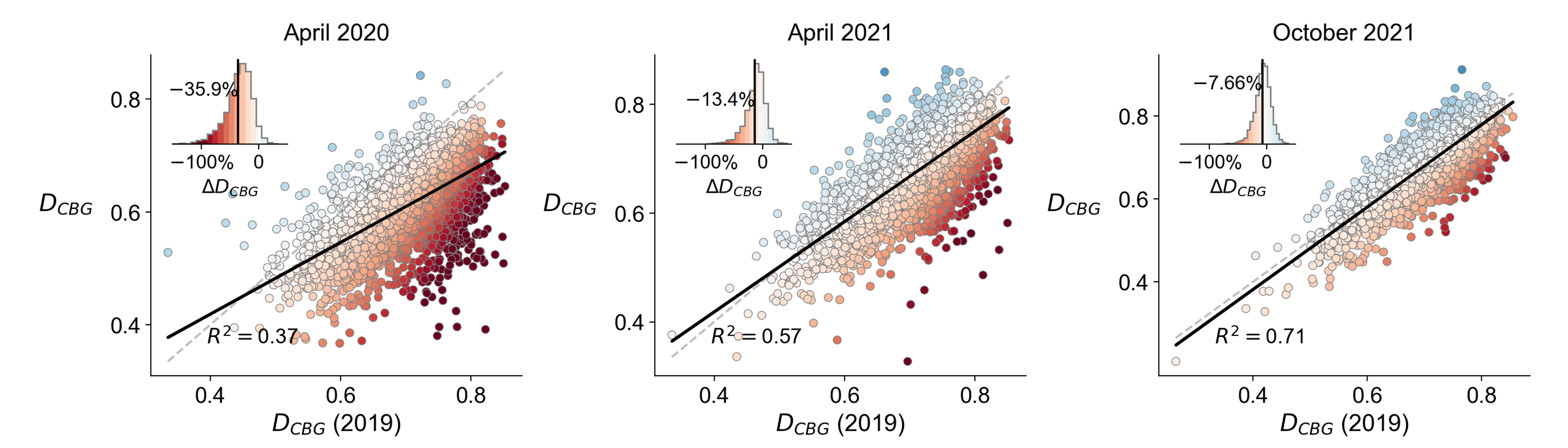}
\caption{Correlation between $D_{CBG}$ in different timings during the pandemic and the corresponding months in 2019.}
\label{fig:cbgcorrelation}
\end{figure}

\begin{figure}
\centering
\includegraphics[width=\linewidth]{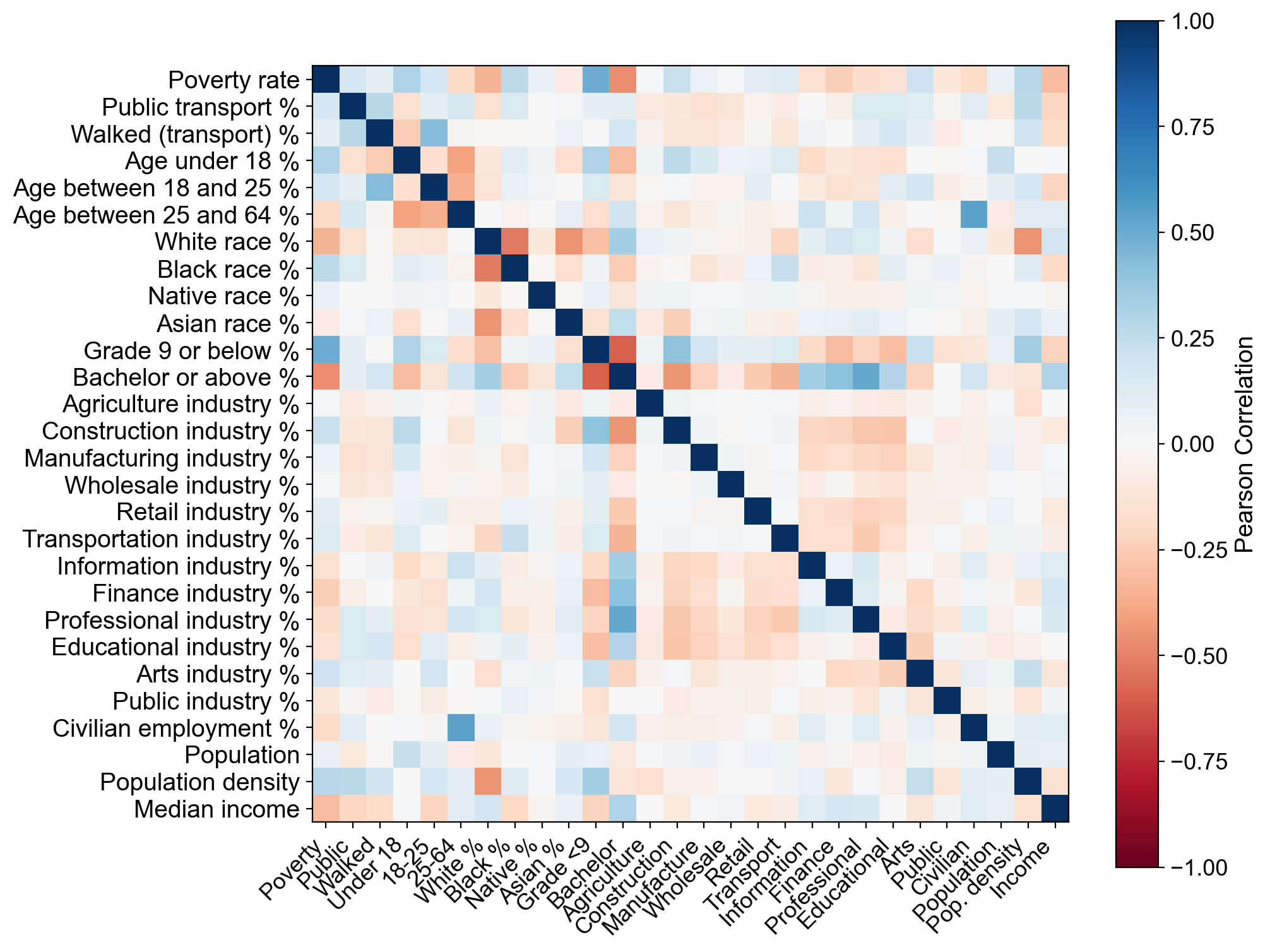}
\caption{Correlation matrix of CBG based residential variables.}
\label{fig:corrs}
\end{figure}

To further understand how the income diversity of encounters decreased heterogeneously across sociodemographic groups during throughout the pandemic, we build simple linear regression models of the form:
\begin{equation*}
    D_{CBG}(t), \Delta D_{CBG}(t) \sim \{R_{CBG}\}+\{P_{CBG}\}+\{M_{CBG}\}
\end{equation*}
where $D_{CBG}(t)$ and $\Delta D_{CBG}(t)$ denote the differences in diversity at time $t$ compared to the same month in the year 2019, and:
\begin{itemize}
    \item $\{R_{CBG}\}$ is the set of all residential variables from the census that describe the demographic, transportation, education, race, employment, wealth, etc. of the Census Block Group. The entire list of these variables can be found in Table \ref{table:summarystats}. 
    \item $\{P_{CBG}\}$ is a vector of variables that indicate the places where individuals living in the CBG spent most of their time in 2019, out of the place subcategories which have at least 100 venues. For each individual, we identify the subcategories where the individual stays more than 0.3\% of their time and obtain a binary vector with the length of 564, which is the number of place subcategories. Then to obtain $\{P_{CBG}\}$ we simply take the average of the vectors of all individuals who are living in the corresponding CBG. The threshold method previously employed in \cite{moro2021mobility} are used for sparse and highly-skewed human data \cite{di2018sequences} to minimize the effect of the noisy and long-tailed distribution of human activities. 
    \item $\{M_{CBG}\}$ is a set of variables that describe the geographical mobility behavior of people living in the corresponding CBG. We use two variables: (i) the radius of gyration of all the places visited by each user, and (ii) the average distance traveled to all places from each individual's home. 
\end{itemize}
The summary statistics of the residential variables are shown in Table \ref{table:summarystats}. Variables that have high correlation amongst eachother, such as '\% of people above the age 65', '\% of people commuting by car', '\% of population between Grades 9 and 12' were removed from the set of variables, and as shown in Figure \ref{fig:corrs}, the correlation among variables are generally low, with the highest magnitude of correlation at $\rho=-0.59$ between '\% with Bachelors degree or higher' with '\% below Grade 9'. We also checked that the variance inflation factor (VIF) are all between 1 to 5, indicating that there is no significant issue of multicollinearity. 
Figure \ref{fig:cbg_div_diff} shows the differences in $\Delta D_{CBG}$ for different periods during the pandemic, and \ref{fig:cbgcorrelation} shows the scatter plots of the income diversity in each CBG compared between before the pandemic and during the pandemic at three time points, in Boston CBSA.

\begin{table}
\centering
\caption{Summary statistics of the residential variables used in the regression models.}
\begin{tabular}{llrrrr}
\toprule
Group & Variable &      Mean &       Std. Dev. &  Min. &       Max. \\
\midrule
Population & $\log_{10}$(Population) &  3.152 &  0.214 &  0.0 &  4.245 \\
& $\log_{10}$(Population density)        &  3.110 &  0.573 &  0.0 &  4.577 \\
\midrule
Wealth & $\log_{10}$(Median Income) &  4.829 &  0.453 &  0.0 &  5.397 \\
 & Poverty rate &  0.088 &  0.108 &  0.0 & 1.000 \\
 \midrule
Means of transportation to work & Public transport          &  0.064 &  0.097 &  0.0 &  1.000 \\
& Walked            &  0.027 &  0.064 &  0.0 &  0.780 \\
\midrule
Age group & Below 18                 &  0.216 &  0.086 &  0.0 &  0.573 \\
& Between 18 and 25 &  0.089 &  0.073 &  0.0 &  0.986 \\
& Between 25 and 64               &  0.551 &  0.091 &  0.0 &  0.955 \\
\midrule
Race & White              &  0.643 &  0.237 &  0.0 &  1.000 \\
& Black              &  0.088 &  0.151 &  0.0 &  1.000 \\
& Native             &  0.005 &  0.017 &  0.0 &  0.507 \\
& Asian              &  0.113 &  0.145 &  0.0 &  0.942 \\
\midrule
Educational attainment & Below grade 9 &  0.058 &  0.084 &  0.0 &  0.650 \\
& Bachelor degree or more      &  0.376 &  0.234 &  0.0 &  1.000 \\
\midrule
Industry employment & Agriculture    &  0.006 &  0.017 &  0.0 &  0.437 \\
& Construction   &  0.066 &  0.065 &  0.0 &  0.705 \\
& Manufacturing  &  0.095 &  0.066 &  0.0 &  0.594 \\
& Wholesale      &  0.030 &  0.034 &  0.0 &  0.378 \\
& Retail         &  0.105 &  0.064 &  0.0 &  0.619 \\
& Transportation &  0.054 &  0.050 &  0.0 &  0.531 \\
& Information    &  0.030 &  0.041 &  0.0 &  0.429 \\
& Finance        &  0.071 &  0.058 &  0.0 &  0.680 \\
& Professional   &  0.140 &  0.082 &  0.0 &  0.853 \\
& Educational    &  0.216 &  0.095 &  0.0 &  0.751 \\
& Arts           &  0.097 &  0.068 &  0.0 &  0.673 \\
& Public         &  0.033 &  0.037 &  0.0 &  0.710 \\
\midrule
Employment status & Civilian in labor force     &  0.669 &  0.104 &  0.0 &  1.000 \\
\midrule
PUMA area & Fixed effects & - & - & - & - \\
\bottomrule
\end{tabular}
\label{table:summarystats}
\end{table}

To evaluate the relative importance of the three groups of variables, we used the approach proposed by Lindeman, Merenda, and Gold (LMG method) \cite{lindeman1980introduction}. The LMG method measures the additional $R^2$ when the variable group is added to the model. Since we have three groups of variables ($A$,$B$,$C$) with six different permutations, thus the contribution of variable group $A$, for example, is:
\begin{equation*}
    LMG(A) = \frac{1}{6} \big( 2R^2(A) + R^2(A|B) + R^2(A|C)+ 2R^2(A|B,C) \big).
\end{equation*}

Figure 3C in the main manuscript shows the relative importance of the three groups of variables for each month, for $D_{CBG}(t)$ and $\Delta D_{CBG}(t)$. Tables \ref{table:regA1} to \ref{table:regA3} and Tables \ref{table:regB1} to \ref{table:regB4} show the full regression results for the selected months for $D_{CBG}$ and $\Delta D_{CBG}$, respectively. All significant variables (with $p<0.01$) in the full model with all variables (\{R,M,P\}) included, are shown for the other models with partial variables. 
The model results for $D_{CBG}$ in Tables \ref{table:regA1} (pre-pandemic), \ref{table:regA2}, and \ref{table:regA3} (during the pandemic) largely agree with the results in \cite{moro2021mobility}, where the residential, mobility, and places variables collectively explain the heterogeneity in diversity well ($R^2 = 0.662$ for October 2021). On the other hand, the differences in diversity $\Delta D_{CBG}$ are less well explained by these variables, where the $R^2$ is at most around $0.3$ during the COVID-19 outbreak periods (April, May 2020 and December 2020 and January 2021), as shown in Figure 3C in the main manuscript. This indicates that the decrease in income diversity during the pandemic (especially during the off-peak months) are relatively homogeneous across all sociodemographic segments.


\begin{table}[!htbp] \centering
\resizebox{\textwidth}{!}{\begin{tabular}{@{\extracolsep{5pt}}lccccccc}
\\[-1.8ex]\hline
\hline \\[-1.8ex]
& \multicolumn{7}{c}{\textit{Dependent variable: $D_{CBG}$ (April 2019)}} \
\cr \cline{2-8}
\\[-1.8ex] & \{R\} & \{M\} & \{P\} & \{R,M\} & \{R,P\} & \{M,P\} & \{R,M,P\} \\
\hline \\[-1.8ex]
  Constant & 0.646$^{***}$ & 0.688$^{***}$ & 0.688$^{***}$ & 0.642$^{***}$ & 0.638$^{***}$ & 0.688$^{***}$ & 0.634$^{***}$ \\
\hline
 \% Age under 18 & -0.011$^{***}$ & & & -0.011$^{***}$ & -0.009$^{***}$ & & -0.009$^{***}$ \\
 \% Age 18-25 & -0.006$^{***}$ & & & -0.006$^{***}$ & -0.005$^{***}$ & & -0.005$^{***}$ \\
 \% Bachelor or more & -0.012$^{***}$ & & & -0.011$^{***}$ & -0.007$^{***}$ & & -0.007$^{***}$ \\
 \% Grade 9 or below & -0.003$^{***}$ & & & -0.003$^{***}$ & -0.003$^{***}$ & & -0.003$^{***}$ \\
 \% Civilian employed & 0.006$^{***}$ & & & 0.006$^{***}$ & 0.005$^{***}$ & & 0.005$^{***}$ \\
 \% Finance industry & -0.003$^{***}$ & & & -0.003$^{***}$ & -0.002$^{***}$ & & -0.002$^{***}$ \\
 \% Manufacturing industry & -0.003$^{***}$ & & & -0.003$^{***}$ & -0.002$^{***}$ & & -0.002$^{***}$ \\
 \% Public industry & 0.002$^{***}$ & & & 0.001$^{***}$ & 0.001$^{***}$ & & 0.001$^{***}$ \\
 \% Wholesale industry & -0.002$^{***}$ & & & -0.002$^{***}$ & -0.002$^{***}$ & & -0.002$^{***}$ \\
 Population density & 0.002$^{***}$ & & & 0.006$^{***}$ & -0.000$^{}$ & & 0.003$^{***}$ \\
 Median income & -0.002$^{***}$ & & & -0.003$^{***}$ & -0.003$^{***}$ & & -0.003$^{***}$ \\
 Population & 0.000$^{}$ & & & -0.001$^{**}$ & 0.003$^{***}$ & & 0.002$^{***}$ \\
 \% Black race & -0.002$^{**}$ & & & -0.002$^{***}$ & -0.001$^{**}$ & & -0.002$^{***}$ \\
 Poverty rate & -0.002$^{***}$ & & & -0.002$^{***}$ & -0.002$^{***}$ & & -0.002$^{***}$ \\
 \% Public transportation & -0.004$^{***}$ & & & -0.003$^{***}$ & -0.004$^{***}$ & & -0.004$^{***}$ \\
\hline
 Traveled distance & & 0.003$^{***}$ & & 0.007$^{***}$ & & 0.004$^{***}$ & 0.006$^{***}$ \\
 Radius of gyration & & -0.006$^{***}$ & & 0.002$^{***}$ & & -0.001$^{}$ & 0.003$^{***}$ \\
 \hline
 Bubble Tea & & & 0.003$^{***}$ & & 0.002$^{***}$ & 0.003$^{***}$ & 0.002$^{***}$ \\
 Sports Bar & & & 0.002$^{***}$ & & 0.002$^{***}$ & 0.002$^{***}$ & 0.002$^{***}$ \\
 Auto Workshop & & & 0.002$^{***}$ & & 0.001$^{***}$ & 0.002$^{***}$ & 0.001$^{***}$ \\
 Dim Sum & & & 0.002$^{***}$ & & 0.001$^{***}$ & 0.002$^{***}$ & 0.001$^{***}$ \\
 Korean & & & 0.001$^{}$ & & 0.001$^{***}$ & 0.001$^{}$ & 0.001$^{***}$ \\
 Piano Bar & & & 0.003$^{***}$ & & 0.001$^{***}$ & 0.003$^{***}$ & 0.001$^{***}$ \\
 Platform & & & 0.001$^{}$ & & 0.001$^{***}$ & 0.001$^{}$ & 0.001$^{***}$ \\
 Pub & & & 0.003$^{***}$ & & 0.001$^{***}$ & 0.003$^{***}$ & 0.001$^{***}$ \\
 Veterinarians & & & 0.002$^{***}$ & & 0.001$^{***}$ & 0.002$^{***}$ & 0.001$^{***}$ \\
 Video Store & & & 0.001$^{***}$ & & 0.001$^{***}$ & 0.001$^{***}$ & 0.001$^{***}$ \\
 Language School & & & -0.003$^{***}$ & & -0.002$^{***}$ & -0.003$^{***}$ & -0.001$^{***}$ \\
 Resort & & & -0.003$^{***}$ & & -0.001$^{***}$ & -0.003$^{***}$ & -0.001$^{***}$ \\
 Basketball & & & -0.004$^{***}$ & & -0.002$^{***}$ & -0.004$^{***}$ & -0.002$^{***}$ \\
 Lodge & & & -0.004$^{***}$ & & -0.002$^{***}$ & -0.004$^{***}$ & -0.002$^{***}$ \\
 Music School & & & -0.002$^{***}$ & & -0.002$^{***}$ & -0.002$^{***}$ & -0.002$^{***}$ \\
 Pilates Studio & & & -0.005$^{***}$ & & -0.002$^{***}$ & -0.005$^{***}$ & -0.002$^{***}$ \\
 South Indian & & & -0.006$^{***}$ & & -0.002$^{***}$ & -0.006$^{***}$ & -0.002$^{***}$ \\
 Wine Bar & & & -0.004$^{***}$ & & -0.002$^{***}$ & -0.004$^{***}$ & -0.002$^{***}$ \\
 Volleyball Court & & & -0.008$^{***}$ & & -0.003$^{***}$ & -0.008$^{***}$ & -0.003$^{***}$ \\
 \hline
 PUMA Fixed Effects & Yes & No & No & Yes & Yes & No & Yes \\
 \hline \\[-1.8ex]
 Observations & 17,824 & 17,824 & 17,824 & 17,824 & 17,824 & 17,824 & 17,824 \\
 $R^2$ & 0.703 & 0.002 & 0.422 & 0.706 & 0.735 & 0.423 & 0.737 \\
 Adjusted $R^2$ & 0.699 & 0.002 & 0.403 & 0.702 & 0.722 & 0.404 & 0.725 \\
\hline
\hline \\[-1.8ex]
\textit{Note:} & \multicolumn{7}{r}{$^{*}$p$<$0.1; $^{**}$p$<$0.05; $^{***}$p$<$0.01} \\
\end{tabular}}
\caption{Regression results for $D_{CBG}$ for April 2019.} 
\label{table:regA1}
\end{table}

\begin{table}[!htbp] \centering
\resizebox{\textwidth}{!}{\begin{tabular}{@{\extracolsep{5pt}}lccccccc}
\\[-1.8ex]\hline
\hline \\[-1.8ex]
& \multicolumn{7}{c}{\textit{Dependent variable: $D_{CBG}$ (April 2020)}} \
\cr \cline{2-8}
\\[-1.8ex] & \{R\} & \{M\} & \{P\} & \{R,M\} & \{R,P\} & \{M,P\} & \{R,M,P\} \\
\hline \\[-1.8ex]
 Constant & 0.565$^{***}$ & 0.605$^{***}$ & 0.605$^{***}$ & 0.563$^{***}$ & 0.561$^{***}$ & 0.605$^{***}$ & 0.558$^{***}$ \\
 \hline
 \% Age under 18 & -0.007$^{***}$ & & & -0.007$^{***}$ & -0.005$^{***}$ & & -0.005$^{***}$ \\
 \% Age 18-25 & -0.004$^{***}$ & & & -0.004$^{***}$ & -0.003$^{***}$ & & -0.003$^{***}$ \\
 \% Age 25-64 & -0.003$^{***}$ & & & -0.004$^{***}$ & -0.003$^{***}$ & & -0.003$^{***}$ \\
 \% Bachelor or more & -0.006$^{***}$ & & & -0.005$^{***}$ & -0.003$^{***}$ & & -0.002$^{**}$ \\
 \% Grade 9 or below & -0.004$^{***}$ & & & -0.004$^{***}$ & -0.003$^{***}$ & & -0.003$^{***}$ \\
 \% Civilian employed & 0.005$^{***}$ & & & 0.004$^{***}$ & 0.004$^{***}$ & & 0.004$^{***}$ \\
 \% Public industry & 0.002$^{***}$ & & & 0.002$^{***}$ & 0.002$^{***}$ & & 0.002$^{***}$ \\
 \% Wholesale industry & -0.001$^{**}$ & & & -0.001$^{***}$ & -0.001$^{***}$ & & -0.001$^{***}$ \\
 Population density & -0.001$^{**}$ & & & 0.003$^{***}$ & -0.002$^{***}$ & & 0.002$^{***}$ \\
 Median income & -0.001$^{***}$ & & & -0.002$^{***}$ & -0.002$^{***}$ & & -0.002$^{***}$ \\
 Poverty ratio & -0.003$^{***}$ & & & -0.002$^{***}$ & -0.003$^{***}$ & & -0.003$^{***}$ \\
 \% Public transportation & -0.006$^{***}$ & & & -0.006$^{***}$ & -0.006$^{***}$ & & -0.005$^{***}$ \\
 \hline
  Traveled distance & & 0.012$^{***}$ & & 0.008$^{***}$ & & 0.007$^{***}$ & 0.008$^{***}$ \\
  \hline
 Bubble Tea & & & 0.002$^{***}$ & & 0.001$^{***}$ & 0.002$^{***}$ & 0.001$^{***}$ \\
 Burgers & & & 0.001$^{***}$ & & 0.001$^{***}$ & 0.001$^{**}$ & 0.001$^{***}$ \\
 Coffee Shop & & & 0.001$^{**}$ & & 0.001$^{***}$ & 0.001$^{**}$ & 0.001$^{***}$ \\
 Discount Store & & & 0.003$^{***}$ & & 0.001$^{***}$ & 0.003$^{***}$ & 0.001$^{***}$ \\
 Theme Park & & & 0.001$^{*}$ & & 0.001$^{***}$ & 0.001$^{}$ & 0.001$^{***}$ \\
 Water Park & & & 0.001$^{*}$ & & 0.001$^{***}$ & 0.001$^{*}$ & 0.001$^{***}$ \\
 Event Space & & & -0.001$^{**}$ & & -0.001$^{***}$ & -0.001$^{**}$ & -0.001$^{***}$ \\
  Historic Site & & & -0.001$^{**}$ & & -0.001$^{***}$ & -0.001$^{**}$ & -0.001$^{***}$ \\
  Apparel & & & -0.002$^{***}$ & & -0.001$^{***}$ & -0.002$^{***}$ & -0.001$^{***}$ \\
 Baseball & & & -0.002$^{***}$ & & -0.001$^{***}$ & -0.002$^{***}$ & -0.001$^{***}$ \\
  Food \& Drink & & & -0.004$^{***}$ & & -0.001$^{**}$ & -0.004$^{***}$ & -0.001$^{***}$ \\
 Language School & & & -0.003$^{***}$ & & -0.002$^{***}$ & -0.002$^{***}$ & -0.001$^{***}$ \\
 Lodge & & & -0.002$^{***}$ & & -0.002$^{***}$ & -0.002$^{***}$ & -0.001$^{***}$ \\
 Music School & & & -0.002$^{***}$ & & -0.001$^{***}$ & -0.002$^{***}$ & -0.001$^{***}$ \\
  Hostel & & & -0.004$^{***}$ & & -0.002$^{***}$ & -0.003$^{***}$ & -0.002$^{***}$ \\
 Pilates Studio & & & -0.005$^{***}$ & & -0.002$^{***}$ & -0.005$^{***}$ & -0.002$^{***}$ \\
 Resort & & & -0.003$^{***}$ & & -0.002$^{***}$ & -0.003$^{***}$ & -0.002$^{***}$ \\
 South Indian & & & -0.004$^{***}$ & & -0.002$^{***}$ & -0.004$^{***}$ & -0.002$^{***}$ \\
 Volleyball Court & & & -0.006$^{***}$ & & -0.003$^{***}$ & -0.006$^{***}$ & -0.003$^{***}$ \\
 \hline
 PUMA Fixed Effects & Yes & No & No & Yes & Yes & No & Yes \\
 \hline \\[-1.8ex]
 Observations & 17,824 & 17,824 & 17,824 & 17,824 & 17,824 & 17,824 & 17,824 \\
 $R^2$ & 0.582 & 0.025 & 0.346 & 0.585 & 0.613 & 0.350 & 0.615 \\
 Adjusted $R^2$ & 0.576 & 0.025 & 0.325 & 0.580 & 0.594 & 0.329 & 0.597 \\
\hline
\hline \\[-1.8ex]
\textit{Note:} & \multicolumn{7}{r}{$^{*}$p$<$0.1; $^{**}$p$<$0.05; $^{***}$p$<$0.01} \\
\end{tabular}}
\caption{Regression results for $D_{CBG}$ for April 2020.} 
\label{table:regA2}
\end{table}

\begin{table}[!htbp] \centering
\resizebox{\textwidth}{!}{\begin{tabular}{@{\extracolsep{5pt}}lccccccc}
\\[-1.8ex]\hline
\hline \\[-1.8ex]
& \multicolumn{7}{c}{\textit{Dependent variable: $D_{CBG}$ (October 2021)}} \
\cr \cline{2-8}
\\[-1.8ex] & \{R\} & \{M\} & \{P\} & \{R,M\} & \{R,P\} & \{M,P\} & \{R,M,P\} \\
\hline \\[-1.8ex]
 Constant & 0.628$^{***}$ & 0.670$^{***}$ & 0.670$^{***}$ & 0.623$^{***}$ & 0.620$^{***}$ & 0.670$^{***}$ & 0.615$^{***}$ \\
 \hline
 \% Age under 18 & -0.012$^{***}$ & & & -0.012$^{***}$ & -0.010$^{***}$ & & -0.010$^{***}$ \\
 \% Age 18-25 & -0.007$^{***}$ & & & -0.007$^{***}$ & -0.007$^{***}$ & & -0.007$^{***}$ \\
 \% Bachelor or more & -0.014$^{***}$ & & & -0.013$^{***}$ & -0.009$^{***}$ & & -0.008$^{***}$ \\
 \% Grade 9 or below & -0.004$^{***}$ & & & -0.004$^{***}$ & -0.005$^{***}$ & & -0.004$^{***}$ \\
 \% Civilian employed & 0.007$^{***}$ & & & 0.007$^{***}$ & 0.006$^{***}$ & & 0.006$^{***}$ \\
 \% Public industry & 0.002$^{***}$ & & & 0.002$^{***}$ & 0.002$^{***}$ & & 0.002$^{***}$ \\
 Population density & 0.002$^{**}$ & & & 0.007$^{***}$ & -0.000$^{}$ & & 0.004$^{***}$ \\
 Median income & -0.003$^{***}$ & & & -0.003$^{***}$ & -0.003$^{***}$ & & -0.003$^{***}$ \\
 Population & -0.000$^{}$ & & & -0.001$^{***}$ & 0.003$^{***}$ & & 0.002$^{***}$ \\
 \% Public transportation & -0.006$^{***}$ & & & -0.005$^{***}$ & -0.006$^{***}$ & & -0.005$^{***}$ \\
 \hline
  Traveled distance & & 0.008$^{***}$ & & 0.008$^{***}$ & & 0.004$^{***}$ & 0.007$^{***}$ \\
 Radius of gyration & & -0.003$^{***}$ & & 0.003$^{***}$ & & 0.000$^{}$ & 0.004$^{***}$ \\
 \hline
 Billiards & & & 0.003$^{***}$ & & 0.002$^{***}$ & 0.003$^{***}$ & 0.002$^{***}$ \\
 Bubble Tea & & & 0.003$^{***}$ & & 0.002$^{***}$ & 0.003$^{***}$ & 0.002$^{***}$ \\
 Library & & & 0.002$^{***}$ & & 0.002$^{***}$ & 0.002$^{***}$ & 0.002$^{***}$ \\
 Motel & & & 0.002$^{**}$ & & 0.002$^{***}$ & 0.001$^{**}$ & 0.002$^{***}$ \\
 Sports Bar & & & 0.003$^{***}$ & & 0.002$^{***}$ & 0.003$^{***}$ & 0.002$^{***}$ \\
 Sushi & & & 0.004$^{***}$ & & 0.002$^{***}$ & 0.004$^{***}$ & 0.002$^{***}$ \\
 Auto Workshop & & & 0.002$^{***}$ & & 0.002$^{***}$ & 0.002$^{***}$ & 0.001$^{***}$ \\
 City Hall & & & 0.000$^{}$ & & 0.001$^{***}$ & 0.000$^{}$ & 0.001$^{***}$ \\
 Golf Driving Range & & & 0.003$^{***}$ & & 0.001$^{***}$ & 0.002$^{***}$ & 0.001$^{***}$ \\
 Shopping Plaza & & & 0.003$^{***}$ & & 0.001$^{***}$ & 0.003$^{***}$ & 0.001$^{***}$ \\
 Water Park & & & 0.001$^{}$ & & 0.001$^{***}$ & 0.001$^{}$ & 0.001$^{***}$ \\
 Basketball & & & -0.004$^{***}$ & & -0.002$^{***}$ & -0.004$^{***}$ & -0.002$^{***}$ \\
 Cycle Studio & & & -0.003$^{***}$ & & -0.002$^{***}$ & -0.003$^{***}$ & -0.002$^{***}$ \\
 Field & & & -0.002$^{***}$ & & -0.002$^{***}$ & -0.002$^{***}$ & -0.002$^{***}$ \\
 Hostel & & & -0.004$^{***}$ & & -0.002$^{***}$ & -0.004$^{***}$ & -0.002$^{***}$ \\
 Lodge & & & -0.004$^{***}$ & & -0.002$^{***}$ & -0.004$^{***}$ & -0.002$^{***}$ \\
 Music School & & & -0.003$^{***}$ & & -0.002$^{***}$ & -0.003$^{***}$ & -0.002$^{***}$ \\
 Pilates Studio & & & -0.005$^{***}$ & & -0.002$^{***}$ & -0.005$^{***}$ & -0.002$^{***}$ \\
 South Indian & & & -0.006$^{***}$ & & -0.002$^{***}$ & -0.006$^{***}$ & -0.002$^{***}$ \\
 Volleyball Court & & & -0.008$^{***}$ & & -0.003$^{***}$ & -0.008$^{***}$ & -0.003$^{***}$ \\
 \hline
 PUMA Fixed Effects & Yes & No & No & Yes & Yes & No & Yes \\
 \hline \\[-1.8ex]
 Observations & 17,824 & 17,824 & 17,824 & 17,824 & 17,824 & 17,824 & 17,824 \\
 $R^2$ & 0.641 & 0.005 & 0.375 & 0.644 & 0.675 & 0.376 & 0.678 \\
 Adjusted $R^2$ & 0.636 & 0.005 & 0.354 & 0.639 & 0.660 & 0.356 & 0.662 \\
\hline
\hline \\[-1.8ex]
\textit{Note:} & \multicolumn{7}{r}{$^{*}$p$<$0.1; $^{**}$p$<$0.05; $^{***}$p$<$0.01} \\
\end{tabular}}
\caption{Regression results for $D_{CBG}$ for October 2021.} 
\label{table:regA3}
\end{table}

\begin{table}[!htbp] \centering
\resizebox{\textwidth}{!}{\begin{tabular}{@{\extracolsep{5pt}}lccccccc}
\\[-1.8ex]\hline
\hline \\[-1.8ex]
& \multicolumn{7}{c}{\textit{Dependent variable: $\Delta D_{CBG}$ (April 2020)}} \
\cr \cline{2-8}
\\[-1.8ex] & \{R\} & \{M\} & \{P\} & \{R,M\} & \{R,P\} & \{M,P\} & \{R,M,P\} \\
\hline \\[-1.8ex]
 Constant & -0.124$^{***}$ & -0.118$^{***}$ & -0.118$^{***}$ & -0.122$^{***}$ & -0.120$^{***}$ & -0.118$^{***}$ & -0.118$^{***}$ \\
 \hline
 \% Age -18 & 0.005$^{***}$ & & & 0.005$^{***}$ & 0.005$^{***}$ & & 0.004$^{***}$ \\
 \% Age 18-25 & 0.003$^{***}$ & & & 0.003$^{***}$ & 0.003$^{***}$ & & 0.003$^{***}$ \\
 \% over Bachelor & 0.008$^{***}$ & & & 0.008$^{***}$ & 0.007$^{***}$ & & 0.007$^{***}$ \\
 \% Manufacturing industry & 0.002$^{***}$ & & & 0.002$^{***}$ & 0.002$^{***}$ & & 0.002$^{***}$ \\
 Population & -0.001$^{}$ & & & -0.001$^{*}$ & -0.004$^{***}$ & & -0.004$^{***}$ \\
 \% Public transportation & -0.004$^{***}$ & & & -0.004$^{***}$ & -0.003$^{***}$ & & -0.003$^{***}$ \\
\hline
 Travel distance & & 0.013$^{***}$ & & 0.003$^{***}$ & & 0.006$^{***}$ & 0.004$^{***}$ \\
 Radius of gyration & & 0.007$^{***}$ & & -0.003$^{***}$ & & -0.001$^{}$ & -0.004$^{***}$ \\
\hline
 Basketball & & & 0.003$^{***}$ & & 0.002$^{***}$ & 0.003$^{***}$ & 0.002$^{***}$ \\
 Housing Development & & & 0.002$^{***}$ & & 0.002$^{***}$ & 0.002$^{***}$ & 0.002$^{***}$ \\
 Non-Profit & & & 0.001$^{**}$ & & 0.002$^{***}$ & 0.001$^{**}$ & 0.002$^{***}$ \\
 Pet Store & & & 0.001$^{***}$ & & 0.002$^{***}$ & 0.001$^{**}$ & 0.002$^{***}$ \\
 Restaurant & & & 0.002$^{***}$ & & 0.002$^{***}$ & 0.002$^{***}$ & 0.002$^{***}$ \\
 Building & & & -0.002$^{***}$ & & -0.002$^{***}$ & -0.002$^{***}$ & -0.002$^{***}$ \\
 Department Store & & & -0.002$^{***}$ & & -0.002$^{***}$ & -0.002$^{***}$ & -0.002$^{***}$ \\
 Funeral Home & & & -0.002$^{***}$ & & -0.002$^{***}$ & -0.002$^{***}$ & -0.002$^{***}$ \\
 Laundromat & & & -0.003$^{***}$ & & -0.002$^{***}$ & -0.003$^{***}$ & -0.002$^{***}$ \\
 Pub & & & -0.002$^{**}$ & & -0.002$^{***}$ & -0.001$^{**}$ & -0.002$^{***}$ \\
 \hline
 PUMA Fixed Effects & Yes & No & No & Yes & Yes & No & Yes \\
\hline \\[-1.8ex]
 Observations & 17,824 & 17,824 & 17,824 & 17,824 & 17,824 & 17,824 & 17,824 \\
 $R^2$ & 0.308 & 0.058 & 0.270 & 0.309 & 0.343 & 0.273 & 0.343 \\
 Adjusted $R^2$ & 0.299 & 0.058 & 0.246 & 0.299 & 0.312 & 0.249 & 0.312 \\
\hline
\hline \\[-1.8ex]
\textit{Note:} & \multicolumn{7}{r}{$^{*}$p$<$0.1; $^{**}$p$<$0.05; $^{***}$p$<$0.01} \\
\end{tabular}}
\caption{Regression results for $\Delta D_{CBG}$ for April 2020.} 
\label{table:regB1}
\end{table}

\begin{table}[!htbp] \centering
\resizebox{\textwidth}{!}{\begin{tabular}{@{\extracolsep{5pt}}lccccccc}
\\[-1.8ex]\hline
\hline \\[-1.8ex]
& \multicolumn{7}{c}{\textit{Dependent variable: $\Delta D_{CBG}$ (May 2020)}} \
\cr \cline{2-8}
\\[-1.8ex] & \{R\} & \{M\} & \{P\} & \{R,M\} & \{R,P\} & \{M,P\} & \{R,M,P\} \\
\hline \\[-1.8ex]
 Constant & -0.081$^{***}$ & -0.082$^{***}$ & -0.082$^{***}$ & -0.080$^{***}$ & -0.076$^{***}$ & -0.082$^{***}$ & -0.074$^{***}$ \\
\hline
 \% Age 25-64 & -0.002$^{***}$ & & & -0.002$^{***}$ & -0.002$^{***}$ & & -0.002$^{***}$ \\
 \% Finance industry & 0.002$^{***}$ & & & 0.002$^{***}$ & 0.002$^{***}$ & & 0.002$^{***}$ \\
 Population density & -0.004$^{***}$ & & & -0.004$^{***}$ & -0.002$^{***}$ & & -0.003$^{***}$ \\
 \% Black race & 0.004$^{***}$ & & & 0.004$^{***}$ & 0.004$^{***}$ & & 0.004$^{***}$ \\
 \% Public transportation & -0.004$^{***}$ & & & -0.004$^{***}$ & -0.004$^{***}$ & & -0.004$^{***}$ \\
\hline
 Radius of gyration & & 0.006$^{***}$ & & -0.002$^{**}$ & & 0.000$^{}$ & -0.002$^{***}$ \\
\hline
 Engineering & & & 0.002$^{***}$ & & 0.002$^{***}$ & 0.002$^{***}$ & 0.002$^{***}$ \\
 Basketball & & & 0.002$^{***}$ & & 0.001$^{***}$ & 0.002$^{***}$ & 0.001$^{***}$ \\
 Coffee Shop & & & 0.000$^{}$ & & 0.001$^{***}$ & 0.000$^{}$ & 0.001$^{***}$ \\
 Medical School & & & 0.000$^{}$ & & 0.001$^{***}$ & 0.000$^{}$ & 0.001$^{***}$ \\
 Music School & & & 0.001$^{**}$ & & 0.001$^{***}$ & 0.001$^{**}$ & 0.001$^{***}$ \\
 Non-Profit & & & 0.001$^{**}$ & & 0.001$^{***}$ & 0.001$^{**}$ & 0.001$^{***}$ \\
 Restaurant & & & 0.002$^{***}$ & & 0.001$^{***}$ & 0.002$^{***}$ & 0.001$^{***}$ \\
 Classroom & & & -0.001$^{**}$ & & -0.001$^{***}$ & -0.001$^{**}$ & -0.001$^{***}$ \\
 Residence Hall & & & -0.002$^{***}$ & & -0.001$^{***}$ & -0.002$^{***}$ & -0.001$^{***}$ \\
 Pub & & & -0.001$^{**}$ & & -0.002$^{***}$ & -0.001$^{**}$ & -0.002$^{***}$ \\
 \hline
 PUMA Fixed Effects & Yes & No & No & Yes & Yes & No & Yes \\
 \hline \\[-1.8ex]
 Observations & 17,824 & 17,824 & 17,824 & 17,824 & 17,824 & 17,824 & 17,824 \\
 $R^2$ & 0.316 & 0.060 & 0.272 & 0.316 & 0.343 & 0.273 & 0.343 \\
 Adjusted $R^2$ & 0.306 & 0.060 & 0.248 & 0.307 & 0.312 & 0.249 & 0.312 \\
\hline
\hline \\[-1.8ex]
\textit{Note:} & \multicolumn{7}{r}{$^{*}$p$<$0.1; $^{**}$p$<$0.05; $^{***}$p$<$0.01} \\
\end{tabular}}
\caption{Regression results for $\Delta D_{CBG}$ for May 2020.} 
\label{table:regB2}
\end{table}

\begin{table}[!htbp] \centering
\resizebox{\textwidth}{!}{\begin{tabular}{@{\extracolsep{5pt}}lccccccc}
\\[-1.8ex]\hline
\hline \\[-1.8ex]
& \multicolumn{7}{c}{\textit{Dependent variable: $\Delta D_{CBG}$ (December 2020)}} \
\cr \cline{2-8}
\\[-1.8ex] & \{R\} & \{M\} & \{P\} & \{R,M\} & \{R,P\} & \{M,P\} & \{R,M,P\} \\
\hline \\[-1.8ex]
 Constant & -0.097$^{***}$ & -0.088$^{***}$ & -0.088$^{***}$ & -0.093$^{***}$ & -0.092$^{***}$ & -0.088$^{***}$ & -0.088$^{***}$ \\
\hline
 Population density & -0.003$^{***}$ & & & -0.005$^{***}$ & -0.002$^{***}$ & & -0.004$^{***}$ \\
 \% Black race & 0.002$^{***}$ & & & 0.003$^{***}$ & 0.002$^{**}$ & & 0.003$^{***}$ \\
 \% Public transportation & -0.005$^{***}$ & & & -0.006$^{***}$ & -0.005$^{***}$ & & -0.005$^{***}$ \\
\hline
 Radius of gyration & & 0.002$^{***}$ & & -0.004$^{***}$ & & -0.001$^{}$ & -0.004$^{***}$ \\
\hline
 Baseball Field & & & 0.002$^{***}$ & & 0.002$^{***}$ & 0.002$^{***}$ & 0.002$^{***}$ \\
 Dentist's Office & & & 0.002$^{***}$ & & 0.001$^{***}$ & 0.002$^{***}$ & 0.001$^{***}$ \\
 Frame Store & & & 0.002$^{***}$ & & 0.001$^{***}$ & 0.002$^{***}$ & 0.001$^{***}$ \\
 Shop & & & 0.001$^{***}$ & & 0.001$^{***}$ & 0.001$^{***}$ & 0.001$^{***}$ \\
 Beer Store & & & -0.002$^{***}$ & & -0.001$^{***}$ & -0.002$^{***}$ & -0.001$^{***}$ \\
 Doctor's Office & & & -0.001$^{***}$ & & -0.001$^{***}$ & -0.001$^{**}$ & -0.001$^{***}$ \\
 Engineering & & & -0.002$^{***}$ & & -0.001$^{***}$ & -0.001$^{***}$ & -0.001$^{***}$ \\
 Bar & & & -0.002$^{***}$ & & -0.001$^{***}$ & -0.002$^{***}$ & -0.001$^{***}$ \\
 Seafood & & & -0.001$^{}$ & & -0.001$^{***}$ & -0.001$^{}$ & -0.001$^{***}$ \\
 Gate & & & -0.001$^{***}$ & & -0.002$^{***}$ & -0.001$^{***}$ & -0.002$^{***}$ \\
 \hline
 PUMA Fixed Effects & Yes & No & No & Yes & Yes & No & Yes \\
\hline \\[-1.8ex]
 Observations & 17,824 & 17,824 & 17,824 & 17,824 & 17,824 & 17,824 & 17,824 \\
 $R^2$ & 0.275 & 0.042 & 0.248 & 0.277 & 0.302 & 0.248 & 0.303 \\
 Adjusted $R^2$ & 0.265 & 0.042 & 0.223 & 0.267 & 0.269 & 0.223 & 0.270 \\
\hline
\hline \\[-1.8ex]
\textit{Note:} & \multicolumn{7}{r}{$^{*}$p$<$0.1; $^{**}$p$<$0.05; $^{***}$p$<$0.01} \\
\end{tabular}}
\caption{Regression results for $\Delta D_{CBG}$ for December 2020.} 
\label{table:regB3}
\end{table}

\begin{table}[!htbp] \centering
\resizebox{\textwidth}{!}{\begin{tabular}{@{\extracolsep{5pt}}lccccccc}
\\[-1.8ex]\hline
\hline \\[-1.8ex]
& \multicolumn{7}{c}{\textit{Dependent variable: $\Delta D_{CBG}$ (January 2021)}} \
\cr \cline{2-8}
\\[-1.8ex] & \{R\} & \{M\} & \{P\} & \{R,M\} & \{R,P\} & \{M,P\} & \{R,M,P\} \\
\hline \\[-1.8ex]
 Constant & -0.070$^{***}$ & -0.080$^{***}$ & -0.080$^{***}$ & -0.070$^{***}$ & -0.071$^{***}$ & -0.080$^{***}$ & -0.070$^{***}$ \\
 \hline
 \% Agriculture industry & 0.002$^{***}$ & & & 0.002$^{***}$ & 0.002$^{***}$ & & 0.002$^{***}$ \\
 \% Professional industry & 0.003$^{***}$ & & & 0.003$^{***}$ & 0.003$^{***}$ & & 0.003$^{***}$ \\
\% Retail industry & 0.003$^{***}$ & & & 0.003$^{***}$ & 0.003$^{***}$ & & 0.003$^{***}$ \\
 \% Black race & 0.004$^{***}$ & & & 0.004$^{***}$ & 0.003$^{***}$ & & 0.003$^{***}$ \\
  \% White race & 0.003$^{***}$ & & & 0.003$^{***}$ & 0.004$^{***}$ & & 0.004$^{***}$ \\
 \% Public transport & -0.005$^{***}$ & & & -0.005$^{***}$ & -0.004$^{***}$ & & -0.004$^{***}$ \\
 \hline 
 Department Store & & & 0.003$^{***}$ & & 0.003$^{***}$ & 0.003$^{***}$ & 0.003$^{***}$ \\
 Asian & & & 0.001$^{***}$ & & 0.002$^{***}$ & 0.001$^{***}$ & 0.002$^{***}$ \\
 Bank & & & 0.001$^{***}$ & & 0.002$^{***}$ & 0.001$^{**}$ & 0.002$^{***}$ \\
 Cosmetics & & & 0.002$^{***}$ & & 0.002$^{***}$ & 0.002$^{***}$ & 0.002$^{***}$ \\
 Grocery Store & & & 0.002$^{***}$ & & 0.002$^{***}$ & 0.002$^{***}$ & 0.002$^{***}$ \\
 Building & & & 0.001$^{**}$ & & 0.001$^{***}$ & 0.001$^{**}$ & 0.001$^{***}$ \\
 Irish & & & -0.001$^{**}$ & & -0.001$^{***}$ & -0.001$^{**}$ & -0.001$^{***}$ \\
 Donburi & & & -0.004$^{***}$ & & -0.002$^{***}$ & -0.004$^{***}$ & -0.002$^{***}$ \\
 \hline
 PUMA Fixed Effects & Yes & No & No & Yes & Yes & No & Yes \\
 \hline \\[-1.8ex]
 Observations & 17,824 & 17,824 & 17,824 & 17,824 & 17,824 & 17,824 & 17,824 \\
 $R^2$ & 0.307 & 0.059 & 0.279 & 0.307 & 0.332 & 0.280 & 0.332 \\
 Adjusted $R^2$ & 0.297 & 0.059 & 0.255 & 0.297 & 0.301 & 0.257 & 0.301 \\
\hline
\hline \\[-1.8ex]
\textit{Note:} & \multicolumn{7}{r}{$^{*}$p$<$0.1; $^{**}$p$<$0.05; $^{***}$p$<$0.01} \\
\end{tabular}}
\caption{Regression results for $\Delta D_{CBG}$ for January 2021.} 
\label{table:regB4}
\end{table}

\begin{figure}
\centering
\includegraphics[width=\linewidth]{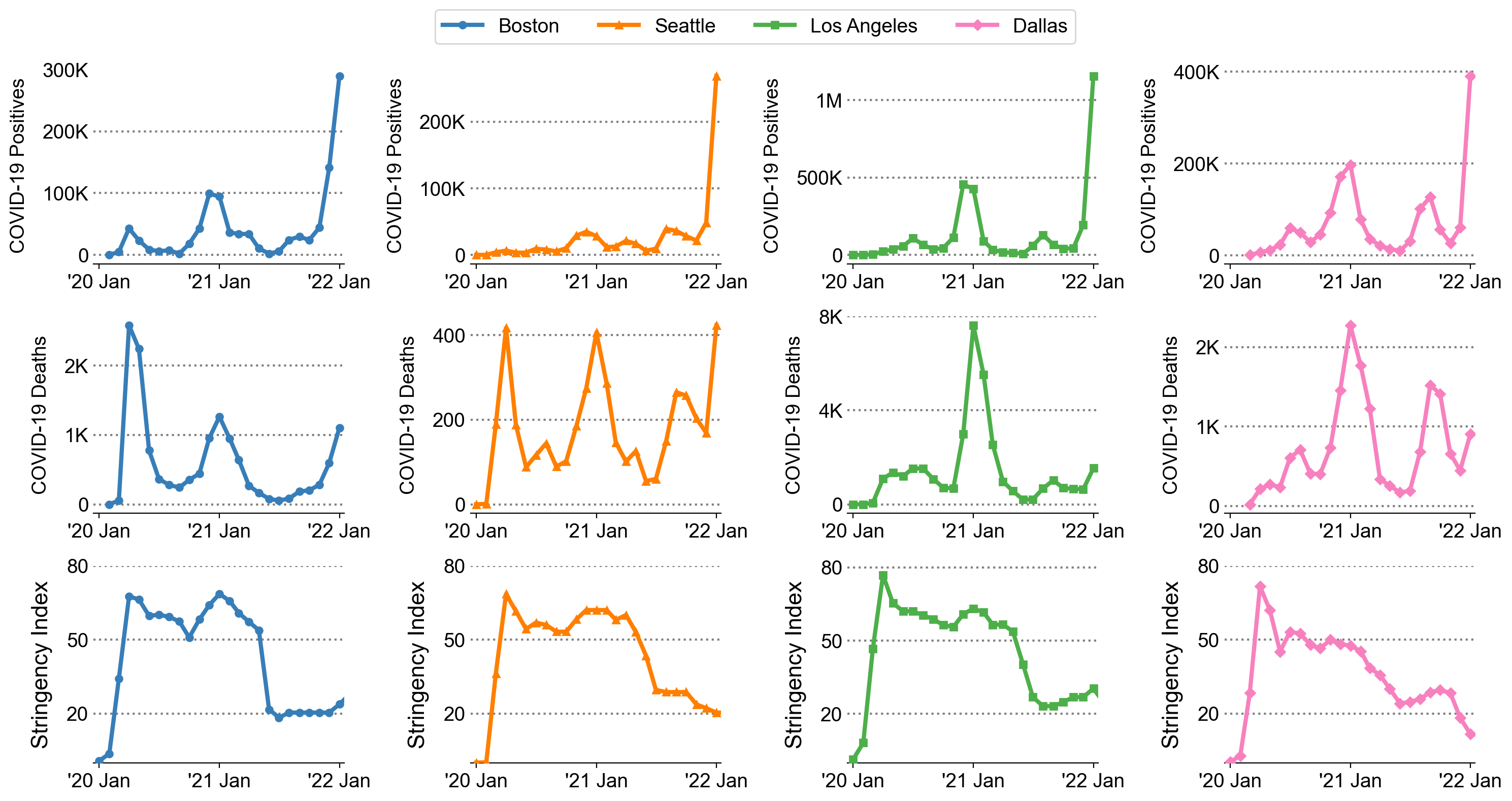}
\caption{Number of monthly COVID-19 cases (top row), COVID-19 deaths (middle row), and stringency index (bottom row) in each of the CBSAs.}
\label{fig:cases_deaths}
\end{figure}

\section{COVID-19 intensity and segregation} \label{sec:stringency}

An interesting aspect of the COVID-19 pandemic was its asynchronicity in terms of outbreaks (number of cases and deaths) and the strictness of implemented policies. 
To further understand the differences in decreased income diversity across CBSAs, we build simple linear regression models with the form:

\begin{equation*}
    \Delta D_{CBSA}(t) \sim Cases_{CBSA}(t) + Deaths_{CBSA}(t) + Cases_{US}(t) + Deaths_{US}(t) + SI_{CBSA}(t),
\end{equation*}
where $Cases_{CBSA}(t)$, $Deaths_{CBSA}(t)$, $Cases_{US}(t)$, and $Deaths_{US}(t)$ denote the number of cases and deaths in the corresponding CBSA and the entire USA on time $t$, which is aggregated monthly. 

Data about the number of cases and deaths in each CBSA and for the entire USA were collected from the New York Times Github page\footnote{\url{https://github.com/nytimes/covid-19-data}}. The data were provided on the county scale and for each day, and were aggregated into monthly values for each CBSA. The number of cases and deaths for the four CBSAs are shown in Figure \ref{fig:cases_deaths}. 

The Oxford Covid-19 Government Response Tracker (OxCGRT) \footnote{\url{https://www.bsg.ox.ac.uk/research/research-projects/covid-19-government-response-tracker}} collects systematic information on policy measures that governments have taken to tackle COVID-19. The different policy responses are tracked since 1 January 2020, cover more than 180 countries and are coded into 23 indicators, such as school closures, travel restrictions, vaccination policy. These policies are recorded on a scale to reflect the extent of government action, and scores are aggregated into a suite of policy indices. 
The stringency index $SI_{CBSA}(t)$ is a composite metric that measures the strictness of COVID-19 policies calculated using data collected in OxCGRT \cite{hale2021global}, and are provided at the state levels for the United States. More specifically, the stringency index takes into  account: 
\begin{enumerate}
    \item closings of schools and universities; scaled from 0 (no measures) to 3 (required closing)
    \item closings of workplaces; scaled from 0 to 3
    \item cancelling of public events; scaled from 0 to 2
    \item limits on gatherings scaled from 0 to 4
    \item closing of public transport scaled from 0 to 2
    \item orders to "shelter-in-place" and otherwise confine to the home scaled from 0 to 3
    \item restrictions on internal movement between cities/regions scaled between 0 and 2
    \item restrictions on international travel scaled from 0 to 4
    \item presence of public info campaigns scaled between 0 and 2
\end{enumerate}
More details are provided in the codebook in the github webpage \footnote{\url{https://github.com/OxCGRT/covid-policy-tracker/blob/master/documentation/codebook.md}}. The stringency index for each CBSA are shown in the bottom row of Figure \ref{fig:cases_deaths}. While all cities had high stringency until late 2020, the rollout of vaccines in early 2021 have significantly lowered the stringency.

\subsection{Model estimation results}
\begin{figure}
\centering
\includegraphics[width=.8\linewidth]{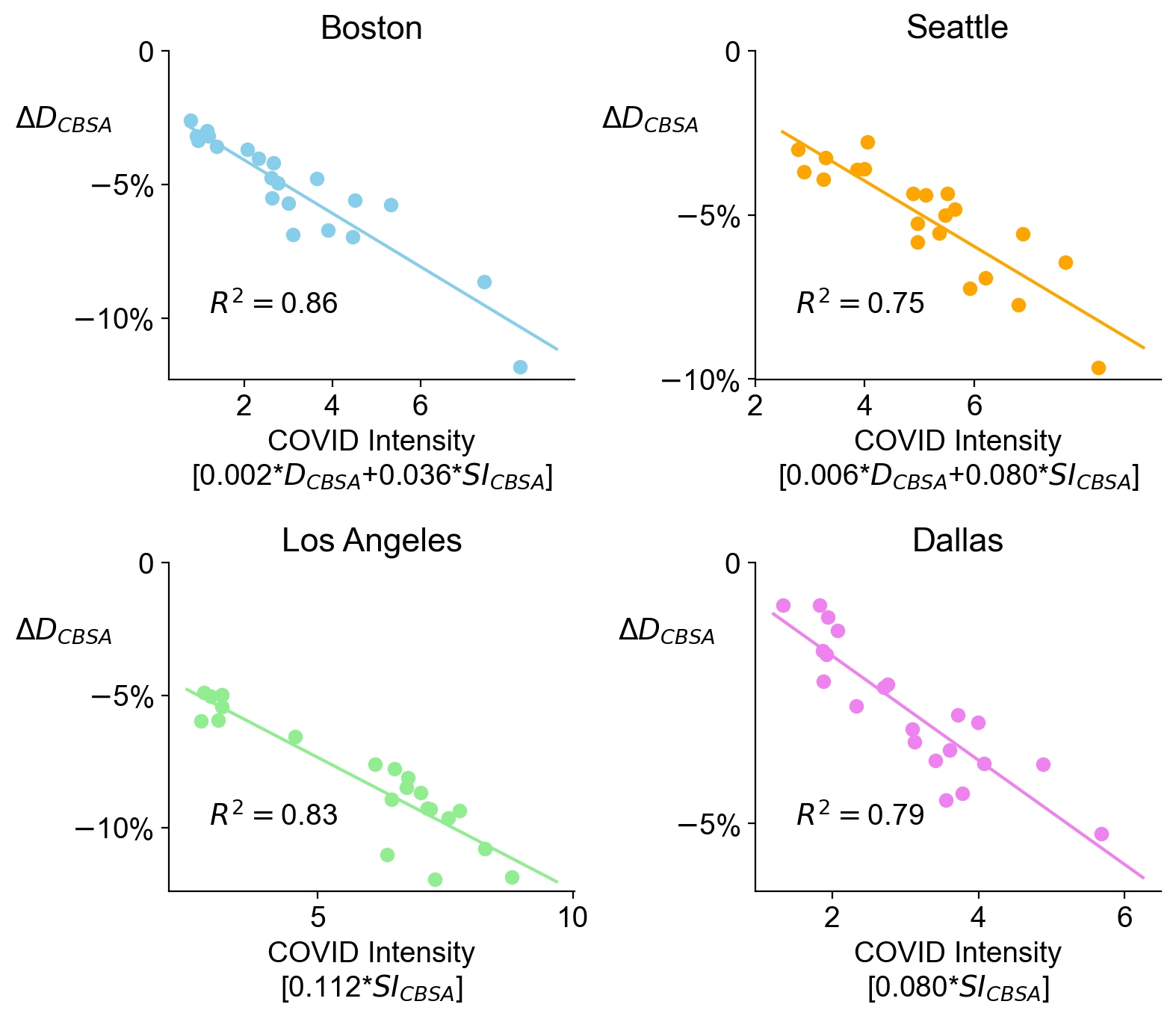}
\caption{Relationship between stringency index and reduction in income diversity in each of the CBSAs.}
\label{fig:stringency_diversity}
\end{figure}

The regression model results for the effects of COVID-19 intensity on income diversity are shown in Table \ref{table:regC1}. We observe that the stringency index is significant for all CBSAs with a negative coefficient, which indicates that stricter the COVID-19 policies, the less diverse urban encounters become. In addition to the stringency index, the number of deaths at the CBSA and federal levels are also significant for Boston and Seattle. Both coefficients are negative, which indicate that when the monthly number of deaths are higher, the less diverse urban encounters become. 
To remove insignificant variables from the model, we tested a more simpler version with the form:
\begin{equation*}
    \Delta D_{CBSA}(t) \sim Deaths_{CBSA}(t) + SI_{CBSA}(t).
\end{equation*}
The model estimation results are shown in Table \ref{table:regC2}. The significance of the variables nor their direction are consistent with the first version of the model. The constants for Boston and Los Angeles are significantly negative, indicating that in the hypothetical scenario where there are zero monthly COVID-19 deaths and zero stringency of policies, the income diversity will have a negative change compared to 2019. Given that the scenario where we completely eliminate COVID-19 cases and deaths as well as social distancing policies in the near future with the coronavirus becoming an endemic disease, this result suggests that there could be a long-lasting effect of the pandemic on the income diversity of urban encounters. 
Regression results when we use only the stringency index is shown in Figure \ref{fig:stringency_diversity}.

\begin{table}
\centering
\begin{tabular}{@{\extracolsep{5pt}}lcccc}
\\[-1.8ex]\hline
\hline \\[-1.8ex]
& \multicolumn{4}{c}{\textit{Dependent variable: $\Delta D_{CBSA} (t)$}} \
\cr \cline{2-5}
\\[-1.8ex] & Boston & Seattle & Los Angeles & Dallas \\
\hline \\[-1.8ex]
 Constant & -2.110$^{***}$ & -0.213$^{}$ & -1.127$^{}$ & -0.027$^{}$ \\
  & (0.542) & (0.960) & (0.886) & (0.665) \\
   Positives (CBSA) & 0.000$^{}$ & 0.000$^{}$ & -0.000$^{}$ & -0.000$^{}$ \\
  & (0.000) & (0.000) & (0.000) & (0.000) \\
 Positives (USA) & -0.000$^{*}$ & -0.000$^{**}$ & -0.000$^{}$ & 0.000$^{}$ \\
  & (0.000) & (0.000) & (0.000) & (0.000) \\
  Deaths (CBSA) & -0.003$^{***}$ & -0.015$^{***}$ & 0.000$^{}$ & 0.001$^{}$ \\
  & (0.000) & (0.003) & (0.000) & (0.001) \\
 Deaths (USA) & 0.000$^{**}$ & 0.000$^{***}$ & -0.000$^{}$ & 0.000$^{}$ \\
  & (0.000) & (0.000) & (0.000) & (0.000) \\
Stringency Index (CBSA) & -0.039$^{***}$ & -0.080$^{***}$ & -0.128$^{***}$ & -0.077$^{***}$ \\
  & (0.011) & (0.017) & (0.014) & (0.016) \\
\hline \\[-1.8ex]
 Observations & 21 & 21 & 21 & 21 \\
 $R^2$ & 0.906 & 0.853 & 0.914 & 0.820 \\
 Adjusted $R^2$ & 0.875 & 0.805 & 0.885 & 0.760 \\
\hline
\hline \\[-1.8ex]
\textit{Note:} & \multicolumn{4}{r}{$^{*}$p$<$0.1; $^{**}$p$<$0.05; $^{***}$p$<$0.01} \\
\end{tabular}
\caption{Regression results for $\Delta D_{CBSA}(t)$ using COVID-19 intensity and policy measures.} 
\label{table:regC1}
\end{table}

\begin{table} \centering
\begin{tabular}{@{\extracolsep{5pt}}lcccc}
\\[-1.8ex]\hline
\hline \\[-1.8ex]
& \multicolumn{4}{c}{\textit{Dependent variable: $\Delta D_{CBSA} (t)$}} \
\cr \cline{2-5}
\\[-1.8ex] & Boston & Seattle & Los Angeles & Dallas \\
\hline \\[-1.8ex]
 Constant & -2.073$^{***}$ & 0.041$^{}$ & -2.343$^{***}$ & 0.215$^{}$ \\
  & (0.500) & (0.727) & (0.682) & (0.460) \\
 Deaths (CBSA) & -0.002$^{***}$ & -0.007$^{***}$ & -0.000$^{}$ & 0.000$^{}$ \\
  & (0.000) & (0.002) & (0.000) & (0.000) \\
 Stringency Index (CBSA) & -0.036$^{***}$ & -0.081$^{***}$ & -0.112$^{***}$ & -0.080$^{***}$ \\
  & (0.012) & (0.014) & (0.014) & (0.010) \\
\hline \\[-1.8ex]
 Observations & 21 & 21 & 21 & 21 \\
 $R^2$ & 0.863 & 0.752 & 0.827 & 0.786 \\
 Adjusted $R^2$ & 0.847 & 0.725 & 0.807 & 0.762 \\
\hline
\hline \\[-1.8ex]
\textit{Note:} & \multicolumn{4}{r}{$^{*}$p$<$0.1; $^{**}$p$<$0.05; $^{***}$p$<$0.01} \\
\end{tabular}
\caption{Regression results for $\Delta D_{CBSA}(t)$ using only COVID-19 local deaths and policy strictness measures.} 
\label{table:regC2}
\end{table}

\subsection{Robustness of results via time series modeling}

\begin{table} \centering
\begin{tabular}{@{\extracolsep{5pt}}lccc}
\\[-1.8ex]\hline
\hline \\[-1.8ex]
& \multicolumn{3}{c}{\textit{$\Delta D_{CBSA} (t)$}} \
\cr \cline{2-4}
\\[-1.8ex] & ADF Statistic & p-value & Stationary? \\
\hline \\[-1.8ex]
 Boston & -13.43 & 4.02$^{-25}$ & Yes  \\
 Seattle & -0.66 & 0.85 & No \\
 Los Angeles & -2.06 & 0.25 & No  \\
 Dallas & -1.06 & 0.72 & No \\
\hline
\hline \\[-1.8ex]
\end{tabular}
\caption{Augmented Dickey Fuller test for $\Delta D_{CBSA}(t)$.} 
\label{table:adtest}
\end{table}

\begin{figure}
\centering
\subfloat[Boston]{\includegraphics[width=\linewidth]{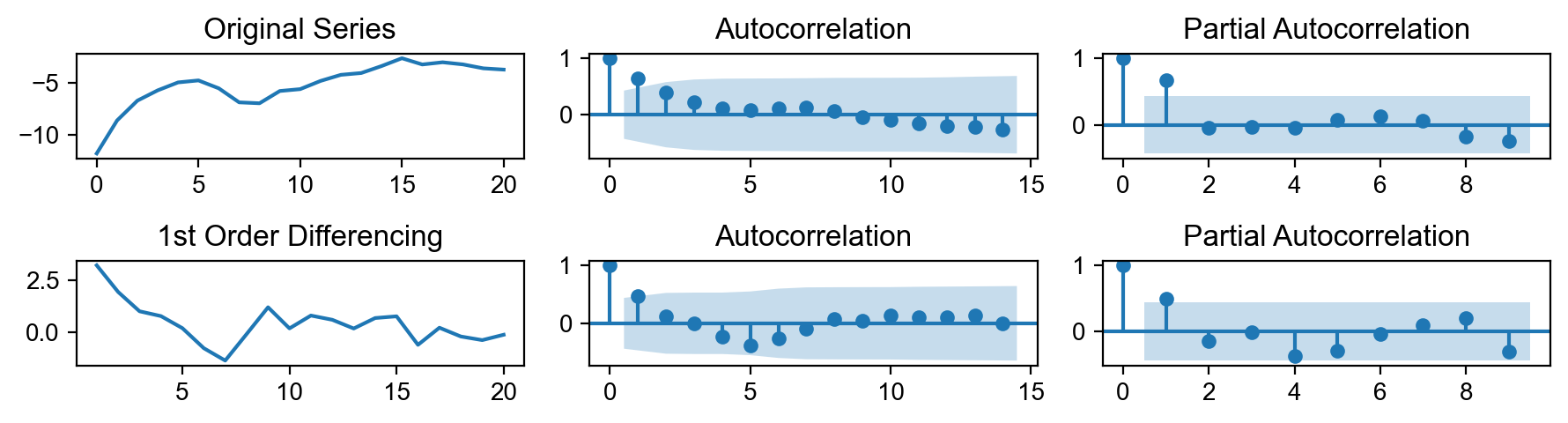}}\\
\subfloat[Seattle]{\includegraphics[width=\linewidth]{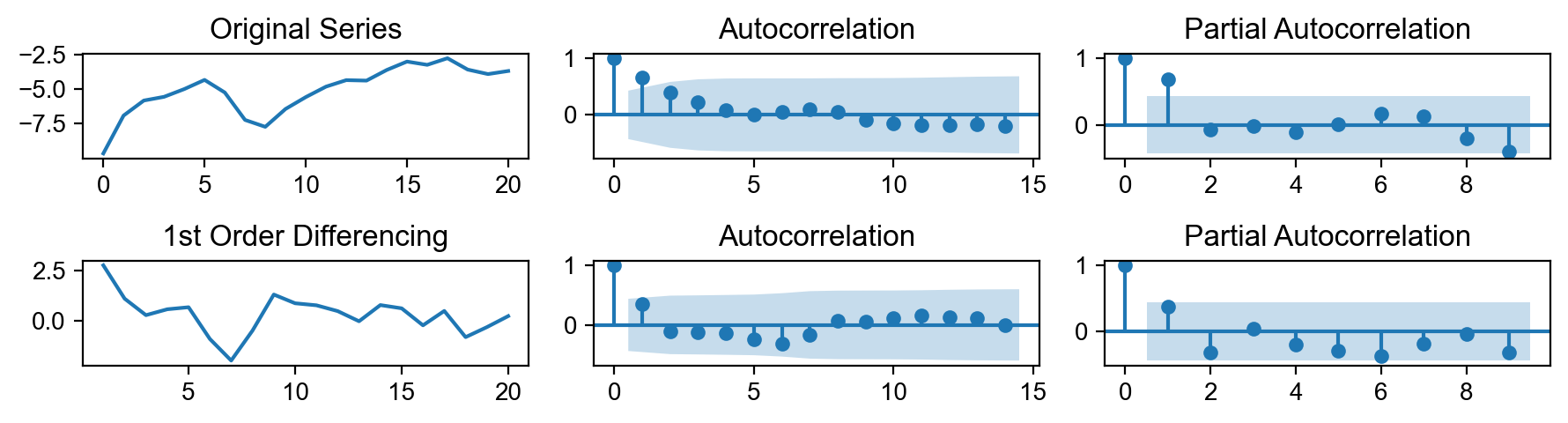}}\\
\subfloat[Los Angeles]{\includegraphics[width=\linewidth]{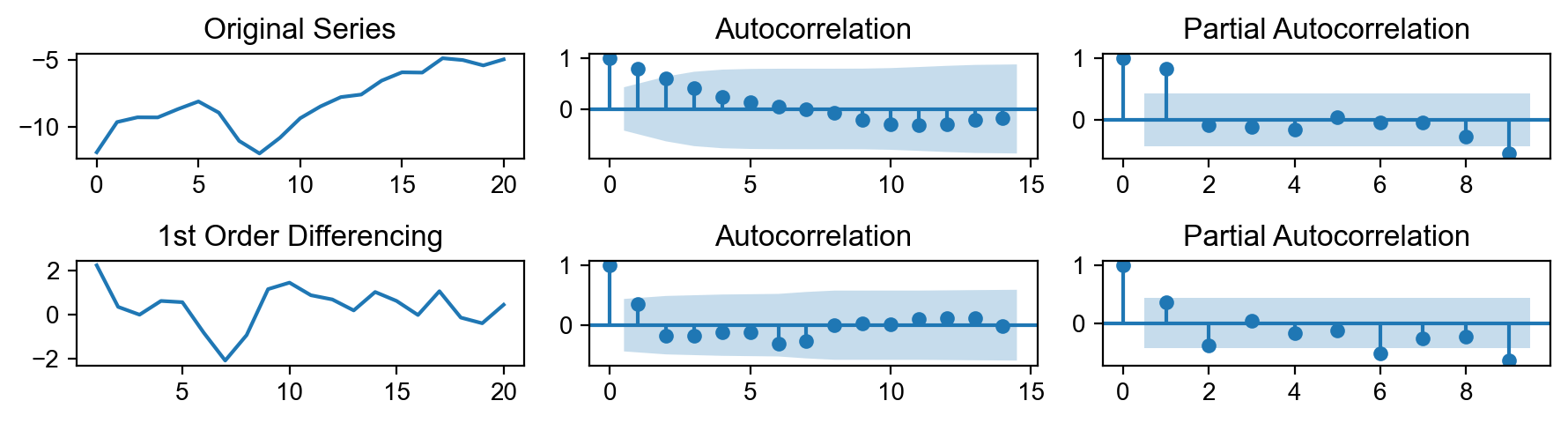}}\\
\subfloat[Dallas]{\includegraphics[width=\linewidth]{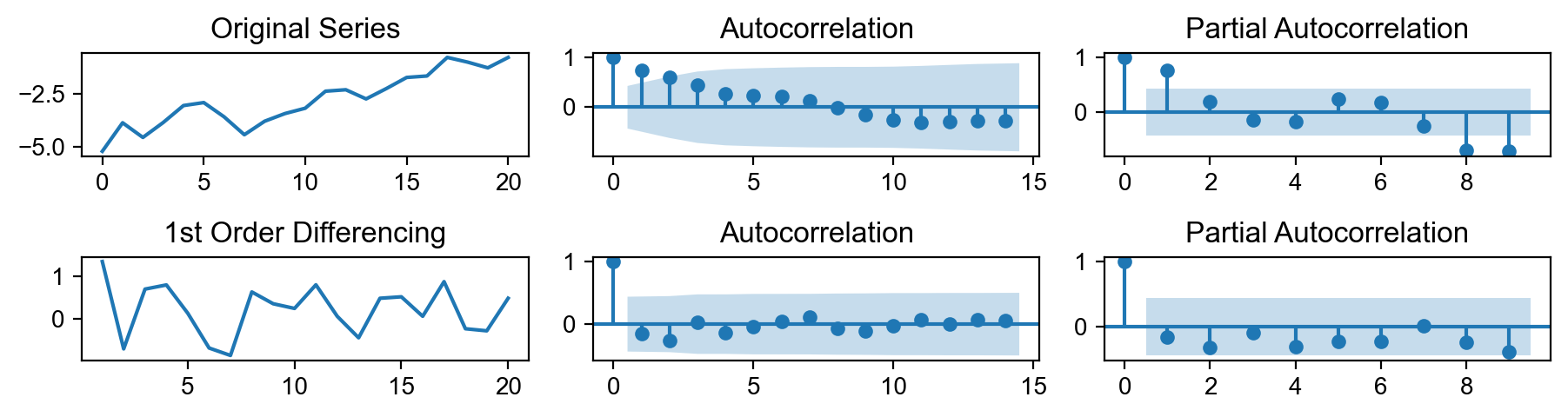}}
\caption{Autocorrelation and partial autocorrelation of original series and 1st order of differencing of $\Delta D_{CBSA}(t)$ for Boston, Seattle, Los Angeles, and Dallas.}
\label{fig:adftest}
\end{figure}

Since the variables used in this model are temporal data, including the decrease in diversity as well as COVID-19 related data, it is important to check stationarity, autocorrelation, and partial autocorrelation, and if applicable test whether such temporal dependencies affect the outcomes of the results. 
To check the stationarity of $\Delta D_{CBSA}(t)$, we conduct the Augmented Dickey Fuller (ADF) test \cite{mushtaq2011augmented}. Table \ref{table:adtest} shows the ADF statistic, p-value, and whether the time series is determined to be stationary or not. The results show that except for Boston, the time series are non-stationary, thus we need to do some differencing. Figure \ref{fig:adftest} shows the autocorrelation and partial autocorrelation of the data $\Delta D_{CBSA}(t)$ under no differencing and 1st order differencing for the four CBSAs. We observe that for all three cities except Boston (which requires no differencing), 1st order differencing is enough to obtain no autocorrelation beyond 1 time step. 

To model the temporal dynamics, we apply an ARIMA$(p,d,q)$ model with covariates (number of local COVID-19 deaths and local stringency index). 
The model parameters $p,d,q$ of the ARIMA model, each corresponding to the autoregressive term (or the lag of the dependent variable, number of differencing needed for stationary time series, and the lagged forecast error term, respectively. 
For Boston, the ADF test shows that no differencing is needed, thus $d=0$. Under no differencing, ARIMA(1,0,0) and ARIMA(0,0,1) were tested for Boston and only the MA term was significant (shown in first column in Table \ref{table:regD1}). 
Using the ARIMA(0,0,1) model for Boston, both the he number of local deaths and the local stringency index were statistically significant with $p<0.01$, indicating robustness of the OLS results in Table \ref{table:regC2}. 
For the other three cities, since $d=1$ was determined using the ADF test, ARIMA(1,1,0), ARIMA(0,1,1), and ARIMA(1,1,1) were modeled and the statistical significance of autoregressive and moving average terms were tested. 
For Seattle, as shown in the second column in Table \ref{table:regD1}, the moving average term showed statistical significance with $p<0.05$ and both the number of local deaths and the local stringency index were also statistically significant with $p<0.05$, indicating robustness of the OLS results in Table \ref{table:regC2}. For Los Angeles and Dallas, both the autoregressive and moving average terms were statistically insignificant, indicating that the dependent variable can be modeled using OLS instead of time series models. To summarize, for Boston and Seattle $\Delta D_{CBSA}$ could be modeled as a moving average process but the coefficients and significance of the independent variables were consistent with the OLS results in Table \ref{table:regC2}. For Los Angeles and Dallas, the temporal components were insignificant, therefore the results in Table \ref{table:regC2} are robust.

\begin{table} \centering
\begin{tabular}{@{\extracolsep{5pt}}lcccc}
\\[-1.8ex]\hline
\hline \\[-1.8ex]
& \multicolumn{4}{c}{\textit{Dep. variable: $\Delta D_{CBSA} (t)$}} \
\cr \cline{2-5}
\\[-1.8ex] & Boston & Seattle & Los Angeles & Dallas \\
Best model & ARIMA(0,0,1) & ARIMA(0,1,1) & None & None \\
\hline \\[-1.8ex]
 Constant & -2.073$^{***}$ & - & - & -  \\
 Deaths (CBSA) & -0.002$^{***}$ & -0.003$^{**}$  & - & - \\
 Stringency Index (CBSA) & -0.036$^{***}$ & -0.070$^{**}$ & - & -   \\
 Autoregressive 1 lag & - & - & - & - \\ 
 Moving average 1 lag & 0.630$^{***}$ & 0.736$^{**}$& - & - \\ 
 $\sigma^2$ & 0.374$^{***}$ & 0.458$^{**}$ & - & - \\ 
 \hline \\[-1.8ex]
 Observations & 21 & 21 & - & - \\
 $AIC$ & 58.849 & 49.934 & - & -  \\
\hline
\hline \\[-1.8ex]
\textit{Note:} & \multicolumn{4}{r}{$^{*}$p$<$0.1; $^{**}$p$<$0.05; $^{***}$p$<$0.01} \\
\multicolumn{5}{r}{Both AR and MA terms were insignificant for Los Angeles and Dallas.} \\
\end{tabular}
\caption{ARIMA regression results for $\Delta D_{CBSA}(t)$ using COVID-19 local deaths and policy strictness measures.} 
\label{table:regD1}
\end{table}

\section{Software}
Analysis was conducted using Python, Jupyter Lab, and the following libraries and software:
\begin{itemize}
    \item \texttt{NumPy} \cite{harris2020array} for general computation on Python.
    \item \texttt{Pandas} \cite{mckinney2011pandas} for loading, transforming, and analyzing data tables.
    \item \texttt{Matplotlib} \cite{hunter2007matplotlib} for creating plots and figures. 
    \item \texttt{GeoPandas} \cite{jordahl2014geopandas} for spatial analysis and plotting map figures. 
    \item \texttt{Statsmodels} \cite{seabold2010statsmodels} for statistical modeling and econometric analysis. 
    \item A Python implementation of the R \texttt{Stargazer} multiple regression model creation tool\footnote{\url{https://github.com/mwburke/stargazer}} was used to create the regression tables.
\end{itemize}

\bibliographystyle{plain}